# The EChO science case

Giovanna Tinetti[1], Pierre Drossart[2], Paul Eccleston[3], Paul Hartogh[4], Kate Isaak[5], Martin Linder[5], Christophe Lovis[6], Giusi Micela[7], Marc Ollivier[8,2], Ludovic Puig[5], Ignasi Ribas[9], Ignas Snellen[10], Bruce Swinyard[3,1].

France Allard[11], Joanna Barstow[12], James Cho[13], Athena Coustenis[2], Charles Cockell[14], Alexandre Correia[15], Leen Decin[16], Remco de Kok[17], Pieter Deroo[18], Therese Encrenaz[2], Francois Forget[19], Alistair Glasse[20], Caitlin Griffith[21], Tristan Guillot[22], Tommi Koskinen[21], Helmut Lammer[23], Jeremy Leconte[24,19], Pierre Maxted[25], Ingo Mueller-Wodarg[26], Richard Nelson[13], Chris North[27], Enric Pallé[28], Isabella Pagano[29], Guseppe Piccioni[30], David Pinfield[31], Franck Selsis[32], Alessandro Sozzetti[33], Lars Stixrude[1], Jonathan Tennyson[1], Diego Turrini[30], Mariarosa Zapatero-Osorio[34].

Jean-Philippe Beaulieu[35], Denis Grodent[36], Manuel Guedel[37], David Luz[38], Hans Ulrik Nørgaard-Nielsen[39], Tom Ray[40], Hans Rickman[41,42], Avri Selig[17], Mark Swain[18].

Marek Banaszkiewicz[41], Mike Barlow[1], Neil Bowles[12], Graziella Branduardi-Raymont[43], Vincent Coudé du Foresto[2], Jean-Claude Gerard[36], Laurent Gizon[4], Allan Hornstrup[45], Christopher Jarchow[4], Franz Kerschbaum[37], Géza Kovacs[45], Pierre-Olivier Lagage[46], Tanya Lim[3], Mercedes Lopez-Morales[9], Giuseppe Malaguti[47], Emanuele Pace[48], Enzo Pascale[27], Bart Vandenbussche[16], Gillian Wright[20], Gonzalo Ramos Zapata[49].

Alberto Adriani[30], Ruymán Azzollini[40], Ana Balado[49], Ian Bryson[20], Raymond Burston[4], Josep Colomé[9], Vincent Coudé du Foresto[2], Martin Crook[3], Anna Di Giorgio[30], Matt Griffin[27], Ruud Hoogeveen[17], Roland Ottensamer[37], Ranah Irshad[3], Kevin Middleton[3], Gianluca Morgante[47], Frederic Pinsard[46], Mirek Rataj[41], Jean-Michel Reess[2], Giorgio Savini[1], Jan-Rutger Schrader[17], Richard Stamper[3], Berend Winter[43].

L. Abe[22], M. Abreu[63], N. Achilleos[1], P. Ade[27], V. Adybekian[62], L. Affer[7], C. Agnor[13], M. Agundez[32], C. Alard[35], J. Alcala[58], C. Allende Prieto[28], F.J. Alonso Floriano[68], F. Altieri[30], C.A. Alvarez Iglesias[28], P. Amado[66], A. Andersen[51], A. Aylward[1], C. Baffa[56], G. Bakos[79], P. Ballerini[29], M. Banaszkiewicz[41], R. J. Barber[1], D. Barrado[34,49], E.J. Barton[1], V. Batista[35], G. Bellucci[30], J.A. Belmonte Avilés[28], D. Berry[63], B. Bézard[2], D. Biondi[30], M. Błęcka[41], I. Boisse[62], B. Bonfond[36], P. Bordé[8], P. Börner[4], H. Bouy[34,49], L. Brown[18], L. Buchhave[51], J. Budaj[65], A. Bulgarelli[47], M. Burleigh[74], A. Cabral[63], M.T. Capria[30], A. Cassan[35], C. Cavarroc[46], C. Cecchi-Pestellini[7], R. Cerulli[30], J. Chadney[26], S. Chamberlain[63], S. Charnoz[46], N. Christian Jessen[44], A. Ciaravella[7], A. Claret[66], R. Claudi[59], A. Coates[43], R. Cole[43], A. Collura[7], D. Cordier[52], E. Covino[58], C. Danielski[8], M. Damasso[33], H.J. Deeg[28], E. Delgado-Mena[62], C. Del Vecchio[56], O. Demangeon[8], A. De Sio[48], J. De Wit[78], M. Dobrijévic[32], P. Doel[1], C. Dominic[61], E. Dorfi[37], S. Eales[27], C. Eiroa[67], M. Espinoza Contreras[28], M. Esposito[28], V. Eymet[32], N. Fabrizio[30], M. Fernández[66], B. Femenía Castella[28], P. Figueira[62], G. Filacchione[30], L. Fletcher[12], M. Focardi[48], S. Fossey[1], P. Fouqué[54], J. Frith[31], M. Galand[26], L. Gambicorti[56], P. Gaulme[8], R.J. García López[28], A. Garcia-Piquer[9], W. Gear[27], J.-C. Gerard[36], L. Gesa[9], E. Giani[56], F. Gianotti[47], M. Gillon[36], E. Giro[59], M. Giuranna[30], H. Gomez[27], I. Gomez-Leal[32], J. Gonzalez Hernandez[28], B. González Merino[28], R. Graczyk[41], D. Grassi[30], J. Guardia[9], P. Guio[1], J. Gustin[36], P. Hargrave[27], J. Haigh[26], E. Hébrard[32], U. Heiter[42], R. L. Heredero[63], E. Herrero[9], F. Hersant[32], D. Heyrovsky[50], M. Hollis[1], B. Hubert[36], R. Hueso[69], G. Israelian[28], N. Iro[55], P. Irwin[12], S. Jacquemoud[53], G. Jones[43], H. Jones[31], K. Justtanont[70], T. Kehoe[15], F. Kerschbaum[37], E. Kerins[75], P. Kervella[2], D. Kipping[78], T. Koskinen[21], N. Krupp[4], O. Lahav[1], B. Laken[28], N. Lanza[29], E. Lellouch[2], G. Leto[29], J. Licandro Goldaracena[28], C. Lithgow-Bertelloni[1], S.J. Liu[30], U. Lo Cicero[7], N. Lodieu[28], P. Lognonné[53], M. Lopez-Puertas[66], M.A. Lopez-Valverde[66], I. Lundgaard Rasmussen[44], A. Luntzer[37], P. Machado[63], C. MacTavish[71], A. Maggio[7], J.-P. Maillard[35], W. Magnes[23], J. Maldonado[67], U. Mall[4], J.-B. Marquette[35], P. Mauskopf[21], F. Massi[56], A.-S. Maurin[32], A. Medvedev[4], C. Michaut[53], P. Miles-Paez[28], M. Montalto[62], P. Montañés Rodríguez[28], M. Monteiro[62], D. Montes[68], H. Morais[15], J. C. Morales[2], M. Morales-Calderón[34,49], G. Morello[1], A. Moro Martín[34,49], J. Moses[80], A. Moya Bedon[34,49], F. Murgas Alcaino[28], E. Oliva[56], G. Orton[18], F. Palla[56], M. Pancrazzi[48], E. Pantin[46], V. Parmentier[22], H. Parviainen[28], K.Y. Peña Ramírez[28], J. Peralta[63], S. Perez-Hoyos[69], R. Petrov[22], S. Pezzuto[30], R. Pietrzak[41], E. Pilat-Lohinger[37], N. Piskunov[42], R. Prinja[1], L. Prisinzano[7], I. Polichtchouk[13], E. Poretti[57], A. Radioti[36], A. A. Ramos[28], T. Rank-Lüftinger[37], P. Read[12], K. Readorn[56], R. Rebolo López[28], J. Rebordão[63], M. Rengel[4], L. Rezac[4], M. Rocchetto[1], F. Rodler[9], V.J. Sánchez Béjar[28], A. Sanchez Lavega[69], E. Sanromá, N. Santos[62], J. Sanz Forcada[34,49], G. Scandariato[7], F.-X. Schmider[22], A. Scholz[40], S. Scuderi[29], J. Sethenadh[4],

S. Shore[60], A. Showman[21], B. Sicardy[2], P. Sitek[41], A. Smith[43], L. Soret[36], S. Sousa[62], A. Stiepen[36], M. Stolarski[41], G. Strazzulla[29], H.M, Tabernero[68], P. Tanga[22], M. Tecsa[12], J. Temple[12], L. Terenzi[47], M. Tessenyi[1], L. Testi[64], S. Thompson[71], H. Thrastarson[77], B.W. Tingley[28], M. Trifoglio[47], J. Martín Torres[34,49], A. Tozzi[56], D. Turrini[30], R. Varley[1], F. Vakili[22], M. de Val-Borro[4], M.L. Valdivieso[28], O. Venot[16], E. Villaver[67], S. Vinatier[2], S. Viti[1], I. Waldmann[1], D. Waltham[73], D. Ward-Thompson[76], R. Waters[17], C. Watkins[13], D. Watson[51], P. Wawer[41], A. Wawrzaszk[41], G. White[72], T. Widemann[2], W. Winek[41], T. Wiśniowski[41], R. Yelle[21], Y. Yung[77], S.N. Yurchenko[1],

[1]: Department of Physics & Astronomy, University College London, WC1E6BT London, UK
[2]: LESIA, Observatoire de Paris, Meudon, France
[3]: STFC Rutherford Appleton Laboratory, Harwell Campus, Didcot OX11 0QX, UK
[4]: Max-Planck-Institut für Sonnensystemforschung, Katlenburg-Lindau, Germany
[5]: European Space Agency-ESTEC, The Netherlands
[6]: Geneva Observatory, Geneva, Switzerland
[7]: INAF: Osservatorio Astronomico di Palermo G.S. Vaiana, Palermo, Italy
[8]: Institut d'Astrophysique Spatiale, Orsay, France
[9]: Institut d'Estudis Espacials de Catalunya (ICE-CSIC), Barcelona, Spain
[10]: Leiden University, Leiden, The Netherlands
[11]: Ecole Normale Superieure, Lyon, France
[12]: Oxford University, Oxford, UK
[13]: Queen Mary University London, London, UK
[14]: Royal Observatory Edinburgh, Edinburgh, UK
[15]: IN, Portugal
[16]: University of Leuven, Belgium
[17]: SRON, Institute for Space Research, The Netherlands
[18]: NASA Jet Propulsion Laboratory, Pasadena, CA, US
[19]: LMD, Jussieu, Paris, France
[20]: STFC UK-ATC, Edinburgh, UK
[21]: University of Arizona, Tucson, AZ, US
[22]: Observatoire de Nice, Nice, France
[23]: IWF, Graz, Austria
[24]: Canadian Institute of Theoretical Astrophysics, University of Toronto, Toronto, CA
[25]: Keele University, UK
[26]: Imperial College, London, UK
[27]: Cardiff University, Cardiff, UK
[28]: Instituto de Astrofisica de Canarias, La Laguna, Tenerife, Spain
[29]: INAF- OAT, Catania, Italy
[30]: INAF-IAPS, Rome, Italy
[31]: University of Hertfordshire, Hatfield, UK
[32]: Université de Bordeaux, Bordeaux, France
[33]: INAF Torino, Italy
[34]: CAB, Madrid, Spain
[35]: Institut d'Astrophysique de Paris, France
[36]: Université de Liège, Belgium
[37]: University of Vienna, Austria
[38]: Universidade de Lisboa, Portugal
[39]: DSRI, Denmark
[40]: Dublin Institute for Advanced Studies, Dublin, Ireland
[41]: Space Research Centre, Polish Academy of Science, Warsaw, Poland
[42]: Department of Physics and Astronomy, Uppsala University, Sweden
[43]: Mullard Space Science Laboratory, University College London, Surrey RH5 6NT, UK
[44]: DTU Space, Denmark
[45]: Konkoly Observatory, Budapest, Hungary
[46]: Centre Energie Atomique – Saclay, France
[47]: INAF – IASF – Bologna, Italy
[48]: Università di Firenze, Firenze, Italy
[49]: INTA, Spain
[50]: Charles University, Czech Republic
[51]: DARK Cosmology Center, Denmark
[52]: Observatoire de Besancon, Besancon, France
[53]: IPGP, France
[54]: LATT, Toulouse, France

[55]: University of Hamburg, Germany
[56]: INAF – Arcetri, Firenze, Italy
[57]: INAF – Brera, Milano, Italy
[58]: INAF – Capodimonte, Napoli, Italy
[59]: INAF – Padova, Italy
[60]: Università di Pisa, Pisa, Italy
[61]: University of Amsterdam, Amsterdam, The Netherlands.
[62]: CAUP, Portugal
[63]: CAAUL, Portugal
[64]: ESO, Garching, Germany
[65]: Slovak Academy of Sciences, Slovakia
[66]: IAA, Spain
[67]: UAM, Spain
[68]: UCM, Spain
[69]: UPV, Spain
[70]: Onsala Space Observatory, Sweden
[71]: Cambridge University, Cambridge, UK
[72]: Open University, Milton Keynes, UK
[73]: Royal Holloway University of London, Surrey, UK
[74]: University of Leicester, Leicester, UK
[75]: University of Manchester, Manchester, UK
[76]: University of Lancaster, Lancaster, UK
[77]: California Institute of Technology, Pasadena, CA, US
[78]: Harvard Smithsonian Center for Astrophysics, Cambridge, MA, US
[79]: Princeton University, NJ, US
[80]: SSI, TX, US

# Abstract

The discovery of almost two thousand exoplanets has revealed an unexpectedly diverse planet population. We see gas giants in few-day orbits, whole multi-planet systems within the orbit of Mercury, and new populations of planets with masses between that of the Earth and Neptune – all unknown in the Solar System. Observations to date have shown that our Solar System is certainly not representative of the general population of planets in our Milky Way. The key science questions that urgently need addressing are therefore: *What are exoplanets made of? Why are planets as they are? How do planetary systems work and what causes the exceptional diversity observed as compared to the Solar System*? The EChO (Exoplanet Characterisation Observatory) space mission was conceived to take up the challenge to explain this diversity in terms of formation, evolution, internal structure and planet and atmospheric composition. This requires in-depth spectroscopic knowledge of the atmospheres of a large and well-defined planet sample for which precise physical, chemical and dynamical information can be obtained.

In order to fulfil this ambitious scientific program, EChO was designed as a dedicated survey mission for transit and eclipse spectroscopy capable of observing a large, diverse and well-defined planet sample within its four-year mission lifetime. The transit and eclipse spectroscopy method, whereby the signal from the star and planet are differentiated using knowledge of the planetary ephemerides, allows us to measure atmospheric signals from the planet at levels of at least $10^{-4}$ relative to the star. This can only be achieved in conjunction with a carefully designed stable payload and satellite platform. It is also necessary to provide broad instantaneous wavelength coverage to detect as many molecular species as possible, to probe the thermal structure of the planetary atmospheres and to correct for the contaminating effects of the stellar photosphere. This requires wavelength coverage of at least 0.55 to 11 μm with a goal of covering from 0.4 to 16 μm. Only modest spectral resolving power is needed, with R~300 for wavelengths less than 5 μm and R~30 for wavelengths greater than this.

The transit spectroscopy technique means that no spatial resolution is required. A telescope collecting area of about 1 $m^2$ is sufficiently large to achieve the necessary spectro-photometric precision: for the Phase A study a 1.13 $m^2$ telescope, diffraction limited at 3 μm has been adopted. Placing the satellite at L2 provides a cold and stable thermal environment as well as a large field of regard to allow efficient time-critical observation of targets randomly distributed over the sky. EChO has been conceived to achieve a single goal: exoplanet spectroscopy. The spectral coverage and signal-to-noise to be achieved by EChO, thanks to its high stability and dedicated design, would be a game changer by allowing atmospheric composition to be measured with unparalleled exactness: at least a factor 10 more precise and a factor 10 to 1000 more accurate than current observations. This would enable the detection of molecular abundances three orders of magnitude lower than currently possible and a fourfold increase from the handful of molecules detected to date. Combining these data with estimates of planetary bulk compositions from accurate measurements of their radii and masses would allow degeneracies associated with planetary interior modelling to be broken, giving unique insight into the interior structure and elemental abundances of these alien worlds.

EChO would allow scientists to study exoplanets both as a population and as individuals. The mission can target super-Earths, Neptune-like, and Jupiter-like planets, in the very hot to temperate zones (planet temperatures of 300 K - 3000 K) of F to M-type host stars. The EChO core science would be delivered by a three-tier survey. The EChO *Chemical Census*: This is a broad survey of a few-hundred exoplanets, which allows us to explore the spectroscopic and chemical diversity of the exoplanet population as a whole. The EChO *Origin*: This is a deep survey of a subsample of tens of exoplanets for which significantly higher signal to noise and spectral resolution spectra can be obtained to explain the origin of the exoplanet diversity (such as formation mechanisms, chemical processes, atmospheric escape). The EChO *Rosetta Stones*: This is an ultra-high accuracy survey targeting a subsample of select exoplanets. These will be the bright "benchmark" cases for which a large number of measurements would be taken to explore temporal variations, and to obtain two and three dimensional spatial information on the atmospheric conditions through eclipse-mapping techniques.

If EChO were launched today, the exoplanets currently observed are sufficient to provide a large and diverse sample. The Chemical Census survey would consist of > 160 exoplanets with a range of planetary sizes, temperatures, orbital parameters and stellar host properties. Additionally, over the next ten years, several new ground- and space-based transit photometric surveys and missions will come on-line (e.g. NGTS, CHEOPS, TESS, PLATO), which will specifically focus on finding bright, nearby systems. The current rapid rate of discovery would allow the target list to be further optimised in the years prior to EChO's launch and enable the atmospheric characterisation of hundreds of planets.

**Table 1. EChO – Exoplanet Characterisation Observatory – Mission Summary**

| | |
|---|---|
| **Key Science Questions to be Addressed** | • Why are exoplanets as they are?<br>• What are the causes for the observed diversity?<br>• Can their formation history be traced back from their current composition and evolution?<br>• How does the Solar System work compared to other planetary systems?<br>• Are planets in the Solar System special in any way? |
| **Science Objectives** | • Detection of planetary atmospheres, their composition and structure<br>• Determine vertical and horizontal temperature structure and their diurnal and seasonal variations<br>• Identify chemical processes at work (thermochemistry, photochemistry, transport quenching)<br>• Constrain planetary interiors (breaking the radius-mass degeneracy)<br>• Quantify the energy budget (albedo, temperature)<br>• Constrain formation and evolution models (evidence for migration)<br>• Detect secondary atmospheres around terrestrial planets (evolution)<br>• Investigate the impact of stellar and planetary environment on exoplanet properties |
| **EChO Core Survey** | • Three-tier survey of 150-300 transiting exoplanets from gas giants to super-Earths, in the very hot to temperate zones of F to M type host stars<br>• Target selection before launch based on ESA science team and community inputs<br>• Chemical Census: statistically complete sample detecting strongest atmospheric molecular features<br>• Origin: retrieval of vertical thermal profiles and abundances of trace gases<br>• Rosetta Stone: high signal-to-noise observations yielding refined molecular abundances, chemical gradients and atmospheric structure; diurnal and seasonal variations; presence of clouds and measurement of albedo<br>• Delivery of a homogeneous catalogue of planetary spectra |
| **EChO Observational Strategy** | • Transit and eclipse spectroscopy with broad, instantaneous, and uninterrupted spectra covering all key molecules<br>• High photometric stability on transit timescales<br>• Required SNR obtained by summing a sufficient number of transits or eclipses<br>• Large instantaneous sky coverage |
| **Payload Telescope** | • Afocal 3-mirror, off-axis Korsch-like system, 1.5 m x 1 m elliptical M1, unobstructed (effective area 1.13 $m^2$), diffraction-limited at 3 μm; <3 μm, 80% encircled energy within diameter of 1.6 arcsec. |
| **Payload Instrument** | • Highly-integrated broadband spectrometer instrument with modular architecture<br>• Common optical train for all spectrometers and the fine guidance system optical module<br>• Continuous wavelength coverage from 0.4 - 11μm in baseline design<br>• Goal wavelength coverage from 0.4 – 16 μm.<br>• Resolving powers of $\lambda/\Delta\lambda$ >300 below 5 μm, and >30 above 5 μm<br>• Passively cooled MCT detectors at ~40K for FGS and science channels < 5μm<br>• Active Ne JT Cooler provides cooling to ~28K for science channels > 5μm |
| **Spacecraft**<br>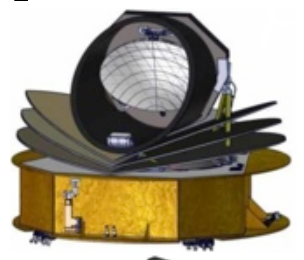<br>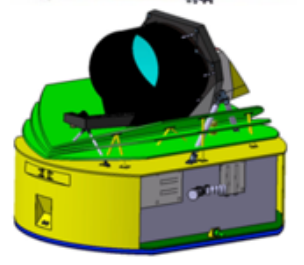 | • Launch mass ~ 1.5 tonnes<br>• Dimensions: Ø 3.6 m x 2.6 m. Designs from the two industrial studies shown to the left.<br>• Pointing requirements: coarse APE of 10 arcsec (3σ); fine APE of 1 arcsec (3σ); PDE of 20 milli-arcseconds (1σ) over 90s to 10hrs; RPE of 50 milli-arcsecond over 90s (1σ)<br>• Attitude control system: reaction wheels and cold gas system complemented by a Fine-Guidance System operating in the visible within the AOCS control loop.<br>• Thermal Control System: Passive cooling via 3 V-grooves to ≤ 47 K<br>• Telecommand, Telemetry and Communication: X-band, 35 Gbit of science data per week transmitted with a High Gain Antenna to a 35 m ESTRACK station |
| **Launcher, Orbit, Mission Phases and Operations** | • Launch from Kourou on a Soyuz-Fregat MT into L2 orbit in 2024 (possible option of launch in 2022)<br>• Nominal mission duration 4 years (goal 6 years)<br>• MOC at ESOC, SOC at ESAC, Instrument Operations and Science Data Centre distributed across consortium members states<br>• 14 hours ground contact/week: 2x2 hours for telecommand uplink and science downlink, remainder for determination of orbital parameters |
| **Data Policy** | • Short proprietary period after nominal SNR is reached, shrinking to 1 month after 3 years |

# 1. Introduction

## 1.1 Exoplanets today

Roughly 400 years ago, Galileo's observations of the Jovian moons sealed the Copernican Revolution, and the Earth was no longer considered the centre of the Universe (*Sidereus Nuncius,* 1610). We are now poised to extend this revolution to the Solar System. The detection and characterisation of exoplanets force the Sun and its cohorts to abdicate from their privileged position as the archetype of a planetary system.

Recent exoplanet discoveries have profoundly changed our understanding of the formation, structure, and composition of planets. Current statistics show that planets are common; data from the Kepler Mission and microlensing surveys indicate that the majority of stars have planets (Fressin et al. 2013; Cassan et al., 2012). Detected planets range in size from sub-Earths to larger than Jupiter (Figure 1). Unlike the Solar System, the distribution of planetary radii appears continuous (Batalha et al., 2013), with no gap between 2 to 4 Earth radii. That is, there appears to be no distinct transition from telluric planets, with a thin, if any, secondary atmosphere, to the gaseous and icy giants, which retain a substantial amount of hydrogen and helium accreted from the protoplanetary disk.

The orbital characteristics among the almost 2000 exoplanets detected also do not follow the Solar System trend, with small rocky bodies orbiting close to a G star and giant gas planets orbiting further out, in roughly circular orbits. Instead giant planets can be found within 1/10 the semi-major axis of Mercury. Planets can orbit host stars with an eccentricity well above 0.9 (e.g. HD 80606b), comparable to Halley's comet. Planets can orbit two mother stars (e.g. Kepler-34b, Kepler-35b, and Kepler-38b): this is not an oddity any more. Planetary systems appear much more diverse than expected. The Solar System template, well explained by our current understanding of planetary formation and evolution, does not seem to be generally applicable.

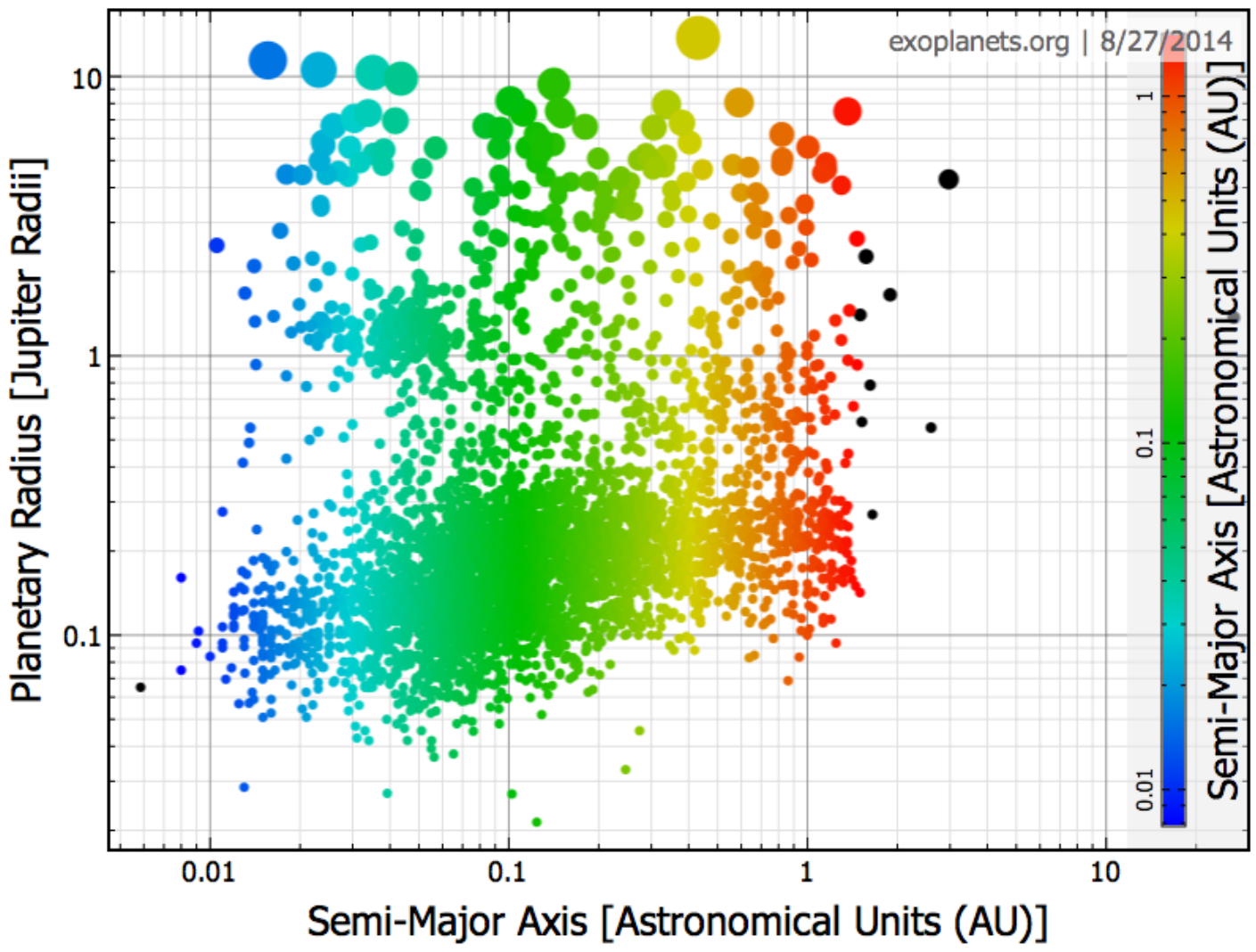

*Figure 1: Currently known exoplanets, plotted as a function of distance to the star and planetary radii (courtesy of exoplanets.org). The graph suggests a continuous distribution of planetary sizes – from sub-Earths to super-Jupiters– and planetary temperatures than span two orders of magnitude.*

The range of orbital parameters and stellar hosts translates into planetary temperatures that span two orders of magnitude. This range of temperatures arises from the range of planet-star proximities, where a year can be less than 6 Earth-hours (e.g. KOI-55b), or over 450 Earth-years (e.g. HR 8799b), and host star temperatures, which can range from 2200 K to 14000 K. Conditions not witnessed in the Solar System lead to exotic planets whose compositions we can only speculate about. Currently, we can only guess that the extraordinarily hot and rocky planets CoRoT-7b, Kepler-10b, Kepler-78b and 55 Cnc-e sport silicate compounds in the gaseous and liquid phases (Léger et al., 2011, Rouan et al., 2011). "Ocean planets" that have densities in between those of giant and rocky planets (Léger et al., 2004, Grasset et al., 2009) and effective temperatures between the triple and critical temperatures of water, i.e. between 273 and 647 K (e.g. GJ 1214b) may have large water-rich atmospheres. The "Mega-Earth", Kepler-10c (Dumusque et al., 2014), is twice the Earth's size but is seventeen times heavier than our planet, making it among the densest planets currently known.

The diversity of currently detected exoplanets not only extends the regime of known conditions, it indicates environments completely alien to the Solar System. *Observations demonstrate that the Solar System is not the paradigm in our Galaxy: one of the outstanding questions of modern astrophysics is to understand why.*

Over the past two decades, primary transit and radial velocity measurements have determined the sizes and masses of exoplanets, thereby yielding constraints on the bulk composition of exoplanets. The missions NASA-K2 and TESS and ESA-Cheops and PLATO, together with ground-based surveys, will increase by a factor of five the number of planets for which we have an accurate measurement of mass and radius. While measurements of the masses and radii of planetary systems have revealed the great diversity of planets and of the systems in which planets originate and evolve, these investigations generate a host of important questions:

(i) *What are the planets' core to atmospheric composition relationships*? The planetary density alone does not provide unique solutions. The degeneracy is higher for super-earths and small Neptunes (Valencia et al., 2013). As an example, it must be noted that a silicate-rich planet surrounded by a very thick atmosphere could have the same mass and radius as an ice-rich planet without an atmosphere (Adams & Seager 2008).

(ii) *Why are many of the known transiting gaseous planets larger than expected*? These planets are larger than expected even when the possibility that they could be coreless hydrogen-helium planets is allowed for (Bodenheimer et al. 2001, Guillot et al. 2006). There is missing physics that needs to be identified.

(iii) *For the gaseous planets, are elements heavier than hydrogen and helium kept inside a central core or distributed inside the planet?* The distribution of heavy elements influences how they cool (Guillot 2005, Baraffe et al. 2008) and is crucial in the context of formation scenarios (Lissauer & Stevenson 2007).

(iv) *How do the diverse conditions witnessed in planetary systems dictate the atmospheric composition*? An understanding of the processes that steer planetary composition bears on our ability to extrapolate to the whole galaxy, and perhaps universe, what we will learn in the solar neighbourhood.

(v) *How does the large range of insolation, planetary spin, orbital elements and compositions in these diverse planetary systems affect the atmospheric dynamics?* This has direct consequences for our ability to predict the evolution of these planets (Cho et al., 2003, 2008).

(vi) *Are planets around low mass, active stars able to keep their atmospheres*? This question is relevant e.g. to the study habitability, as given the meagre energy output of M dwarfs, their habitable zones are located much closer to the primary than those of more massive stars (e.g.~ 0.03 AU for stars weighting one tenth of the Sun) (Lammer, 2013).

We cannot fully understand the atmospheres and interiors of these varied planetary systems by simple analogy with the Solar System, nor from mass and radii measurements alone. As shown by the historical investigations of planets in our own Solar System, these questions are best addressed through spectroscopic measurements. However, as shown by the historical path taken in astronomy, a large sample and range of planetary atmospheres are needed to place the Solar System in an astronomical context. Spectroscopic measurements of a large sample of planetary atmospheres may divulge their atmospheric chemistry, dynamics, and interior structure, which can be used to trace back to planetary formation and evolution.

In the past decade, pioneering results have been obtained using transit spectroscopy with Hubble, Spitzer and ground-based facilities, enabling the detection of a few of the most abundant ionic, atomic and molecular species and to constrain the planet's thermal structure (e.g. Charbonneau et al., 2002; Vidal-Madjiar et al., 2003; Knutson et al., 2007; Swain et al., 2008; Linsky et al., 2010; Snellen et al., 2010, 2014; Majeau et al., 2012). The infrared range, in particular, offers the possibility of probing the neutral atmospheres of exoplanets. In the IR the molecular features are more intense and broader than in the visible (Tinetti et al., 2007b) and less perturbed by clouds, hence easier to detect. On a large scale, the IR transit and eclipse spectra of hot-Jupiters seem to be dominated by the signature of water vapour (e.g. Barman 2007, Beaulieu et al. 2010; Birkby et al., 2013; Burrows et al. 2007, Charbonneau et al. 2008; Crouzet et al. 2012, 2014; Danielski et al. 2014; Deming et al. 2013; Grillmair et al. 2008; Kreidberg et al., 2014b, McCullough et al. 2014; Swain et al. 2008, 2009; Tinetti et al. 2007, 2010, Todorov et al., 2014), similarly, the atmosphere of hot-Neptune HAT-P-11b appears to be water-rich (Fraine et al., 2014). The data available for other warm Neptunes, such as GJ 436b, GJ 3470b are suggestive of cloudy atmospheres and do not always allow a conclusive identification of their composition (Stevenson et al. 2010; Beaulieu et al. 2011; Knutson et al. 2011; Morello et al., 2015; Fukui et al. 2013; Ehrenreich et al, 2014).

The analysis of the transmission and day-side spectra for the transiting 6.5 $M_{Earth}$ super-Earth GJ 1214b suggests either a metal-rich or a cloudy atmosphere (Bean et al. 2010; Berta et al., 2012; Kreidberg et al., 2014, Stevenson et al., 2014).

Despite these early successes, the data available are still too sparse to provide a consistent interpretation, or any meaningful classification of the planets analysed. The degeneracy of solutions embedded in the current transit observations (Swain et al., 2009; Madhusudhan and Seager, 2009; Lee et al., 2012; Line et al., 2013; Waldmann et al., 2014) inhibits any serious attempt to estimate the elemental abundances. New and better quality data are needed for this purpose.

Although these and other data pertaining to extrasolar planet atmospheres are tantalising, uncertainties originating in the narrow-band spectra and sparsity/non simultaneity of the data and, in some cases, low signal to noise ratio, mean that definitive conclusions concerning atmospheric abundances cannot be made today. Current data do not allow one to discriminate between different formation and evolution scenarios for the observed planets.

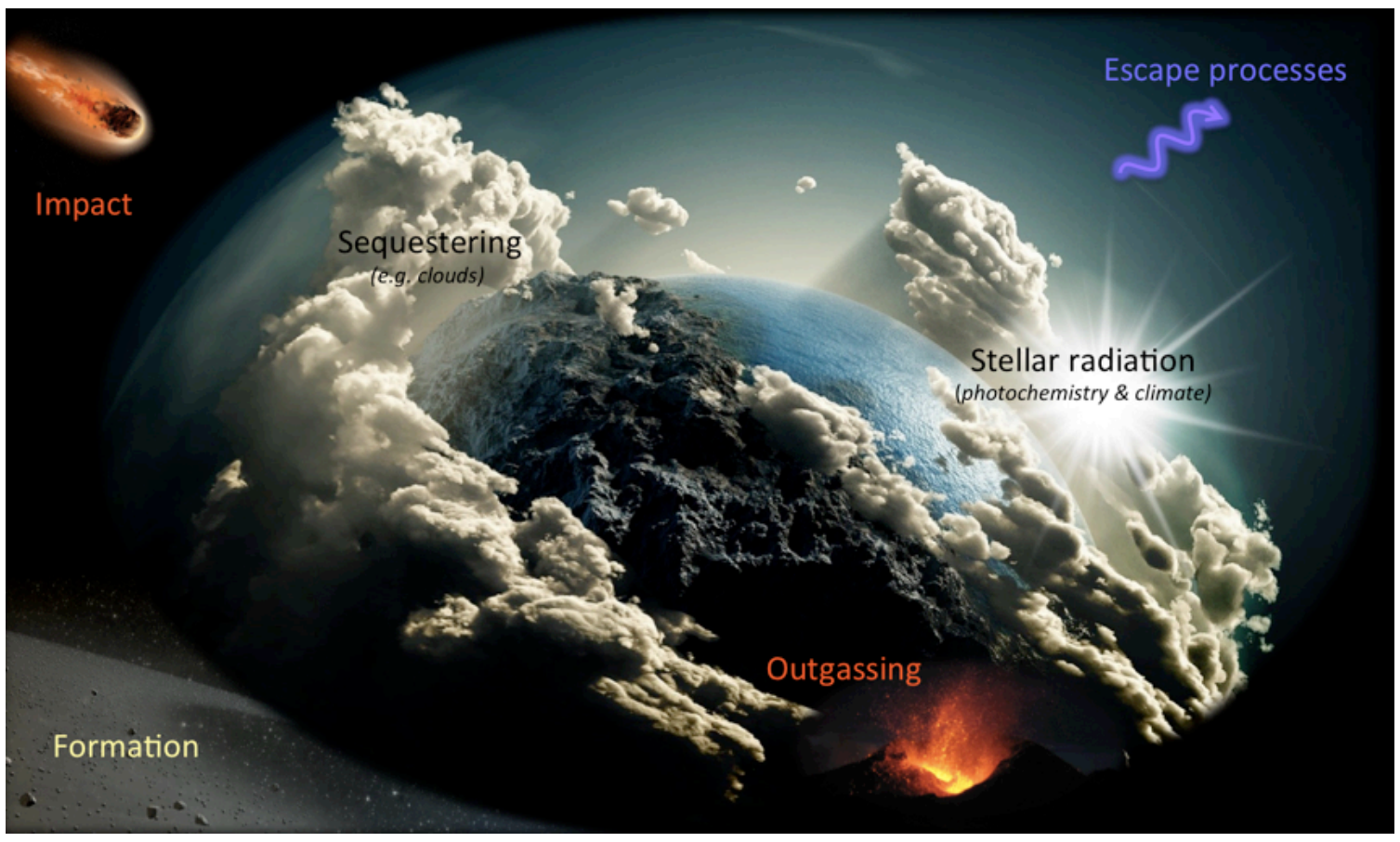

*Figure 2: Key physical processes influencing the composition and structure of a planetary atmosphere. While the analysis of a single planet cannot establish the relative impact of all these processes on the atmosphere, by expanding observations to a large number of very diverse exoplanets, we can use the information obtained to disentangle the various effects.*

The Exoplanet Characterisation Observation (EChO) is a dedicated space-borne telescope concept whose characteristics are summarised in Table 1.The spectral coverage and stability to be achieved by an EChO-like mission would be a game changer, allowing atmospheric compositions to be measured with unparalleled exactness: statistically speaking, at least a factor 10 more precisely and a factor 10 to 1000 more accurately than current observations. This would enable the detection of molecular abundances three orders of magnitude smaller than currently possible. We would anticipate at least a fourfold increase from the handful of molecules currently detected today. Each of these molecules tells us a story, and having access to a larger number means understanding aspects of these exotic planets that are today completely ignored. Combining these data with estimates of planetary bulk compositions from accurate measurements of their radii and masses will allow degeneracies associated with planetary interior modelling to be broken (Adams et al 2008, Valencia et al., 2013), giving unique insight into the interior structure and elemental abundances of these alien worlds.

### *1.1.1 Major classes of planetary atmospheres: what should we expect?*

EChO would address the fundamental questions "*what are exoplanets made of?*" and "*how do planets form and evolve*?" through direct measurement of bulk and atmospheric chemical composition. EChO can observe super-Earths, Neptune-like and Jupiter-like exoplanets around stars of various masses. These broad classes of planets are all expected to have very different formation, migration and evolution histories that will be imprinted on their atmospheric and bulk chemical signatures. Many theoretical studies have tried to understand and model the various processes controlling the formation and evolution of planetary atmospheres, with some success for the Solar System. However, such atmospheric evolution models need confirmation and tight calibrations from observations. In Figure 3 we show the predicted bulk atmospheric compositions as a function of planetary temperature and mass (Leconte, Forget & Lammer, 2014; Forget &

Leconte, 2014) and we briefly describe in the following paragraphs the possible origins of the various scenarios.

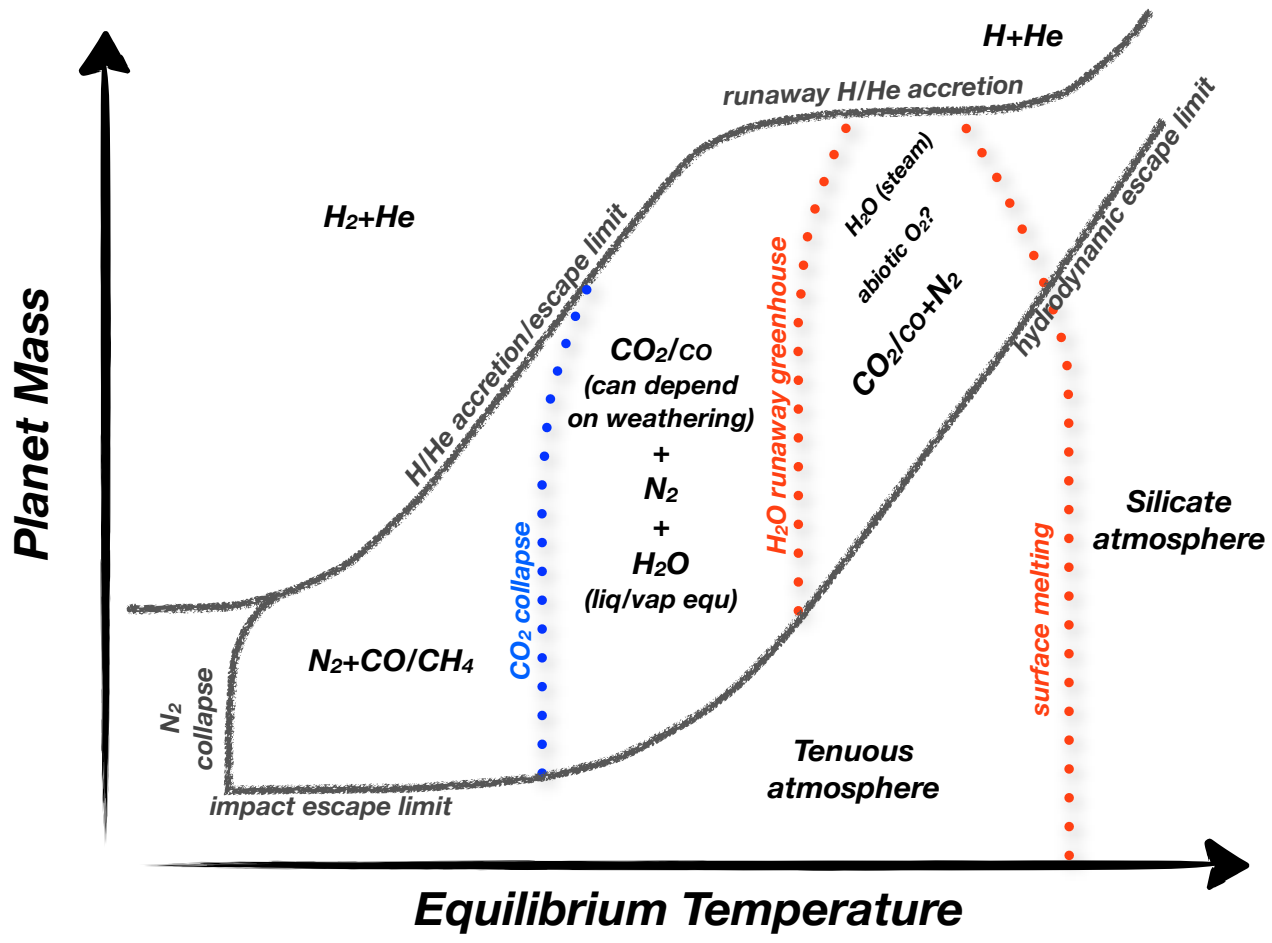

*Figure 3: Schematic summary of the various classes of atmospheres as predicted by Leconte, Forget & Lammer (2014). Only the expected dominant species are indicated, other (trace) gases will be present. Each line represents a transition from one regime to another, but these "transitions" need tight calibrations from observations. Interestingly, many atmospheric-regime transitions occur in the high-mass/high-temperature, domain, which is exactly where EChO is most sensitive.*

**H/He dominated** − Hydrogen and helium being the lightest elements and the first to be accreted, they can most easily escape. The occurrence of H/He dominated atmospheres should thus be limited to objects more massive than the Earth. Because giant planets play a pivotal role in shaping planetary systems (e.g. Tsiganis et al. 2005, Turrini, Nelson, Barbieri, 2014), determining precisely their internal structure and composition is essential to understand how planets form. In particular, the abundances of high-Z elements compared to the stellar values and the relative ratios of the different elements (e.g. C, N, S) represent a window on the past histories of the extrasolar systems hosting the observed planets.

In the Solar System, none of the terrestrial planetary bodies managed to accrete or keep their primordial H/He envelope, not even the coldest ones which are less prone to escape. The presence of a large fraction of primordial nebular gas in the atmosphere of warm to cold planets above a few Earth masses should be fairly common. However, being more massive than that is by no means a sufficient condition: some objects have a bulk density similar to the Earth up to 8-10 $M_{Earth}$. Possibly planets forming on closer orbits can accrete less nebular gas (Ikoma & Hori, 2012), or hotter planets exhibit higher escape rates.

**Thin silicate atmospheres** − For very hot or low mass objects (lower part of Figure 3), the escape of the lightest elements at the top of the atmosphere is a very efficient process. Bodies in this part of the diagram are thus expected to have tenuous atmospheres, if any. Among the most extreme examples, some rocky exoplanets, such as CoRoT- 7 b or 55 Cnc e, are so close to their host star that the temperatures reached on the dayside are sufficient to melt the surface itself. As a result some elements, usually referred to as "refractory", become more volatile and can form a thin "silicate" atmosphere (Léger et al., 2011). Depending on the composition of the crust, the most abundant species should be, by decreasing abundance, Na, K, $O_2$, O and SiO. In addition, silicate clouds could form.

**$H_2O/CO_2/N_2$ atmospheres** − In current formation models, if the planet is formed much closer to –or even beyond– the snow line[1], the water content of the planetesimals could be significantly large and tens to thousands of Earth oceans of water could be accreted (Elkins-Tanton, 2011). This suggests the existence of a vast population of planets with deep oceans (aqua-planets) or whose bulk composition is dominated by water (Ocean planets (Léger et al., 2004)). Another source of volatiles are the planetesimals that accrete to form the bulk of the planet itself. These will be the major sources of carbon compounds (mainly $CO_2$ and possibly $CH_4$), water (especially if they formed beyond the snow line), and, to a lesser extent, $N_2/NH_3$ and other trace gases. In the case of rocky planets, their low gravity field leads to $H_2$ escape. On a much longer, geological timescale, the volatiles that remained trapped in the mantle during the solidification can be released through volcanic outgassing. Along with $H_2O$ and $CO_2$, this process can bring trace gases to the surface, such as $H_2S$, $SO_2$, $CH_4$, $NH_3$, HF, $H_2$, CO and noble gases. On Earth and Mars, there is strong evidence that this secondary outgassing has played a major role in shaping the present atmosphere (Forget & Leconte, 2014).
Water vapour has a tendency to escape, as illustrated by the atmospheric evolutions of Mars and Venus. This certainly happened to the terrestrial planets in our Solar System. In Venus' and Mars' atmospheres the D/H ratio is between 5 and 200 times the Solar ratio, suggesting water on the surface was lost through time. Also their global atmospheric composition, with mostly $CO_2$ and a few percent of $N_2$, are similar. The surface pressures and temperatures are very different, though, as a result of their different initial masses and evolutions. The Earth is an exception in the Solar System, with the conversion of $CO_2$ in the water oceans to $CaCO_3$ and the large abundance of $O_2$ (and its photodissociation product $O_3$) as a consequence of the appearance of life (Lovelock 1975; Rye & Holland 1998).

[1] *Snow line*: distance from a central protostar at which ice grains can form. This occurs at temperatures of ~ 150-170 K

Within each of the above planet taxonomic classes, the stochastic nature of planetary formation and evolution will be reflected in significant variations in the measured abundances, providing important information about the diverse pathways experienced by planets that reside within the same broad class. Our Solar System only provides one or two particular examples, if any, for each of the aforementioned planetary classes. It is therefore impossible to understand the "big picture" on this basis. This is where extrasolar planets are an invaluable asset. This means that, even before being able to characterise an Earth-like planet in the habitable zone, we need to be able to characterise giant planets' atmospheres and exotic terrestrial planet atmospheres in key regimes that are mostly unheard of in the Solar System. Thus, the first observations of exoplanet atmospheres, whatever they show, will allow us to make a leap forward in our understanding of planetary formation, chemistry, evolution, climates and, therefore, in our estimation of the likelihood of life elsewhere in the universe. Only a dedicated transit spectroscopy mission can tackle such an issue.

## 1.2 The case for a dedicated mission from space

EChO has been designed as a dedicated survey mission for transit and eclipse spectroscopy capable of observing a large, diverse and well-defined planet sample within its four years mission lifetime. The transit and eclipse spectroscopy method, whereby the signal from the star and planet are differentiated using knowledge of the planetary ephemerides, allows us to measure atmospheric signals from the planet at levels of at least $10^{-4}$ relative to the star. This can only be achieved in conjunction with a carefully designed stable payload and satellite platform.

It is also necessary to have a broad instantaneous wavelength coverage to detect as many molecular species as possible, to probe the thermal structure of the planetary atmospheres and to correct for the contaminating effects of the stellar photosphere. Since the EChO investigation include planets with temperatures spanning from ~ 300K up to ~3000K, this requires a wavelength coverage ~ 0.55 to 11 μm with a goal of covering from 0.4 to 16 μm. Only modest spectral resolving power is needed, with R~100 for wavelengths less than 5 μm and R~30 for wavelengths greater than this.

The transit spectroscopy technique means that no angular resolution is required. A telescope collecting area of about 1 $m^2$ is sufficiently large to achieve the necessary spectro-photometric precision: for this study the telescope has been assumed 1.13 $m^2$, diffraction limited at 3 μm. Placing the satellite at L2 provides a cold and stable thermal environment as well as a large field of regard to allow efficient time-critical observation of targets randomly distributed over the sky. EChO was designed to achieve a single goal: exoplanet spectroscopy.

It is important to realise that a statistically significant number of observations must be made in order to fully test models and understand which are the relevant physical parameters. This requires observations of a large sample of objects, generally on long timescales, which can only be done with a dedicated instrument like EChO, rather than with multi-purpose telescopes such as the James Web Space Telescope (JWST) or the European Extremely Large Telescope (E-ELT). Another significant aspect of the search relates to the possibility to discover unexpected "Rosetta Stone" objects, i.e. objects that definitively confirm or inform theories. This requires wide searches that are again possible only through dedicated instruments. EChO would allow planetary science to expand beyond the narrow boundaries of our Solar System to encompass our Galaxy. EChO would enable a paradigm shift by identifying the main constituents of hundred(s) of exoplanets in various mass/temperature regimes, we would be looking no longer at individual cases but at populations. Such a universal view is critical if we truly want to understand the processes of planet formation and evolution and how they behave in various environments.

# 2. EChO Science Objectives

In this section we explain the key science objectives addressed by EChO, and how we would tackle these questions through the observations provided by EChO, combined with modeling tools and laboratory data

## 2.1 Key science questions addressed by EChO

EChO has been conceived to address the following fundamental questions:

- Why are exoplanets as they are?
- What are the causes for the observed diversity?
- Can their formation and evolution history be traced back from their current composition?

EChO would provide spectroscopic information on the atmospheres of a large, select sample of exoplanets allowing the composition, temperature (including profile), size and variability to be determined at a level never previously attempted. This information can be used to address a wide range of key scientific questions relative to exoplanets:

- *What are they made of?*
- *Do they have an atmosphere?*
- *What is the energy budget?*
- *How were they formed?*
- *Did they migrate and, if so, how?*
- *How do they evolve?*
- *How are they affected by starlight, stellar winds and other time-dependent processes?*
- *How do weather conditions vary with time?*

And of course:

- *Do any of the planets observed have habitable conditions?*

These objectives, tailored for gaseous and terrestrial planets, are detailed in the next sections and summarised in Figure 4 and Table 2.

In the next sections we also explain how these questions can be tackled through the observations provided by EChO, combined with modelling tools and auxiliary information from laboratory data and preparatory observations with other facilities prior to the EChO launch.

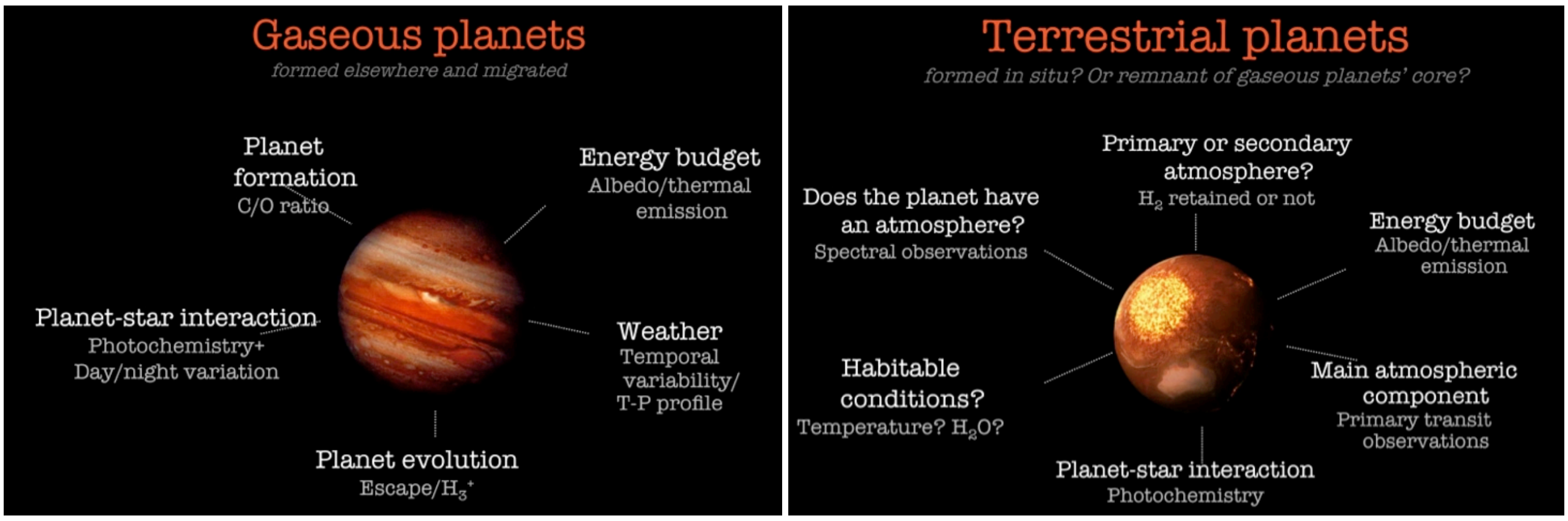

*Figure 4: Key questions for gaseous & rocky planets that will be addressed by EChO (Tinetti et al. 2013).*

| Planet type | Scientific question | Observable | Observational strategy | Survey Type |
|---|---|---|---|---|
| **Gaseous** | Energy budget | Incoming and outgoing | Stellar flux + planetary albedo and thermal | Chemical |
| **planets** | | radiation | emission with VIS and IR photometry during eclipses | Census |
| | Planetary interior | a. Density<br>b. Hints from atmospheric composition? | a. Transit spectra<br>b. Transit and eclipse spectra | Chemical Census |
| | Chemical processes:<br>Thermochemistry?<br>Transport + quenching?<br>Photochemistry? | a. Chemistry of planets around different stars & different temperatures<br>b. Day/night chemical variations<br>c. Vertical mixing ratios | a. Transit and eclipse spectra of planets around different stars & different temps.<br>b. Relative abundances of minor molecular species (HCN, $NH_3$, $C_2H_2$, etc.) | Origin |
| | Dynamics:<br>Time scale of horizontal and vertical mixing | a. Vertical thermal profile<br>b. Horizontal gradients<br>c. Diurnal variations<br>d. Temporal variability, seasonal/inter-seasonal variations… | a. IR eclipse spectra<br>b. IR Eclipse mapping<br>c. IR orbital phase lightcurves<br>d. Repeated observations & use of chemical species as tracers (e.g. $CH_4$, $NH_3$, $CO_2$, and HCN etc) | Origin & Rosetta Stone |
| | Formation:<br>Core accretion or gravitation instability? | a. Planetary density<br>b. C/O ratio | a. Transit + mass from Radiative Velocity<br>b. Relative abundances of carbon versus oxygen-bearing molecules | Origin |
| | Migration:<br>Any evidence of the initial conditions? | a. Comparison star/planet metallicity (C/O, O/H, C/H..)<br>b. Chemistry of planets around different stars. | a. Relative abundances of carbon-, oxygen-, bearing molecules, etc.<br>b. Transit and eclipse spectra of planets around different stars & different T | Origin |
| | 2D and 3D maps | Exoplanet image at multiple wavelengths | Ingress and egress eclipse spectra<br>Orbital phase-curves | Rosetta Stone |
| | Evolution:<br>Escape processes | $H_3^+$ detection and ionospheric temperature measurement | Transit and eclipse spectra | Origin |
| **Terrestrial planets** | Energy budget<br>Albedo & Temperature | Incoming and outgoing radiation | Stellar flux + planetary albedo and thermal emission with VIS and IR photometry during eclipses | Chemical Census |
| | Is there an atmosphere? | Featureless spectrum or not | Transit spectra at multiple wl (IR in particular) to constrain the scale height | Chemical Census |
| | Primary or secondary atmosphere? | Hydrogen rich atmosphere? | Transit spectra at multiple wl (IR in particular) to constrain the scale height | Chemical Census |
| | Main atmospheric component | Scale height | Transit spectra at multiple wl (IR in particular) to constrain the scale height | Chemical Census |
| | Planetary interior | a. Density<br>b. Hints from atmospheric composition? | a. Transit + mass from Radial Velocity<br>b. Transit and eclipse spectra | Chemical Census |
| | Formation:<br>Formed in situ? Migrated? Core of a | a. Density<br>b. Is there an atmosphere?<br>c. Primary ($H_2$-rich) or | a. Transit + mass from Radial Velocity<br>b. c. d. Transit and | Chemical Census |
| | giant planet? Frequency of Venus-like, Mercury-like, Ocean planets.. | secondary atmosphere?<br>d. Atmospheric composition? | eclipse spectra | |
| **Temperate terrestrial planets** | Habitability | a. Temperature<br>b. Chemical composition ($H_2O$? $CO_2$? $O_3$?) | a. Eclipse measurements<br>b. Transit or eclipse measurements at low resolution. | Challenging, need a late M star, bright in the IR |

*Table 2: Traceability matrix*

## 2.2 Terrestrial planets (predominantly solid)

Several scenarios may occur for the formation and evolution of terrestrial-type planets (see 1.1.1 and Figure 3). To start with, these objects could have formed *in situ*, or have moved from their original location because of dynamical interaction with other bodies, or they could be remnant cores of more gaseous objects which have migrated in. Having a lower mass, their atmospheres could have evolved quite dramatically from the initial composition, with lighter molecules, such as hydrogen, escaping more easily. Impacts with other bodies, such as asteroids or comets, or volcanic activity might also alter significantly the composition of the primordial atmosphere. EChO can confirm the presence or absence of a substantial atmosphere enveloping terrestrial planets. On top of this, EChO can detect the composition of their atmospheres ($CO_2$, SiO, $H_2O$ etc.), so we can test the validity of current theoretical predictions (section 1.1.1 and Figure 3). In particular:

(i) A very thick atmosphere (several Earth masses) of heavy gas, such as carbon dioxide, ammonia, water vapour or nitrogen, is not realistic because it requires amounts of nitrogen, carbon, and oxygen with respect to silicon much higher than all the stellar ratios detected so far. If EChO detects an atmosphere which is not made of hydrogen and helium, the planet is almost certainly from the terrestrial family, which means that the thickness of the atmosphere is negligible with respect to the planetary radius. In that case, theoretical works provided by many authors in the last decade (Léger et al., 2004; Valencia et al., 2006, 2007; Adams & Seager, 2008; Grasset et al., 2009) can be fully exploited in order to characterise the inner structure of the planet (Figure 5).

(ii) If an object exhibits a radius that is bigger than that of a pure water world (water being the least dense, most abundant material except for H/He) of the same mass, this tells us that at least a few % of the total mass of the planet is made of low density species, most likely $H_2$ and He. The fact that many objects less massive than Neptune are in this regime shows that it is possible to accrete a large fraction of gas down to 2-3 $M_{Earth}$, the mass of Kepler-11 f (Figure 5). EChO can test this hypothesis by probing the presence of $H_2$, He and $H_2O$ through primary transit spectroscopy (Figure 5).

(iii) A major motivation for exoplanet characterisation is to understand the probability of occurrence of habitable worlds, i.e. suitable for surface liquid water. While EChO may reveal the habitability of one or more planets – temperate super-Earths around nearby M-dwarfs are within reach of EChO's capabilities – its major contribution to this topic results from its capability to detect the presence of atmospheres on many terrestrial planets even outside the habitable zone and, in many cases, characterise them.

## 2.3 The intermediate family (Neptunes and Sub-Neptunes)

Planets with masses between the small solid terrestrial and the gas giants planets are key to understanding the formation of planetary systems (Guillot & Stixrude 2014). The existence of these intermediate planets close to their star, as found by radial velocity and transit surveys (see Figure 1), already highlights the shortcomings of current theoretical models.

(i) Standard planet formation scenarios predict that embryos of sufficient mass (typically above 5 $M_{Earth}$) should retain some of the primordial hydrogen and helium from the protoplanetary disc. With EChO's primary transit spectroscopic measurements, we may probe which planets possess a hydrogen helium atmosphere and directly test the conditions for planet formation (Figure 5).

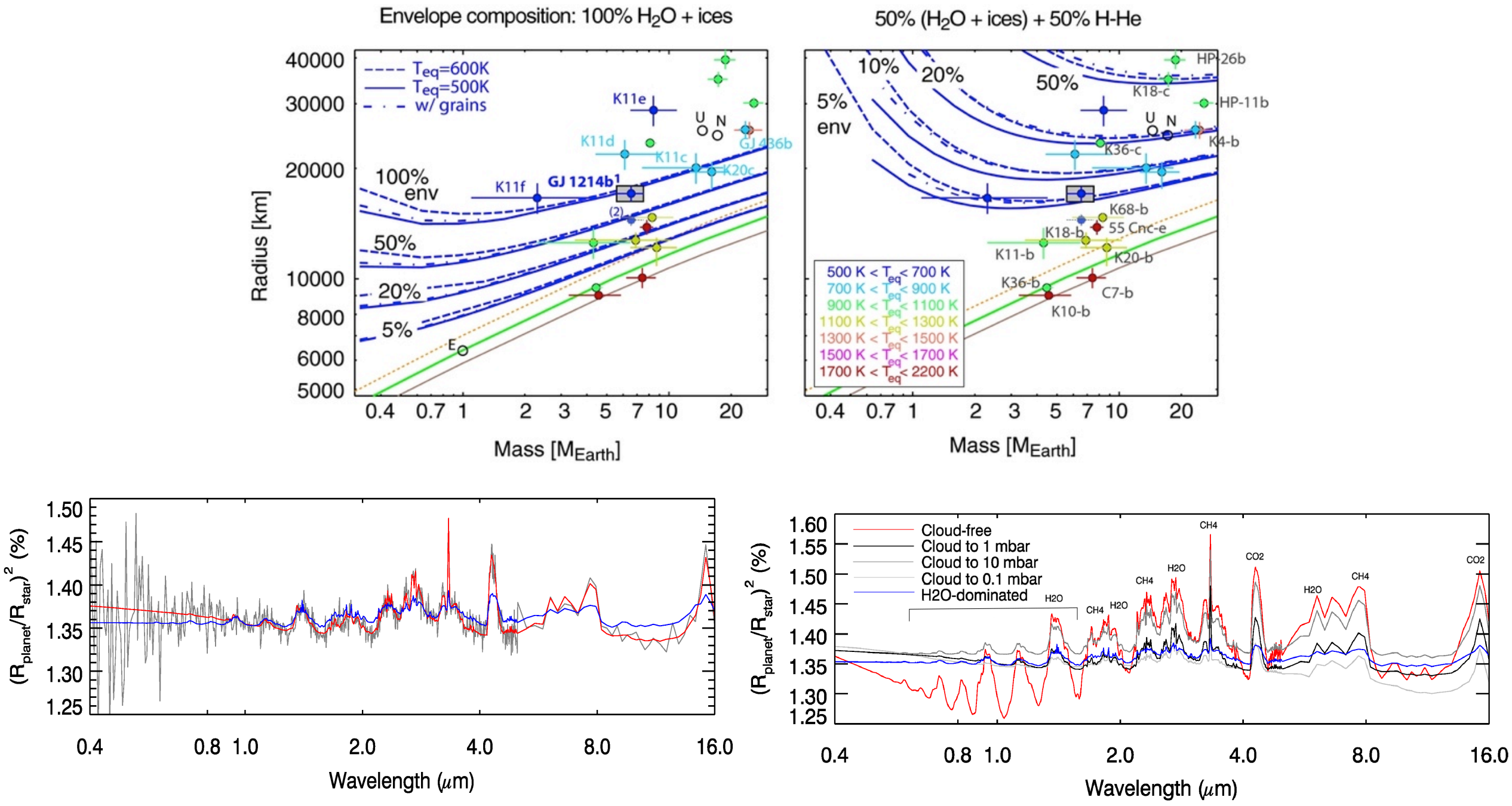

*Figure 5: Top: Mass–radius relationships for Ocean planets and sub-Neptunes and degeneracy of interpretation (Valencia et al., 2013). Two envelope compositions are shown: 100% $H_2O$/ices (left) and with 50% ($H_2O$/ices)+50% H/He (right): they both explain the densities of the planets identified with blue dots. Bottom: Synthetic spectra between 0.4 and 16 μm of the super-Earth GJ1214b for a range of atmospheric scenarios (Barstow et al., 2013a). Left: Two retrieval fits to a noisy EChO-like simulated spectrum, with a $H_2$-He rich atmosphere (red) and a 50% $H_2O$ model (blue). EChO is able to distinguish between the two competing scenarios. Right: simulations of cloudy and cloud-free atmospheres. EChO's broad wavelength range and sensitivity would enable the identification of different molecular species and type of clouds.*

(ii) The only two intermediate solar system planets that we can characterise –Uranus and Neptune– are significantly enriched in heavy elements, in the form of methane. The reason for this enrichment is unclear: is it due to upward mixing, early or late delivery of planetesimals? EChO would guarantee these measurements in many planets, thereby providing observations that are crucial to constrain models.

(iii) We do not know where to put the limits between solid, liquid and gaseous planets. While EChO cannot measure directly the phase of a planet as a whole, the determination of its size and of the composition of its atmosphere will be key to determining whether its interior is solid, partially liquid, or gaseous.

## 2.4 Gaseous exoplanets

Giant planets are mostly made of hydrogen and helium and are expected to be always in gaseous form. Unlike solid planets, they are relatively compressible and the progressive loss of heat acquired during their formation is accompanied by a global contraction. Inferring their internal composition thus amounts to understanding how they cool. The dominance of hydrogen and helium implies that the degeneracy in composition (i.e. uncertainty on the mixture of ices/rocks/iron) is much less pronounced than for solid planets, so that the relevant question concerns the amounts of all elements other than hydrogen and helium, i.e. heavy elements, that are present. A fundamental question is by how much are these atmospheres enriched in heavy elements compared to their parent star. Such information will be critical to:

- understand the early stage of planetary and atmospheric formation during the nebular phase and the immediately following few millions years (Turrini, Nelson, Barbieri, 2014)
- test the effectiveness of the physical processes directly responsible of their evolution.

We detail below the outstanding questions to be addressed by an EChO-like mission and how these can be achieved.

### 2.4.1 *The chemistry of gaseous planets' atmospheres*

(i) *The relative importance of thermochemical equilibrium, photochemistry, and transport-induced quenching* in controlling the atmospheric composition of gaseous exoplanets largely depends on the thermal structure of the planets. Transport-induced quenching of disequilibrium species allows species present in the deep atmosphere of a planet to be transported upward in regions where they should be unstable, on a time scale shorter than the chemical destruction time. The disequilibrium species are then "quenched" at observationally accessible atmospheric levels. In the solar system, this is the case, in particular, for CO in the giant planets, as well as $PH_3$ and $GeH_4$ in Jupiter and Saturn (Encrenaz, 2004). Another key process, which also leads to the production of disequilibrium species, is photochemistry (Yung & DeMore, 1999). The energy delivered by the absorption of stellar UV radiation can break chemical bonds and lead to the formation of new species. In the solar system, the photochemistry of methane is responsible for the presence of numerous hydrocarbons in the giant planets. In the case of highly irradiated hot Jupiters, these disequilibrium species are expected to be important. In some of the known hot-Jupiters, $CH_4$ and $NH_3$ are expected to be enhanced with respect to their equilibrium abundances due to vertical transport-induced quenching. These species should be dissociated by photochemistry at higher altitude, leading, in particular, to the formation of $C_2H_2$ and HCN on the day side (Moses et al., 2011, Venot et al., 2012). EChO can address these open questions, by deriving the abundances of both key and minor molecular species, with mixing ratios down to $10^{-5}$ to $10^{-7}$ (Figure 6), temporally and spatially resolved in the case of very bright sources (see 2.3.2.3).

(ii) *Chemistry and dynamics* are often entangled. Agúndez et al. (2012, 2014) showed that for hot-Jupiters, for instance, the molecules CO, $H_2O$, and $N_2$ and $H_2$ show a uniform abundance with height and longitude, even including the contributions of horizontal or vertical mixing. For these molecules it is therefore of no relevance whether horizontal or vertical quenching dominates. The vertical abundance profile of the other major molecules $CH_4$, $NH_3$, $CO_2$, and HCN shows, conversely, important differences when calculated with the horizontal and vertical mixing. EChO spectroscopic measurements of the dayside and terminator regions would provide a key observational test to constrain the range of models of the thermochemical, photochemical and transport processes shaping the composition and vertical structure of these atmospheres.

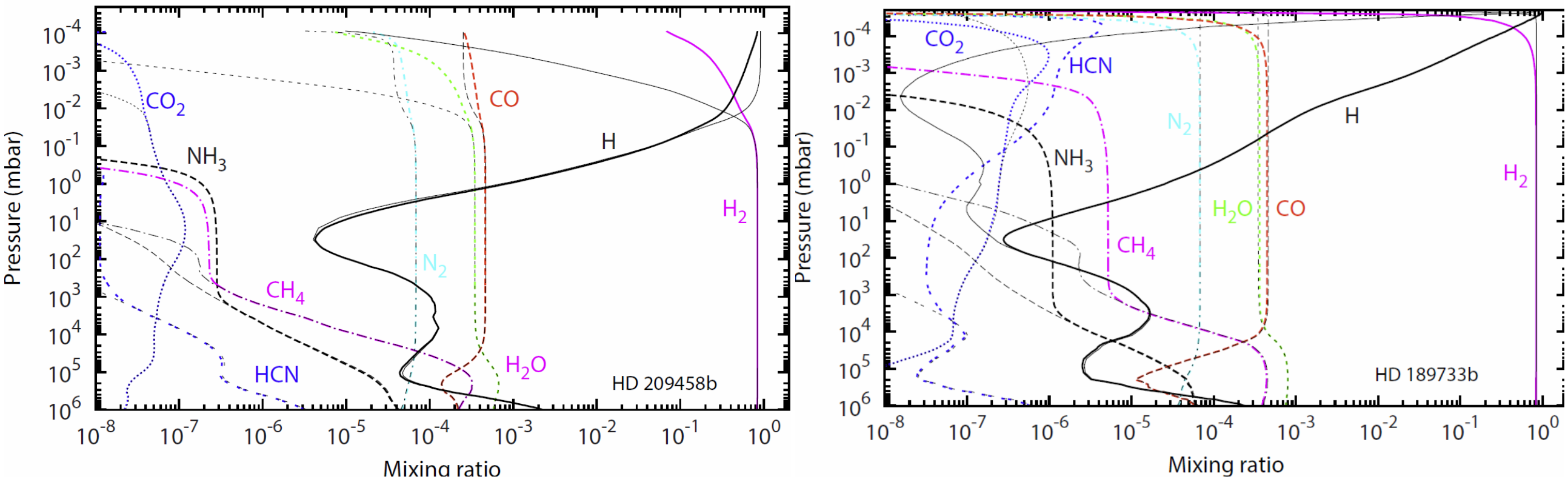

*Figure 6: Steady-state composition of HD 209458b (left) and HD 189733b (right) calculated with a non-equilibrium model (colour lines), compared to the thermodynamic equilibrium (thin black lines) (Venot et al. 2012). For HD 189733b, one can clearly notice the higher sensitivity to photolyses and vertical mixing, with all species affected, except the main reservoirs, $H_2$, $H_2O$, CO, and $N_2$. Since the atmosphere of HD209458b is hotter, it is mostly regulated by thermochemistry. The EChO Origin survey would measure these differences by deriving the abundances of both key and minor molecular species, with mole fractions down to $10^{-5}$ to $10^{-7}$ (see Section 3.2.2 and 3.2.3).*

### 2.4.2 *Energy Budget: heating and cooling processes*

(i) *Albedo and thermal emission*. The spectrum of a planet is composed mainly of reflected stellar light and thermal emission from the planet; the measurement of the energy balance is an essential parameter in quantifying the energy source of dynamical activity of the planet (stellar versus internal sources). The Voyager observations of the Giant Planets in the Solar System have allowed an accurate determination of the energy budget by measuring the Bond albedo of the planets (Jupiter: Hanel et al., 1981; Saturn: Hanel et al, 1983; Uranus: Pearl et al, 1990; Neptune: Pearl & Conrath, 1991). EChO extends these methods to exoplanets: the reliable determination of the spectrum in reflected versus thermal range will provide a powerful tool for classifying the dynamical activity of exoplanets.

(ii) *Non-LTE emissions.* Observation of the $CH_4$ non-LTE emission on the day side of Jupiter and Saturn (Encrenaz et al, 1996; de Graauw et al, 1997) is an important new tool to sound the upper atmosphere levels around the homopause (typically at the microbar level for giant planets), the layer separating the turbulent mixing from the diffusive layers where molecules are separated by their molecular weight. This region is an important transition between the internal dynamical activity and the radiatively controlled upper atmosphere, with the breaking of gravity waves identified as an important mechanism responsible of high thermospheric temperatures in giant planets. Swain et al. (2010) and Waldmann et al. (2012) identified an unexpected spectral feature near 3.25 μm in the atmosphere of the hot-Jupiter HD 189733b which was found to be inconsistent with LTE conditions holding at pressures typically sampled by infrared measurements. They proposed that this feature results from non-LTE emission by $CH_4$, indicating that non-LTE effects may need to be considered, as is also the case in our Solar System for Jupiter and Saturn as well as for Titan. EChO can conclusively unveil the nature of this feature and address the same question for many hot gaseous planets, making use of the improved observing conditions from space.

(iii) $H_3^+$ emission (3.5-4.1 μm). Of particular interest in the study of gas giants within our own solar system are emissions of $H_3^+$ which dominate their emissions between 3 and 4 μm. $H_3^+$ is a powerful indicator of energy inputs into the upper atmosphere of Jupiter (Maillard & Miller, 2011), suggesting a possible significance in exoplanet atmospheres as well. As the unique atmospheric constituent radiatively active, $H_3^+$ plays a major role in regulating the ionospheric temperature. Simulations by Yelle (2004) and Koskinen et al. (2007) have investigated the importance of $H_3^+$ as a constituent and IR emitter in exoplanet atmospheres. A finding of these calculations is that close-orbiting extrasolar planets (0.2 AU) may host relatively small abundances only of $H_3^+$ due to the efficient dissociation of $H_2$, a parent molecule in the creation path of $H_3^+$. As a result, the detectability of $H_3^+$ may depend on the distance of the planet from the star. EChO can test this hypothesis by detecting or setting an upper limit on the $H_3^+$ abundance in many giant planets.

(iv) Clouds may modify the albedo and contribute to the green-house effect, therefore their presence can have a non-negligible impact on the atmospheric energy budget. If present, clouds will be revealed by EChO through transit and eclipse spectroscopy in the VIS-NIR**.** Clouds show, in fact, distinctive spectroscopic signatures depending on their particle size, shape and distribution (see Figures 14, 19). Current observations in the VIS and NIR with Hubble and MOST have suggested their presence in some of the atmospheres analysed (e.g. Rowe et al., 2008; Sing et al., 2011; Demory et al., 2013; Kreidberg et al., 2014; Knutson et al, 2014). We do not know, though, their chemical composition, how they are spatially distributed and whether they are a transient phenomenon or not. Further observations over a broad spectral window and through time are needed to start answering these questions (see most recent work done for brown dwarfs (Apai et al., 2013)).

### *2.4.3 Spatial and temporal variability: weather, climate and exo-cartography*

(i) *Temporal variability*: Tidally synchronised and unsynchronised gaseous planets are expected to possess different flow and temperature structures. Unencumbered by complicating factors, such as physical topography and thermal orography, the primary difference will be in the amplitude and variability of the structures. An example is shown in Figure 7 for the case of HD 209458b, a synchronised hot-Jupiter. The state-of-the-art, high-resolution simulation shows giant, tropical storms (cyclones) generated by large-amplitude planetary waves near the substellar point. Once formed, the storms move off poleward toward the nightside, carrying with them heat and chemical species, which are observable, and which dissipate to repeat the cycle, in this case, after a few planet rotations (Cho et al., 2003, 2008). Storms of such size and dynamism are characteristic of synchronized planets, much more so than unsynchronized ones. There are other even more prominent periodicities (e.g., approximately 1.1, 2.1, 4.3, 8.3, 15 and 55 planet rotations), all linked to specific dynamical features. Through its excellent temporal coverage of individual objects (i.e. tens of repeated observations as part of the Rosetta Stone survey, see Section 3.2.2), EChO can well distinguish the two different models and type of rotation.

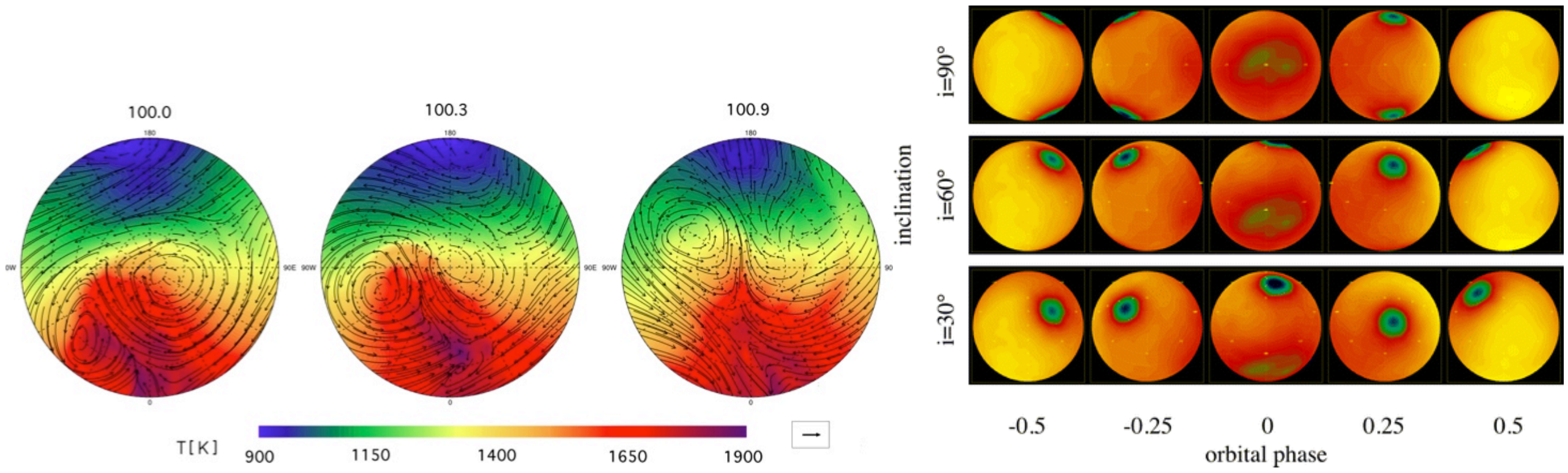

*Figure 7: Left: Giant storms on a synchronized, gaseous planet. Wind vectors superimposed on temperature map over approximately one planet rotation period, viewed from the north pole. Synchronized planets experience intense irradiation from the host star (at lon = 0 point), exciting large-amplitude planetary waves and active storms that move off to the night side (top half in each frame). The storms dissipate and regenerate with a distinct period of a few planet rotations (Cho et al., 2003, 2008). Other dynamically-induced periodicities are present on synchronized planets. The periodicities can be used to distinguish synchronized and unsynchronized planets, among other things. Right: Simulated phase variations for a hot-Jupiter with different inclinations (Rauscher et al., 2008).*

(ii) <u>*Horizontal thermal structure*</u>: phase curves, spherical harmonics & eclipse mapping. Longitudinal variations in the thermal properties of the planet cause a variation in the brightness of the planet with orbital phase (Figure 7). This orbital modulation has been observed in the IR in transiting (Knutson et al., 2007, 2008) and non-transiting systems (Harrington et al., 2006, Crossfield et al., 2010). One of the great difficulties in studying extrasolar planets is that we cannot directly resolve the surfaces of these bodies, as we do for planets in our solar system. The use of occultations or eclipses to spatially resolve astronomical bodies, has been used successfully for stars in the past. Most recently Majeu et al. (2012) and de Wit et al. (2012) derived the two-dimensional map of the hot-Jupiter HD189733b in the IR. Majeu et al. (2012) combined 7 observations at 8 μm with Spitzer-IRAC and used two techniques: slice mapping & spherical harmonic mapping (see Figure 12). Both techniques give similar maps for the IR dayside flux of the planet. EChO can provide phase curves and 2D-IR maps recorded simultaneously at multiple wavelengths, for several gaseous planets, an unprecedented achievement outside the solar system. These curves and maps will allow one to determine horizontal and vertical, thermal and chemical gradients and exo-cartography (Figure 8).

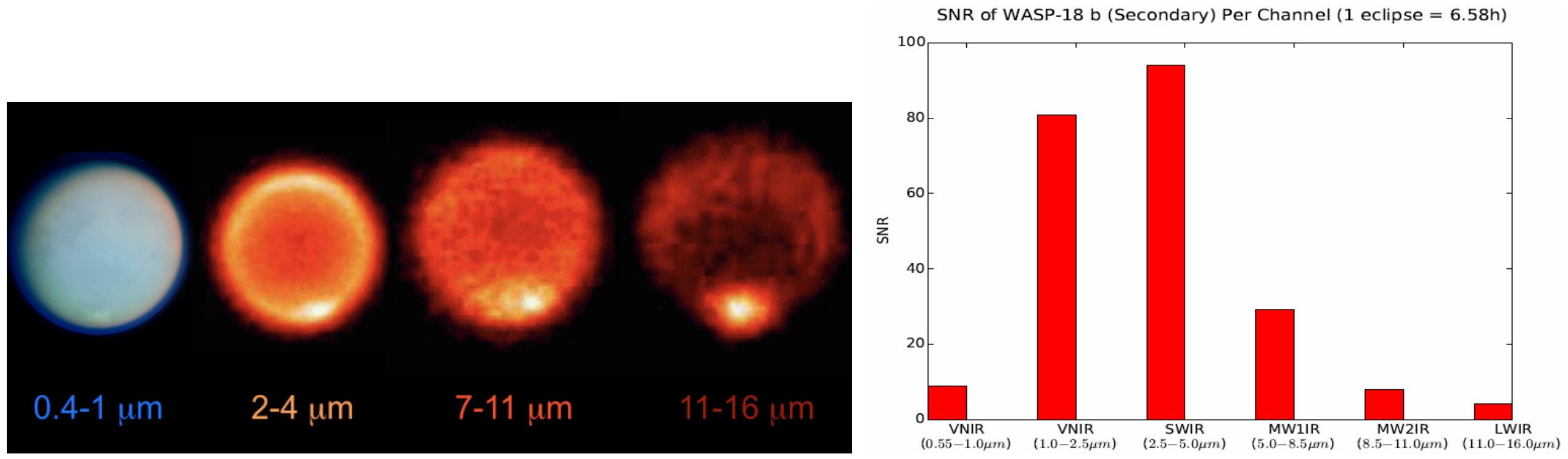

*Figure 8: Left: Demonstration of possible results from exo-cartography of a planet at multiple photometric bands. Right: simulations of EChO performances for the planet WASP-18b: the SNR in one eclipse is high enough at certain wavelength to allow one to resolve spatially the planet through eclipse mapping.*

### *2.4.4 Planetary interior*

Although EChO has been conceived to measure the characteristics of planetary atmospheres it can also be crucial in improving our knowledge of planetary interiors (Guillot & Stixrude, 2014). EChO can measure with exquisite accuracy the depth of the primary transit and therefore the planetary size. But the major improvements for interior models will come from the ability to characterise the atmosphere in its composition, dynamics and structure. As described in the previous sections, this can be achieved by a combination of observations of transits and of observations of the planetary lightcurve during a full orbital cycle.

EChO directly contributes to the understanding of the interiors of giant exoplanets through the following measurements:

(i) Measurements on short time scales. A few hours of continuous observations of the transit or eclipse reveal the abundances of important chemical species globally on the terminator or on the dayside. The comparison of these measurements with the characteristics of the star and of the planet, in particular the stellar metallicity and the mass of heavy elements required to fit the planetary size is key in the determination of whether the heavy elements are mixed all the way to the atmosphere or mostly present in the form of a central core.

(ii) Measurements on long time scales. A half or a full planetary orbit, i.e. hours/days of continuous observations, lead to a very accurate description of the atmospheric dynamics (wind speed, vertical mixing from disequilibrium species), atmospheric structure (vertical and longitudinal temperature field, presence of clouds) and variability. This is essential to estimate the depth at which the atmosphere becomes well mixed and therefore the heat that is allowed to escape.

### *2.4.5 Chemical composition of gaseous planets: a pointer to planet formation and migration history*

Formation and migration processes play fundamental roles in determining planetary bulk and atmospheric compositions that ultimately reflect the chemical structure and fractionation within nascent protoplanetary discs. For the purpose of illustration, Turrini, Nelson & Barbieri (2014) have considered a number of simplified planetary accretion and migration scenarios within discs with Solar chemical abundance. They show that models of accretion onto planetary cores can lead to final envelope C/O values that range from less than 0.54 up to 1, and correlate with where and how the planet forms and migrates in a predictable manner. EChO can provide much needed observational constraints on the C/O values for many gaseous planets. In the following paragraphs we outline how key formation and migration processes may lead to diverse chemical signatures.

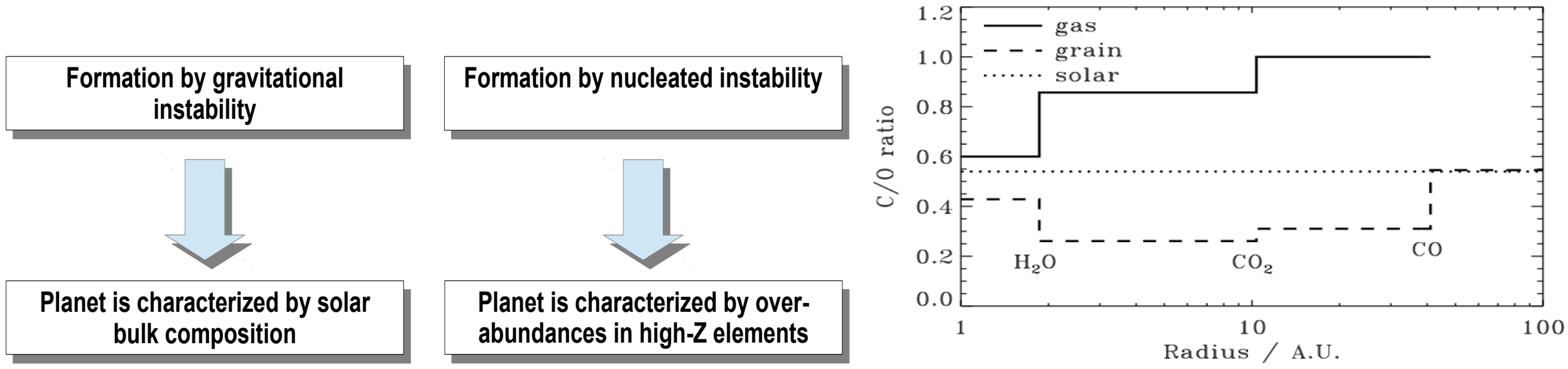

*Figure 9: Left: expected differences in the atmospheric composition due to different formation scenarios. Right: Locations of the ice-lines and their influence on the C/O ratios for the gas and solids (adapted from Oberg et al 2011).*

(i) Giant planet formation via *gravitational instability* that occurs during the earliest phases of protoplanetary disc evolution will result initially in planets with bulk and atmospheric abundances reflecting that of the protoplanetary disc. Recent studies show that formation is followed by rapid inward migration on time scales ~ $10^3$ yr (Baruteau et al 2011, Zhu et al 2012), too short for significant dust growth or planetesimal formation to arise between formation and significant migration occurring. Migration and accompanying gas/dust accretion should therefore maintain initial planetary abundances if protoplanetary discs possess uniform elemental abundances. Post-formation enrichment may occur through bombardment from neighbouring planetesimals or star-grazing comets, but this enrichment will occur in an atmosphere with abundances that are essentially equal to the stellar values, assuming these reflect the abundances present in the protoplanetary disc.

In its simplest form, the *core accretion model* of planet formation begins with the growth and settling of dust grains, followed by the formation of planetesimals that accrete to form a planetary core. Growth of the core to a mass in excess of a few Earth masses allows for the settling of a significant gaseous envelope from the surrounding nebula. Halting growth at this point results in a super-Earth or Neptune-like planet. Continued growth through gas and planetesimal accretion leads to a gas giant planet. A key issue for determining the atmospheric abundances in a forming planet is the presence of ice-lines at various distances from the central star, beyond which volatiles such as water, carbon dioxide and carbon monoxide freeze-out onto grains and are incorporated into planetesimals. Figure 9 shows the effect of ice-lines associated with these species on the local gas- and solid-phase C/O ratios in a protoplanetary disc with solar C/O ratio ~ 0.54. A $H_2O$ ice-line is located at 1-3 AU, a $CO_2$ ice-line at ~10 AU, and a CO ice-line at ~ 40 AU (Oberg et al 2011)). Interior to the $H_2O$ ice-line, carbon- and silicate-rich grains condense, leading to a gas-phase C/O ~ 0.6 (due to the slight overabundance of oxygen relative to carbon in these refractory species). The atmospheric abundances of a planet clearly depend on where it forms, the ratio of gas to planetesimals accreted at late times, and the amount of accretion that occurs as the planet migrates. As a way of illustrating basic principles, we note that a planet whose core forms beyond the $H_2O$ ice-line, and which then accretes gas but no planetesimals interior to 2 AU as it migrates inward will have an atmospheric C/O ~ 0.54. Additional accretion of planetesimals interior to 2 AU would drive C/O below 0.54. Similarly, a planet that forms a core and accretes all of its gas beyond the $CO_2$ ice-line at 10 AU before migrating inward without further accretion will have an envelope C/O ~ 1. Clearly a diverse range of atmospheric C/O values are possible. More realistic N-body simulations of planet formation that include migration, gas accretion and disc models with the chemical structure shown in Figure 10 have been performed recently by Coleman & Nelson (2013). These show a range of final C/O values for short-period planets, as illustrated by the example run shown in Figure 10.

(ii) Gas disc-driven migration is only one plausible mechanism by which planets can migrate. The large eccentricities (and obliquities) of the extrasolar planet population suggest that planet-planet gravitational scattering ("Jumping Jupiters") may be important (Weidenschilling & Marzari 1996; Chatterjee et al 2008), and this is likely to occur toward the end of the gas disc lifetime, when its ability to damp orbital eccentricities is diminished. When combined with tidal interactions with the central star, planet-planet scattering onto highly eccentric orbits can form short-period planets that have not migrated toward the central star while accreting from the protoplanetary disc. These planets are likely to show chemical signatures that reflect this alternative formation history, being composed of higher volatile fractions if they form exterior to the $H_2O$ ice line. Measurements of bulk and atmospheric chemical compositions by EChO will provide important clues regarding the full diversity of the formation and migration pathways that were followed by the observed planetary sample.

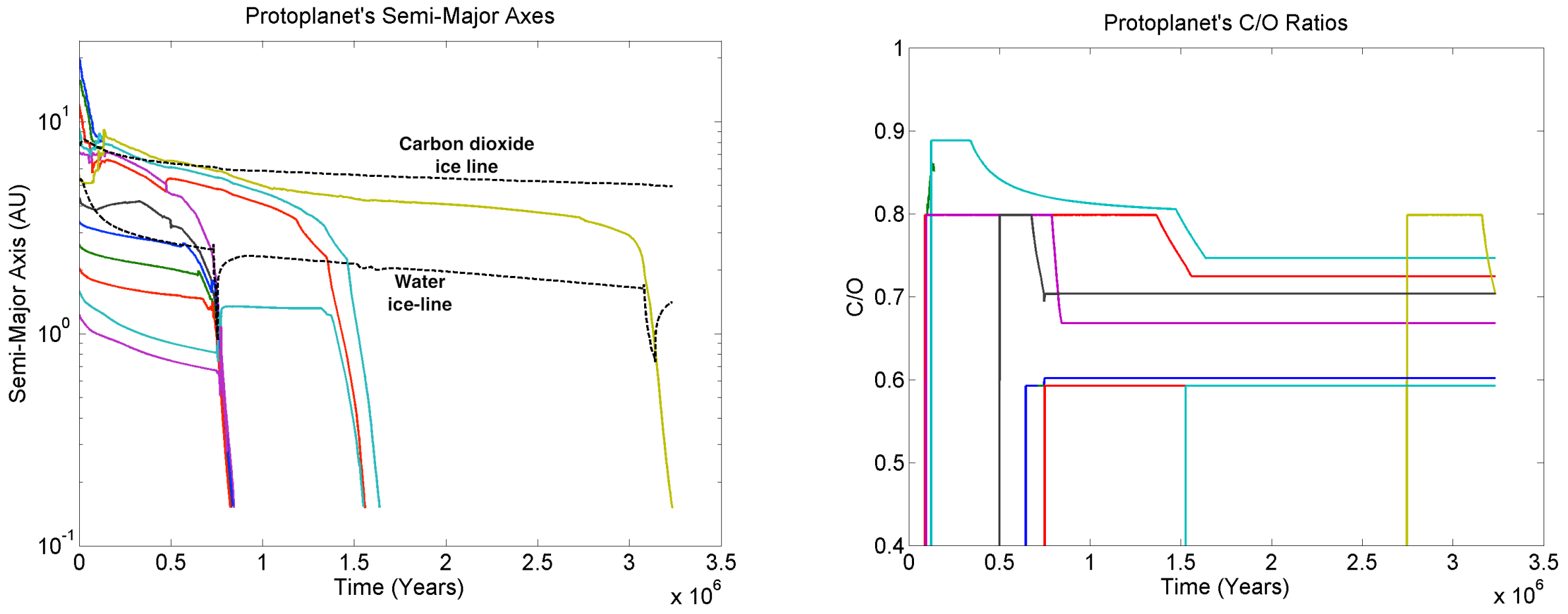

*Figure 10: Left panel: migration trajectories of forming planets. Right panel: Corresponding C/O ratios of planetary envelopes as they accrete and migrate. Note the initially high C/O ratios of planets forming beyond $CO_2$ ice-line and reductions in C/O as planets migrate inward where the local disc gas C/O ratio is close to the solar value of ~ 0.54. Images taken from Coleman & Nelson (2013).*

# 3. EChO observational techniques

In this section we detail the observational techniques and strategies that EChO may adopt to maximise the scientific return.

The transit and eclipse spectroscopy allow us to measure atmospheric signals from the planet at levels of at least $10^{-4}$ relative to the star. Analysis techniques to decorrelate the planetary signal from the astrophysical and instrumental noise are presented.
A broad instantaneous wavelength coverage is essential to detect as many chemical species as possible, to probe the thermal structure of the planetary atmospheres and to correct for the contaminating effects of the stellar photosphere.
The EChO core science may be optimised by a three-tier survey, distinguished by the SNR and the resolving power of the observations. Those are tailored to achieve well defined scientific objectives and might need to be revised at a later stage, closer to launch, to account for the new developments and achievements of the field.

## 3.1 Transits, eclipses and phase-curves

EChO will probe the atmospheres of extrasolar planets using temporal variations to separate out planet light from the star – a technique that has grown to be incredibly powerful over the last decade. It makes use of (a) planet transits, (b) secondary eclipses, and (c) planet phase variations (Figure 11).

(i) *Transit spectroscopy*: When a planet moves in front of its host star, starlight filters through the planet's atmosphere. The spectral imprint of the atmospheric constituents can be distilled from the spectrum of the host star by comparing in-transit with out-of-transit spectra (Seager & Sasselov, 2001; Brown, 2001; Tinetti et al., 2007a). Transit spectroscopy probes the high-altitude atmosphere at the day/night terminator region of the planet. The absorption signals mainly depend on the temperature and the mean molecular weight of the atmosphere, and on the volume mixing ratio of the absorbing gas. If present, clouds can be detected mainly in the VIS.

(ii) *Eclipse spectroscopy*: On the opposite side of the orbit, the planet is occulted by the star (the eclipse), and therefore temporarily blocked from our view. The difference between in-eclipse and out-of-eclipse observations provides the planet day-side spectrum. In the near- and mid-infrared, the radiation is dominated by thermal emission, modulated by molecular features (Deming et al., 2005; Charbonneau et al., 2005). This is highly dependent on the vertical temperature structure of the atmosphere, and probes the atmosphere at higher pressure-levels than transmission spectroscopy. At visible wavelengths, the planet's spectrum is dominated by Rayleigh and/or Mie scattering of light from the host star (e.g. De Kok & Stam, 2012). For the latter, clouds can play an important role.

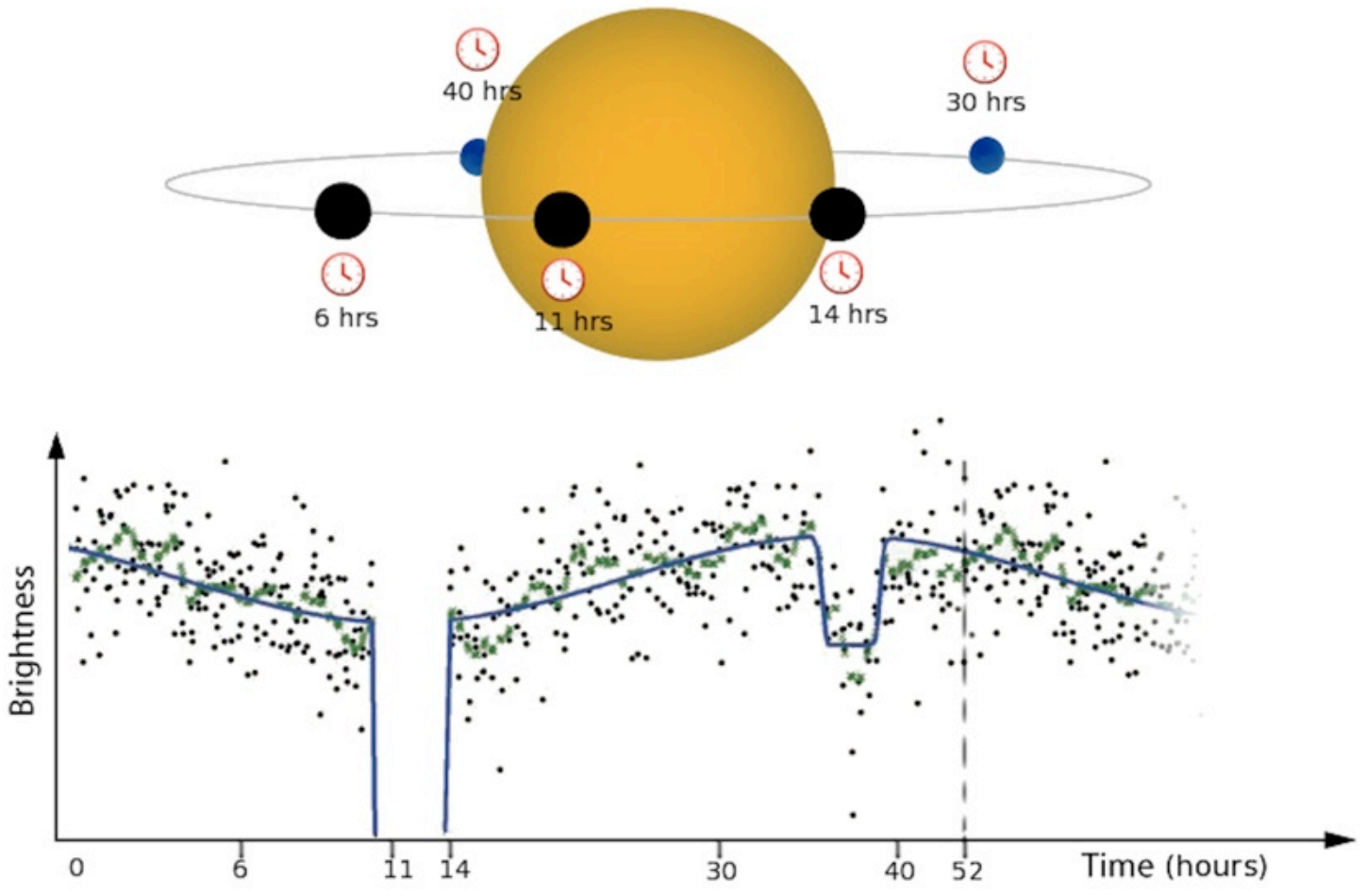

*Figure 11: Optical phase curve of planet HAT-P-7b observed by Kepler (Borucki et al., 2009) showing the transit, eclipse, and variations in brightness of system due to the varying contribution from the planet's day and night-side as function of orbital phase.*

(iii) *Planet phase variations*: During a planet's orbit, varying parts of the planet's day- and night-side are seen. By measuring the minute changes in brightness as a function of orbital phase, the longitudinal brightness distribution of a planet can be determined (Knutson et al., 2007; Borucki et al., 2009; Snellen et al., 2009). On the one hand, such observations are more challenging since the time-scales over which the planet contributions vary are significantly longer than for transit and eclipse spectroscopy. On the other hand, this method can also be applied to non-transiting planet systems (Harrington et al., 2006; Crossfield et al., 2010). Phase variations in the IR are important in understanding a planet's atmospheric dynamics and redistribution of absorbed stellar energy from their irradiated day-side to the night-side. Phase variations in the VIS are very useful to infer the cloud distribution (Demory et al., 2013).

(iv) *Exoplanet mapping and meteorological monitoring*: The combination of the three prime observational techniques utilized by EChO provides us with information from different parts of the planet atmosphere; from the terminator region via transit spectroscopy, from the day-side hemisphere via eclipse spectroscopy, and from the unilluminated night-side hemisphere using phase variations. In addition, eclipses can be used to spatially resolve the day-side hemisphere. During ingress and egress, the partial occultation effectively maps the photospheric emission region of the planet (Rauscher et al., 2007). Figure 12 illustrates possible results from eclipse mapping observations (Majeau et al., 2012). In addition, an important aspect of EChO is the repeated observations of a number of key planet targets in both transmission and secondary eclipse mode. This will allow the monitoring of global meteorological variations in the planetary atmospheres (see Section 2.4.3).

All three techniques have already been used very successfully from the optical to the near- and mid-infrared, showing molecular, atomic absorption and Rayleigh scattering features in transmission (Charbonneau et al., 2002; Vidal-Madjar et al., 2003; Redfield et al., 2007; Knutson et al., 2007, 2014; Tinetti et al., 2007, 2010; Snellen et al., 2008; Swain et al., 2008; Beaulieu et al., 2009, 2010; Linsky et al., 2010; Sing et al., 2011; Berta et al., 2012; Crouzet et al., 2012, 2014; Deming et al., 2013) and/or emission spectra (Charbonneau et al., 2008; Grillmair et al., 2008; Swain et al., 2009a,b; Stevenson et al., 2010; Todorov et al., 2014; Kreidberg et al., 2014b) of a few of the brightest and hottest transiting gas giants, using the Hubble and Spitzer space telescopes. In addition, infrared phase variations have been measured at several wavelengths using Spitzer, showing only a relatively small temperature difference (300 K) between the planet's day and night-side - implying an efficient redistribution of the absorbed stellar energy (Knutson et al., 2007). These same observations show that the hottest (brightest) part of this planet is significantly offset with respect to the sub-stellar point, indicative of a longitudinal jet-stream transporting the absorbed heat to the night-side.

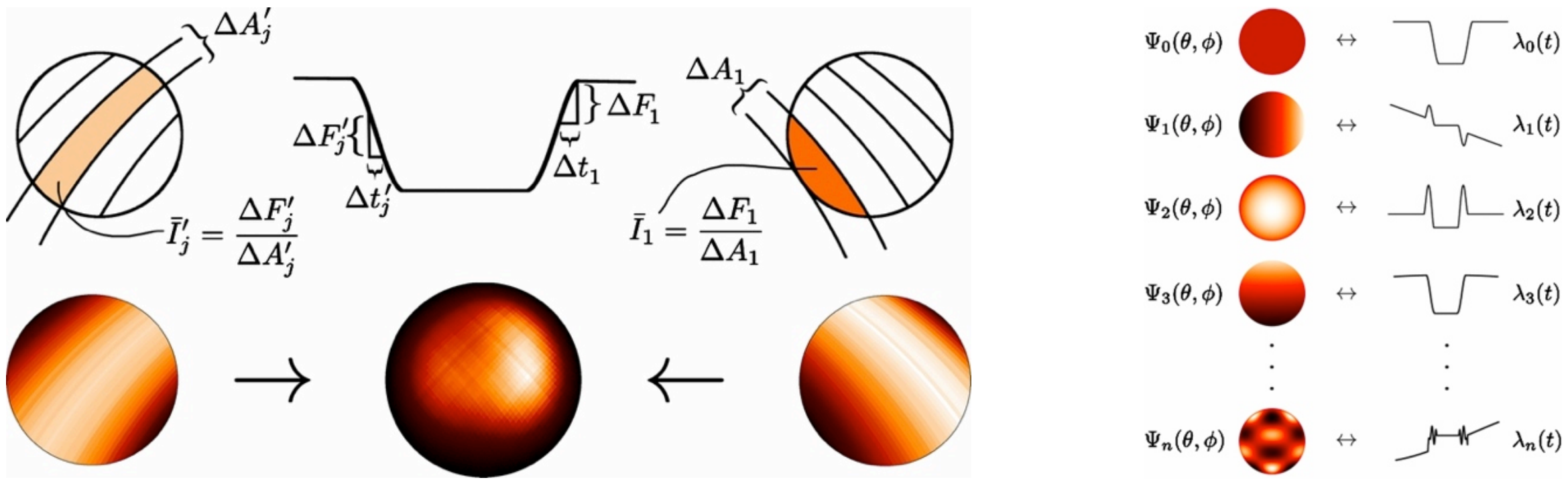

*Figure 12: Two techniques to resolve spatially the planet. Right: spherical harmonics. Left: slice mapping with ingress and egress maps as well as a combined map of HD189733b at 8 µm. These were achieved with Spitzer (Majeau et al., 2012). See also (Parmentier et al., 2014, De Witt et al., 2012).*

## 3.2 EChO observational strategy

To maximise the science return, EChO would study exoplanets both as a population & as individual objects. We describe in the following sections how EChO would achieve its objectives.

### *3.2.1 EChO spectral coverage & resolving power*

To maximise the scientific impact achievable by EChO, we need to access all the molecular species expected to play a key role in the physics and chemistry of planetary atmospheres. It is also essential that we can observe planets at different temperatures (nominally from 300 K to 3000 K, Figure 13) to probe the differences in composition potentially linked to formation and evolution scenarios. A broad wavelength coverage is therefore required to:

- Measure both albedo and thermal emission to determine the planetary energy budget.

- Capture the variety of planets at different temperatures (Tessenyi et al., 2012).
- Detect the variety of chemical components present in exoplanet atmospheres (Tessenyi et al., 2013) .
- Guarantee redundancy (i.e. molecules detected in multiple bands of the spectrum) to secure the reliability of the detection – especially when multiple chemical species overlap in a particular spectral range (Tessenyi et al., 2013; see Tables 4, 5, 6).
- Enable an optimal retrieval of the chemical abundances and thermal profile, Figure 17 (Barstow et al., 2013).

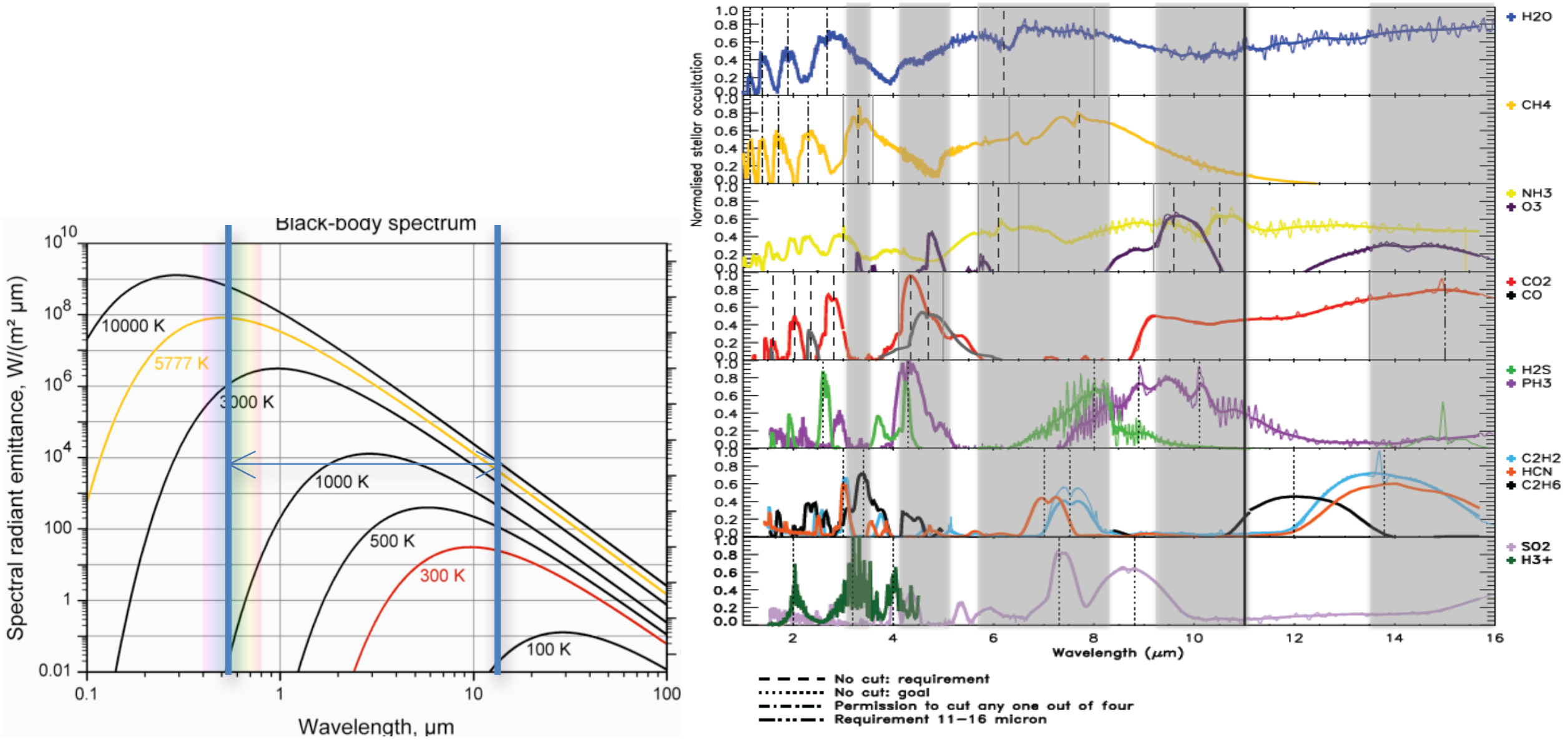

*Figure 13: Left: Blackbody curves corresponding to different temperatures: the colder the temperature, the longer the wavelengths where the Planckian curves peak. The two blue lines show optimal wavelength range to characterise planets from 300 K to 3000 K. Right: molecular signatures in the 1-16 μm range at the required and goal spectral resolving power proposed for EChO. Dashed lines indicate the key spectral features. Grey bands indicate the protected spectral windows, i.e. where no split between spectrometer channels should occur.*

This means covering the largest wavelength range feasible given the temperature limits (i.e. from the visible to the Mid-IR, ~0.4 to 16 μm). Some spectral regions are more critical than others, as it is explained in the following paragraphs (Tinetti, Encrenaz, Coustenis, 2013; Encrenaz et al., 2014).

(i) The wavelength coverage 0.55-11 μm is critical for EChO, as it guarantees that ALL the key chemical species ($H_2O$, $CH_4$, CO, $CO_2$, $NH_3$) and all other species (Na, K, $H_2S$, $SO_2$, SiO, $H_3^+$, $C_2H_2$, $C_2H_4$, $C_2H_6$, $PH_3$, HCN etc.) can be detected, if present, in all the exoplanet types observed by EChO, with the exception of $CO_2$ and $C_2H_6$ in temperate planetary atmospheres (see Figure 13).

Molecular species such as $H_2O$, $CH_4$, $CO_2$, CO, $NH_3$ are key to understand the chemistry of those planets: the broad wavelength coverage guarantees that these species can be detected in multiple spectral bands, even at low SNR, optimising their detectability in atmospheres at different temperatures. Redundancy (i.e. molecules detected in multiple bands of the spectrum) significantly improves the reliability of the detection, especially when multiple chemical species overlap in a particular spectral range. Redundancy in molecular detection is also necessary to allow the retrieval of the vertical thermal structure and molecular abundances. The wavelength range 0.55-11 μm guarantees the retrieval of molecular abundances and thermal profiles, especially for gaseous planets, with an increasing difficulty in retrieving said information for colder atmospheres (Barstow et al., 2014).

For hot planets, opacities in the visible range are dominated by metallic resonance lines (Na at 0.59 μm, K at 0.77 μm, and possibly weaker Cs transitions at 0.85 and 0.89 μm). TiO, VO and metal hydrides are also expected by analogy to brown dwarfs (Sharp & Burrows, 2007).

(ii) The target wavelength coverage of 0.55-16 μm guarantees that $CO_2$ and $C_2H_6$ can be detected in temperate planetary atmospheres. It also offers the possibility of detecting additional absorption features for HCN, $C_2H_2$, $CO_2$ and $C_2H_6$ for all other planets and improves the retrieval of thermal profiles (Barstow et al. 2014).

(iii) The target wavelength coverage of 0.4-11 μm might improve the detection of Rayleigh scattering in hot and warm gaseous planets if clouds are not present. In a cloud-free atmosphere, the continuum in the UV-VIS is given by Rayleigh scattering on the blue side, i.e. for wavelengths shorter than 1 μm (Rayleigh scattering varies as $1/\lambda^4$). If there are clouds or hazes with small-size particles, those should be detectable in the visible even beyond 0.55 μm (see Figure 19).

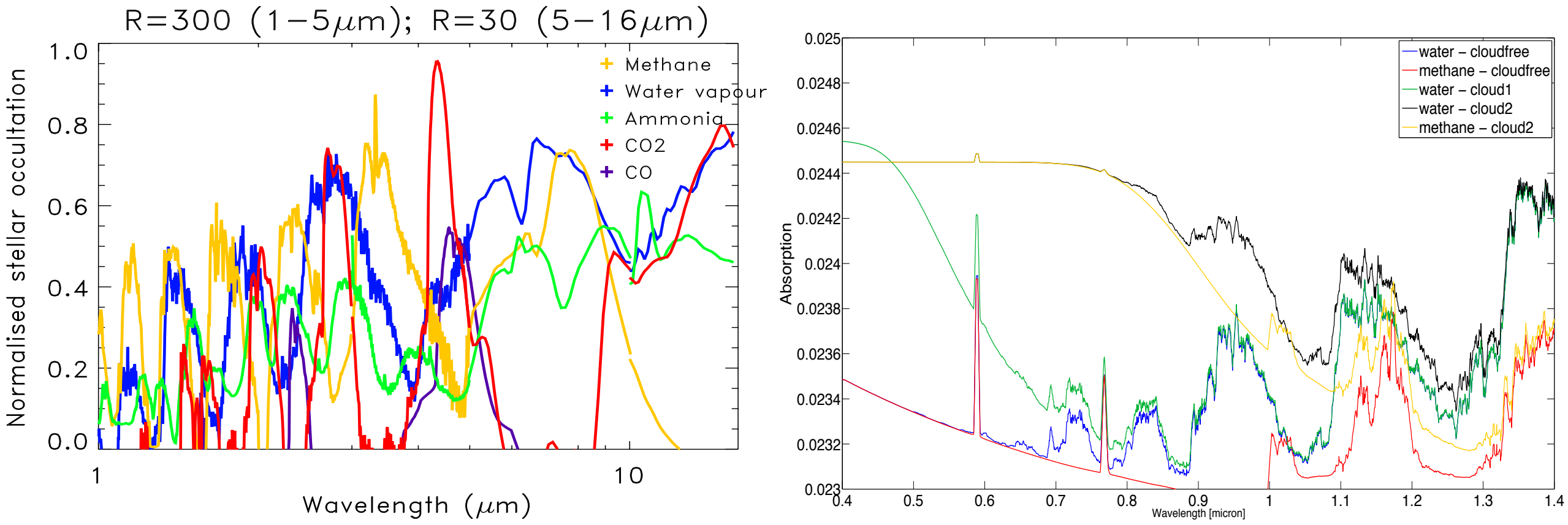

*Figure 14: Simulated transmission spectra of a gaseous exoplanet at 800 K (Hollis et al., 2013). The atmospheric absorption is normalised to 1; typically the fraction of stellar flux absorbed by the atmosphere of a hot planet is $10^{-4}$-$10^{-3}$. The spectra were generated at a resolving power R=300 for $\lambda < 5$ μm and R =30 for $\lambda > 5$ μm (left). Right: transmission spectra of cloud-free and cloudy atmosphere of a gaseous planet (Hollis et al., 2013). Particle size, shape, distribution and the pressure of the atmospheric layer where clouds/hazes form cause changes in the spectra in the VIS-NIR (Liou, 2002).*

(iv) A spectral resolving power of R = 300 for $\lambda < 5$ μm will permit the detection of most molecules at any temperature. At $\lambda > 5$ μm, R = 30 is enough to detect the key molecules at hot temperatures, due to broadening of their spectral signatures. For temperate planets, R = 30 at longer wavelengths is also an optimal solution, given there are fewer photons (Tinetti, Encrenaz, Coustenis, 2013).

In Figure 14 left, two values (300 and 30) are used for the spectral resolving power of the simulated transmission spectra. In addition to the main candidate absorbers ($H_2O$, $CH_4$, $NH_3$, CO, $CO_2$), Figure 13 shows the contributions from HCN, $O_3$, $H_2S$, $PH_3$, $SO_2$, $C_2H_2$, $C_2H_6$ and $H_3^+$. Among those, $H_3^+$ around 2 μm and 3-4 μm is detectable with a resolving power of > 100.

While R=30 enables the detection of most of the molecules absorbing at $\lambda > 5$ μm, especially at higher temperatures, we would lose the possibility of resolving the $CO_2$, HCN and other hydrocarbon Q-branches, for which R>100 is needed. The current instrument design allows a spectral resolving power between the two.

In the visible, for cloud-free atmospheres, a resolving power of ~ 100 is still sufficient for identifying the resonance lines of Na and K, but not to resolve the centre of the lines. For the star, Hα can be easily identified at 0.656 μm.

### *3.2.2 EChO's three surveys*

An optimised way to capture the EChO science case is through three survey tiers. These are briefly described below and summarised in Table 3 and in Figure 15.

## Chemical Census

- For all planetary cases (hundreds of planets), this tier will measure the planetary albedos and bulk temperatures.
- For all planetary cases (hundreds of planets), this tier enables the detection of the strongest features in the measured spectra. These include the presence of clouds or hazes, and the major atomic and molecular species (e.g. Na, K, $CH_4$, CO, $CO_2$, $NH_3$, $H_2O$, $C_2H_2$, $C_2H_6$, HCN, $H_2S$ and $PH_3$), provided the atomic/molecular abundances are large enough (e.g. mixing ratios ~ $10^{-6}/10^{-7}$ for $CO_2$, $10^{-4}/10^{-5}$ for $H_2O$), see Tables 3, 4, 5.

- For the temperate super-Earths, we also show that with R =30 and SNR=5, $O_3$ can be detected with an abundance of $10^{-7}$ at 9.6 μm, see Table 6.

## Origin survey

A subsample of the Chemical Census (tens of planets). The Origin tier allows:

- Higher degree of confidence in the detection of key molecular features in multiple bands (see Tables 3, 4, 5, 6, Figure 18) enabling the retrieval of the vertical thermal profile (Figures 17, 18)
- Measurement of the abundances of trace gases (see Tables 3, 4, 5) constraining the current proposed scenarios for the chemical and physical processes for exoplanet atmospheres (see Section 2.4).
- Allow determination of the C/O ratio and constrain planetary formation/migration scenarios (see Section 2.4.5)
- Constrain the type of clouds and cloud parameters when condensates are present (thickness, distribution, particle size, cloud-deck pressure).

## Rosetta Stones

Benchmark cases, which we plan to observe in great detail to understand an entire class of objects. For these planets we can observe:

- Weak spectral features for which the highest resolving power and SNR are needed.

Among Rosetta Stones, a good candidate for the Exo-Meteo & Exo-Maps survey, is a planet whose requirements for the Chemical Census can be achieved in one transit or eclipse. Gaseous planets such as HD 189733b, HD 209458b, or GJ 436b are the most obvious candidates for this type of observations today, meaning we can observe:

- Temporal variability, i.e. Exo-Meteo (weather, Section 2.4)
- Spatial resolution, i.e. Exo-Maps (2D and 3D maps, Section 2.4)

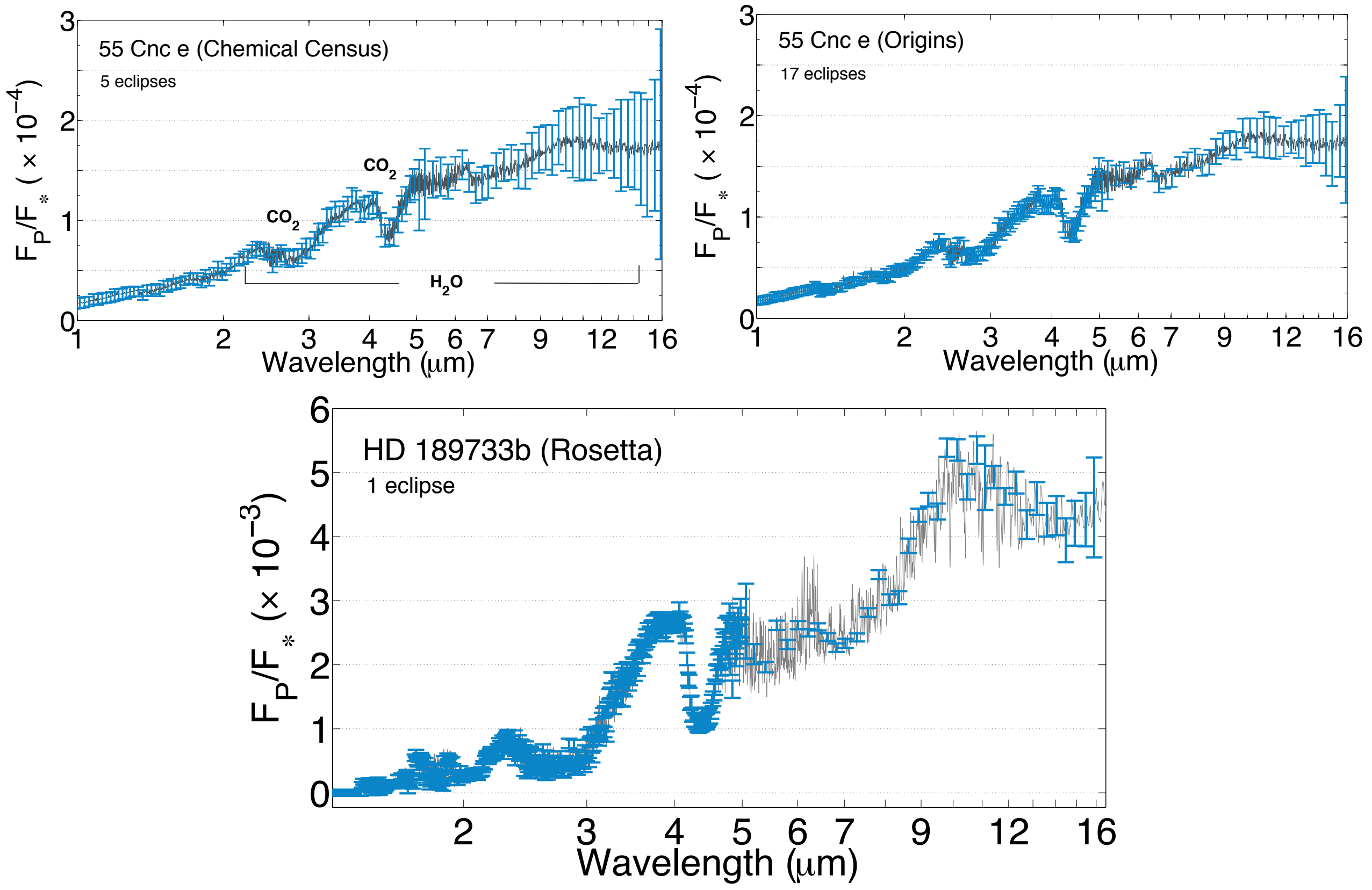

*Figure 15: EChOSim simulations (see Section 3.4) of transmission and emission spectra as observed by EChO with different survey programs. The transits or eclipses needed are reported in the figure. Top: emission spectra of super-Earth 55 Cnc e with Chemical Census and Origin surveys. The spectral features of $CO_2$ and water vapour are detectable in Chemical Census, their abundances and thermal profile retrievable in Origin. Bottom: emission spectrum of hot Jupiter HD 189733b (Rosetta Stones program). The key gases are retrievable very precisely, see Figures 17 & 18.*

| ***Tier*** | ***Key science objectives*** | ***Observables & derived products*** | ***Observational strategy*** |
|---|---|---|---|
| ***Chemical Census*** ***(Survey)*** | • *Exploring the diversity of exoplanet atmospheres*<br>• *What are exoplanets made of?* | - Presence of most abundant atmospheric components, e.g. $H_2O$, $CH_4$, CO, $CO_2$, $NH_3$ etc.<br>- Albedo and thermal emission<br>- Presence of clouds/hazes | A sample of planets (hundreds) which is representative of the local volume (super-earths, Neptunes & Jupiters, with a range of temperatures, orbital and stellar parameters).<br>**R~50 for $\lambda < 5\mu m$**<br>**R~30 for $\lambda > 5\mu m$**<br>**SNR~5**<br>Transits or eclipses until the required R & SNR is reached to detect most abundant atmospheric molecules. |
| ***Origin*** ***(Deep survey)*** | a. *Understanding the origin of exoplanet diversity & the physical mechanisms in place*<br>b. *How do planet form and evolve?* | - Molecular abundances of both key components and trace gases in the atmosphere,<br>- vertical thermal profiles,<br>- constraints on clouds/albedo. | A subset (tens) of the planets analysed through the Chemical Census tier, with a prevalence of Neptunes and Jupiters.<br>**R~100 for $\lambda < 5\mu m$**<br>**R~30 for $\lambda > 5\mu m$**<br>**SNR~10**<br>Transits + eclipses until the required SNR & R are reached to retrieve molecular abundances for most trace gases and vertical thermal profiles. |
| ***Weather, Exo-maps & Rosetta Stones*** ***(Ultra deep survey)*** | *A very detailed and exhaustive study of a select sample of benchmark cases.* | - Very precise molecular abundances of key components and trace gases,<br>- vertical and horizontal thermal profiles and chemical gradients,<br>- spatial and temporal variability, | A select sample chosen among the most favourable exoplanets in their own category (typically 10 or 20). For the Exo-Meteo and Exo-Maps, exoplanets whose stars are very bright should be selected (e.g. HD 189733b).<br>**R~300 for $\lambda < 5\mu m$**<br>**R~30 for $\lambda > 5\mu m$**<br>**SNR~20** |
| | | - orbital modulations,<br>- constraints on clouds/albedo. | Many repeated obs. of transits and/or eclipses +<br>orbital lightcurves +<br>eclipse mapping. |

*Table 3: Summary of EChO's three tiers: objectives addressed and observational strategies adopted.*

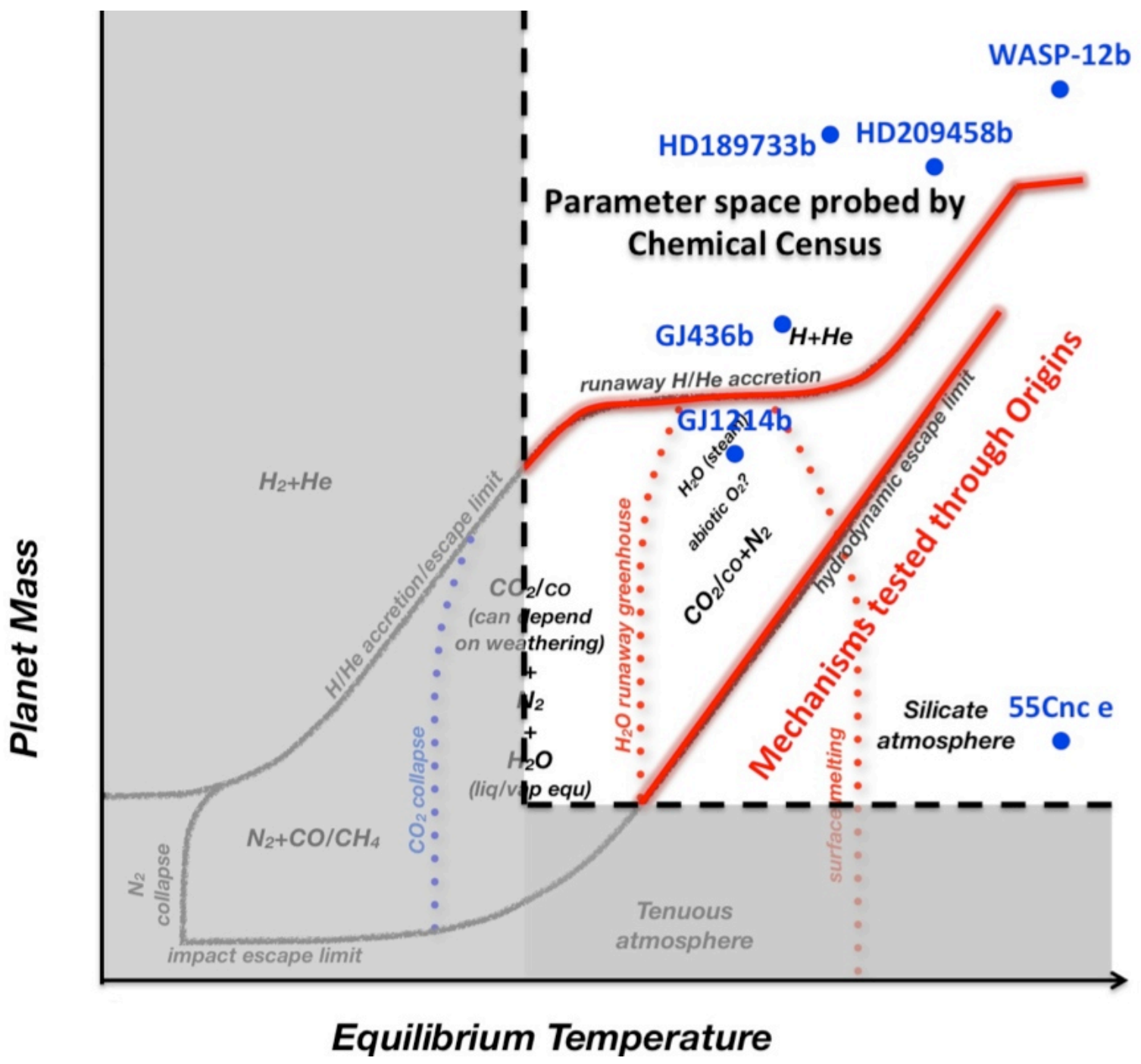

*Figure 16: Parameter space probed by the Chemical Census, i.e. a large number of planets with masses ranging from ~ 5 Earth Masses to very massive Jupiters, and temperatures spanning two orders of magnitude, i.e. from temperate, where water can exist in a liquid phase, to extremely hot, where iron melts. A few known planets, benchmark cases representative of classes of objects, are shown in the diagram to orientate the reader. These are excellent objects to study as Rosetta Stones. Key physical processes responsible of transitions among classes of exoplanets are identified: these mechanisms can be tested through the Origin survey.*

### *3.2.3 Optimal SNR & information retrieved*

Most of the science objectives detailed in Section 3.2, are based on the assumption that EChO can retrieve the molecular composition and the thermal structure of a large number of exoplanet atmospheres at various levels of accuracy and confidence, depending on the scientific question and target selected.

We consider here the goal wavelength coverage assumed for EChO, i.e. 0.4 to 16 μm, and investigate the key molecular features present in a range of planetary atmospheres with a temperature between ~300 K and 3000 K. In a planetary spectrum, as measured through a transit or an eclipse, the molecular features appear as departures from the continuum. At a fixed temperature-pressure profile, the absorption depth or emission features will depend only on the abundance of the molecular species. Tables 4 to 6 show the minimum abundance detectable for a selected molecule absorbing in a planetary atmosphere, as a function of wavelength and observing tier, i.e. Chemical Census, Origin, Rosetta Stones (see Table 3). We show here the results for three planetary cases: warm Neptune, hot and temperate super-Earth. The spectral resolving power is lowered to R=20 in the 5 to 16 μm spectral interval for the temperate super-Earth, being the most challenging planet type that EChO might observe. For simulations on hot and temperate Jupiters see (Tessenyi et al., 2013).

As shown by Tables 4, 5 and 6, for most planetary cases, the Chemical Census tier is enough to detect the very strongest spectral features for the most abundant molecules, whereas the Origin tier can reveal most molecules with mixing ratios of $10^{-6}$ or lower, often at multiple wavelengths, which is excellent for constraining the type of chemistry or the C/O ratio. The robustness of these results was tested by exploring the sensitivity to parameters such as the vertical thermal profile, the mean molecular weight of the atmosphere and the relative water abundances: the main conclusions remain valid except for the most extreme cases (Tessenyi et al., 2013). Should clouds/hazes be present, having multiple absorption bands available greatly help the molecular detection. In general, small cloud particles affect mainly the short wavelengths (i.e. VIS and NIR), while the atmosphere becomes more transparent at longer wavelengths (Liou, 2002).

| | $CH_4$ | | $CO$ | | $CO_2$ | | | $NH_3$ | | | $H_2O$ | | |
|---|---|---|---|---|---|---|---|---|---|---|---|---|---|
| Obs. Mode | 3.3$\mu m$ | 8$\mu m$ | 2.3$\mu m$ | 4.6$\mu m$ | 2.8$\mu m$ | 4.3$\mu m$ | 15$\mu m$ | 3$\mu m$ | 6.1$\mu m$ | 10.5$\mu m$ | 2.8$\mu m$ | 5-8$\mu m$ | 11-16$\mu m$ |
| Rosetta St. | $10^{-7}$ | $10^{-6}$ | $10^{-4}$ | $10^{-6}$ | $10^{-7}$ | $10^{-7}$ | $10^{-7}$ | $10^{-7}$ | $10^{-6}$ | $10^{-7}$ | $10^{-6}$ | $10^{-6}$ | $10^{-5}$ |
| Origins | $10^{-7}$ | $10^{-6}$ | $10^{-3}$ | $10^{-5}$ | $10^{-6}$ | $10^{-7}$ | $10^{-6}$ | $10^{-6}$ | $10^{-6}$ | $10^{-6}$ | $10^{-6}$ | $10^{-5}$ | $10^{-4}$ |
| Ch. Census | $10^{-7}$ | $10^{-5}$ | $10^{-3}$ | $10^{-4}$ | $10^{-6}$ | $10^{-7}$ | $10^{-5}$ | $10^{-5}$ | $10^{-5}$ | $10^{-5}$ | $10^{-5}$ | $10^{-5}$ | $10^{-4}$ |

| | $HCN$ | | | $PH_3$ | | $C_2H_6$ | | $H_2S$ | | | $C_2H_2$ | | |
|---|---|---|---|---|---|---|---|---|---|---|---|---|---|
| Obs. Mode | 3$\mu m$ | 7$\mu m$ | 14$\mu m$ | 4.3$\mu m$ | 10$\mu m$ | 3.3$\mu m$ | 12.2$\mu m$ | 2.6$\mu m$ | 4.25$\mu m$ | 8$\mu m$ | 3$\mu m$ | 7.5$\mu m$ | 13.7$\mu m$ |
| Rosetta St. | $10^{-7}$ | $10^{-5}$ | $10^{-7}$ | $10^{-7}$ | $10^{-6}$ | $10^{-6}$ | $10^{-6}$ | $10^{-5}$ | $10^{-4}$ | $10^{-4}$ | $10^{-7}$ | $10^{-5}$ | $10^{-7}$ |
| Origins | $10^{-6}$ | $10^{-5}$ | $10^{-6}$ | $10^{-7}$ | $10^{-6}$ | $10^{-5}$ | $10^{-5}$ | $10^{-5}$ | $10^{-4}$ | $10^{-3}$ | $10^{-7}$ | $10^{-4}$ | $10^{-6}$ |
| Ch. Census | $10^{-6}$ | $10^{-4}$ | $10^{-5}$ | $10^{-7}$ | $10^{-5}$ | $10^{-5}$ | $10^{-5}$ | $10^{-4}$ | $10^{-3}$ | - | $10^{-7}$ | $10^{-3}$ | $10^{-5}$ |

*Table 4: Examples of average detectable abundances for a warm-Neptune (e.g. GJ 436b) for the three tiers (Tessenyi et al., 2013). The molecular abundance is expressed as mixing ratio.*

| | $H_2O$ | | | $CO_2$ | | |
|---|---|---|---|---|---|---|
| Obs. Mode | 2.8$\mu m$ | 5 - 8$\mu m$ | 11 - 16$\mu m$ | 2.8$\mu m$ | 4.3$\mu m$ | 15$\mu m$ |
| Rosetta St. | $10^{-4}$ | $10^{-4}$ | $10^{-4}$ | $10^{-5}$ | $10^{-7}$ | $10^{-5}$ |
| Origins | $10^{-4}$ | $10^{-3}$ | $10^{-3}$ | $10^{-5}$ | $10^{-6}$ | $10^{-4}$ |
| Ch. Census | $10^{-3}$ | - | - | $10^{-4}$ | $10^{-5}$ | - |

*Table 5: Examples of average detectable abundances for a hot super-Earth around a G-type star (e.g. 55 Cnc e) for the three tiers. The molecular abundance is expressed as mixing ratio.*

| | $H_2O$ | | $CO_2$ | $NH_3$ | | $O_3$ | |
|---|---|---|---|---|---|---|---|
| SNR | 5 - 8 $\mu m$ | 11 - 16 $\mu m$ | 15 $\mu m$ | 6 $\mu m$ | 11 $\mu m$ | 9.6 $\mu m$ | 14.3 $\mu m$ |
| 10 | $10^{-6}$ | $10^{-4}$ | $10^{-6}$ | $10^{-6}$ | $10^{-6}$ | $10^{-7}$ | $10^{-5}$ |
| 5 | $10^{-6}$ | $10^{-4}$ | $10^{-6}$ | $10^{-5}$ | $10^{-6}$ | $10^{-7}$ | $10^{-5}$ |

*Table 6: Examples of average molecular detectability for a temperate super-Earth (~ 320 K) around a late M for fixed SNR and R=20. The molecular abundance is expressed as mixing ratio.*

Similar conclusions were reached through simulations with the NEMESIS (Non-linear optimal Estimator for Multivariate spectral analysis) radiative transfer and retrieval tool (Barstow et al., 2013; 2013a). NEMESIS was used to explore the potentials of the proposed EChO payload to solve the retrieval problem for a range of $H_2$-He planets orbiting different stars and Ocean planets such as GJ 1214b.

NEMESIS results show that EChO should be capable of recovering all gases in the atmosphere of a hot-Jupiter to within 2-sigma for all tiers. However, we see differences in the retrieved T-p profile between the Chemical Census, Origin and Rosetta Stone tiers. As expected, for the Chemical Census the spectral resolution is too low to fully break the degeneracy between temperature and gas mixing ratios, so the retrieved profile is less accurate. This is not the case for Origin and Rosetta Stone (Figure 17). Examples of spectral fits for the Rosetta case are also shown in Figure 17. The temperature prior chosen does not affect the retrieval or the spectral fit.

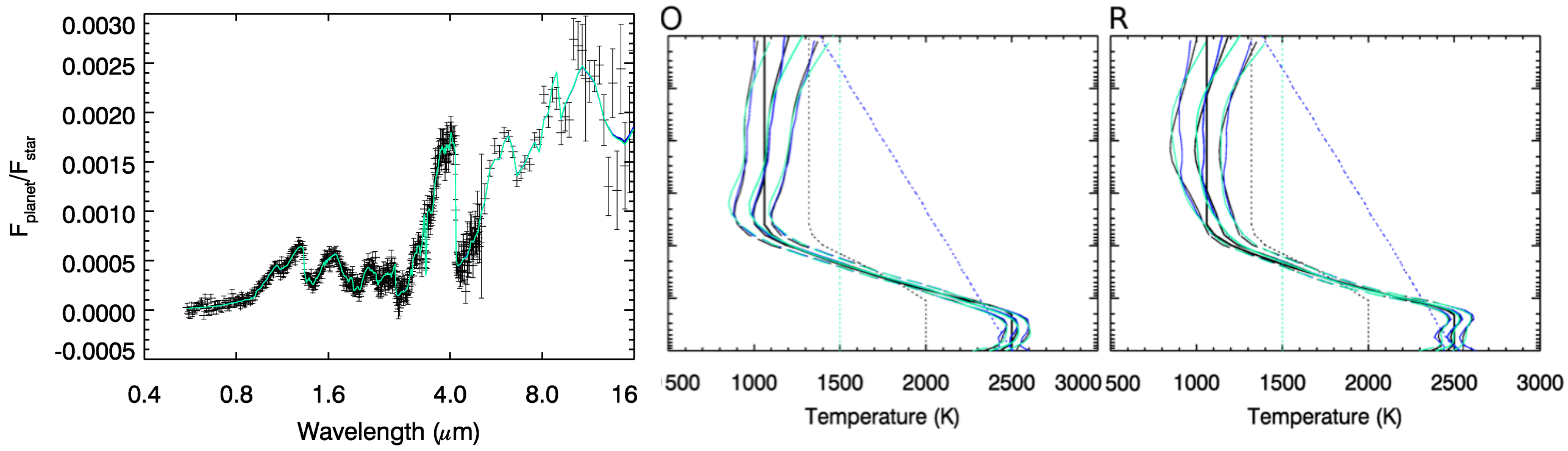

*Figure 17: Left: Eclipse spectra for a hot-Jupiter observed in Rosetta Stone program. The fitted spectra colours correspond to different temperature priors, as on the right. The temperature prior used does not affect the resultant spectral fit. Right: Temperature retrievals of a hot-Jupiter from eclipse observations (L-R: Origin, Rosetta Stone). The three different temperature priors used are shown by dotted lines; the thick black line is the input profile, and the three retrieved profiles are shown by the thin solid lines. The retrieval error is shown by the dashed lines.*

Similar results were obtained for the hot-Jupiter's transit spectra and for the hot-Neptune's transit and eclipse spectra (Figure 18; Barstow et al., 2014). In primary transit, it is not possible to retrieve independently the T-p profile due to the limited sensitivity to temperature, but by performing multiple retrievals with different assumed T-p profiles and comparing the goodness-of-fit of the resulting spectra, we can obtain the constraints needed. In Figure 18, the different colours correspond to retrievals using different model T-p profiles, with the best fit being provided by the input temperature profile, as expected. From this, we can correctly infer the temperature and gaseous abundances from primary transit.

As well as constraining the temperature of hot Jupiters and Neptunes, with a few tens of eclipses we can obtain sufficient signal-to-noise to allow a retrieval of the stratospheric temperature of super-Earths atmospheres, such as GJ 1214b, which has not been achieved to date (Barstow et al., 2013a). An independent constraint on the temperature will be valuable for interpreting the better-studied transit spectrum of GJ 1214b, which will also be significantly improved in quality by EChO observations (see Figure 5).

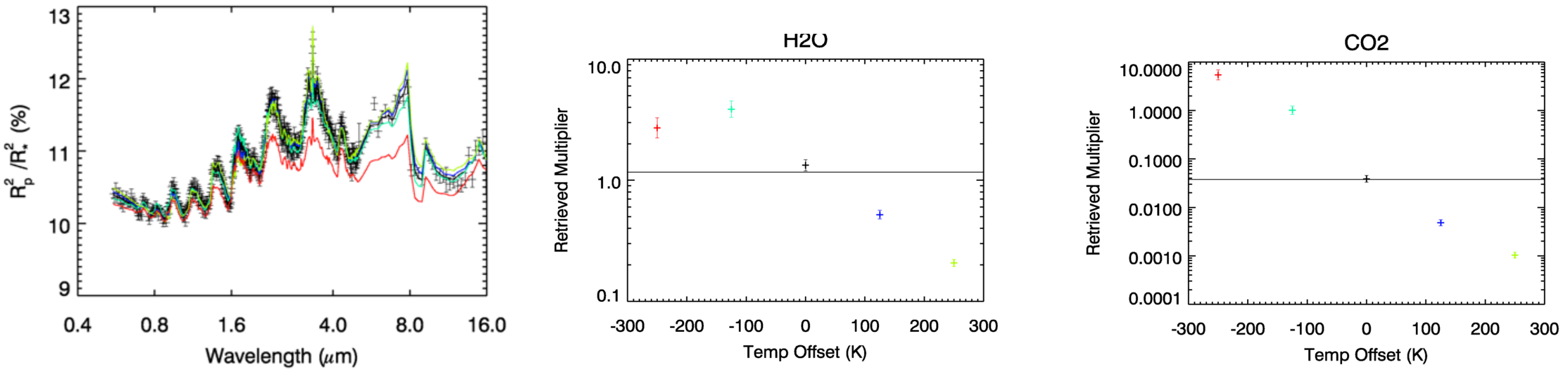

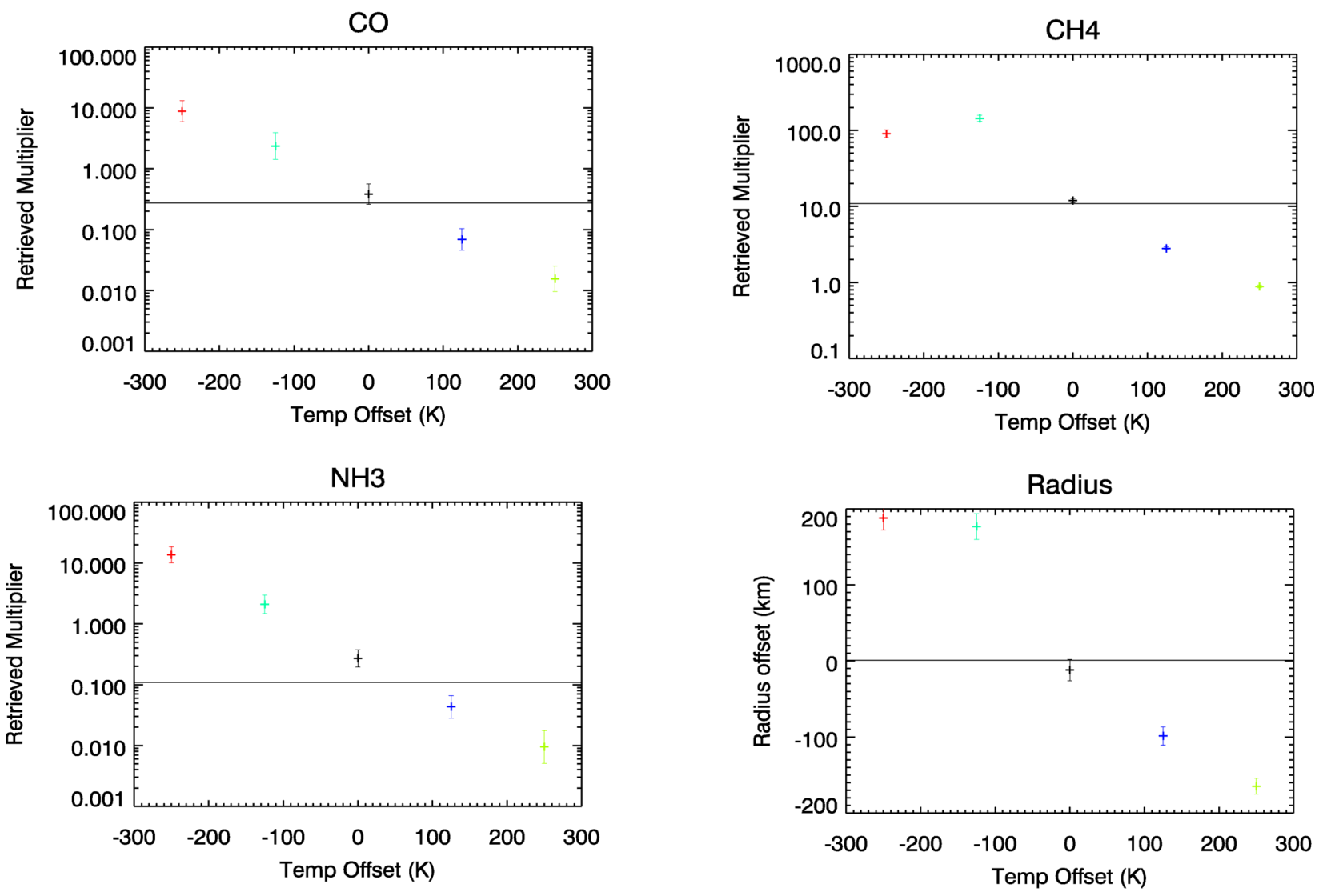

*Figure 18: Retrieved results for a hot Neptune transit spectrum observed in the Rosetta Stone tier.*

*Colours correspond to different reduced $\chi^2$: red=17.1, green=2.0, black=1.2 (best fit), blue=1.6, yellow=2.8. For a good retrieval the reduced $\chi^2$ should be close to 1. The best fit is the black one, for which the temperature and gases are correctly retrieved.*

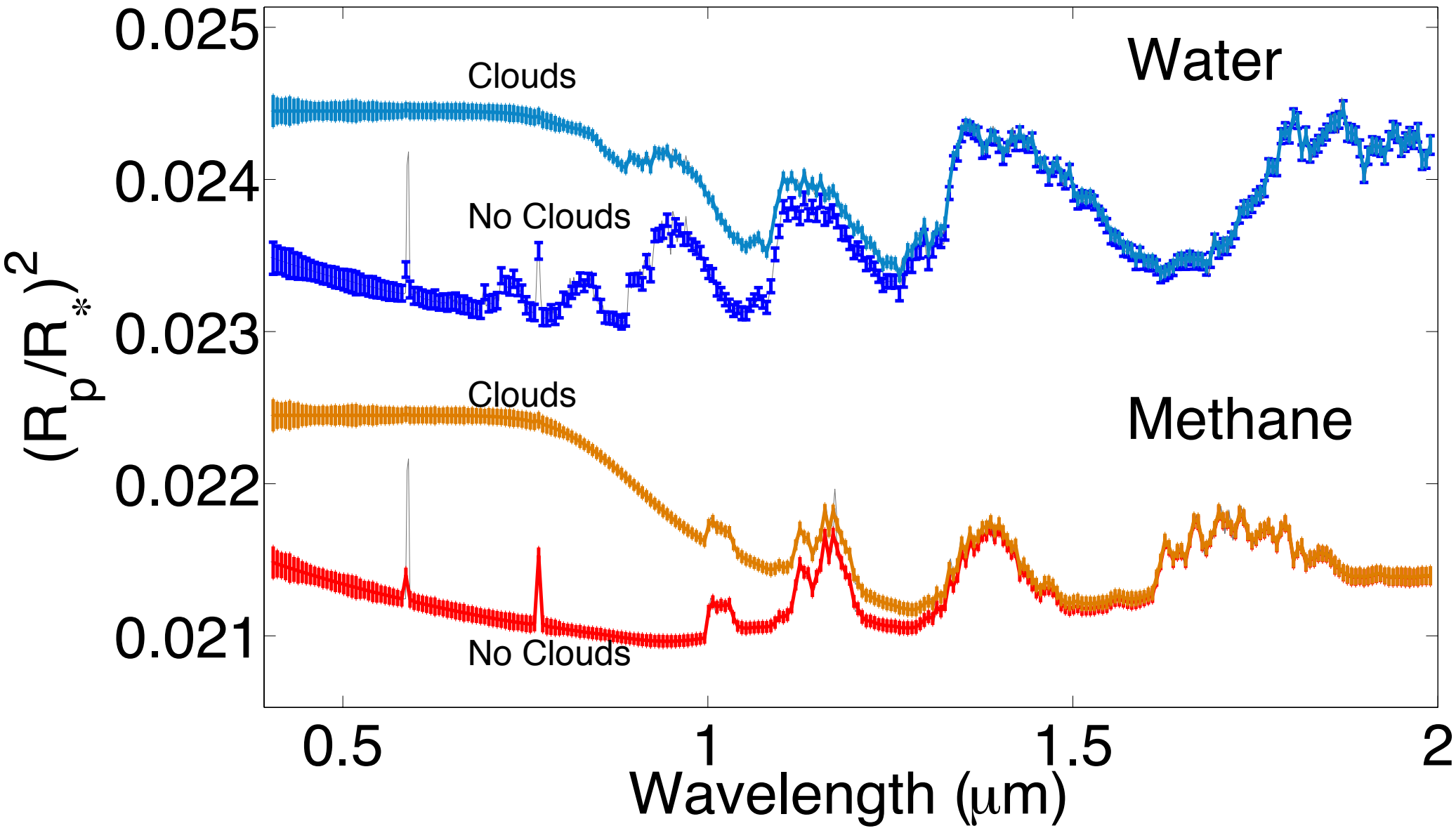

*Figure 19: Examples of cloudy and clear-sky gaseous planet spectra with different molecular compositions as observed by EChO. The performances obtainable with EChO allow the detection of clouds/hazes and their characteristics, as well as the extraction of the molecular abundances. For clarity, we have included an offset between the methane-rich and the water-rich spectra.*

## 3.3 Laboratory data for EChO

### *3.3.1 Linelists*

Interpreting exoplanetary spectra requires access to appropriate laboratory spectroscopic data, as does the construction of associated radiative transport and atmospheric models. These objects may reach temperatures up to about 3000 K meaning that billions of transitions are required for an accurate model (Yurchenko & Tennyson 2014; Yurchenko et al., 2014). A dedicated project is in progress to provide comprehensive sets of line lists for all the key molecules expected to important in exoplanet atmospheres (both hydrogen-rich gas giants and oxygen-rich terrestrial-like atmospheres). The ExoMol project (www.exomol.com) aims at providing complete lists for the 30 most important species (including methane, water, ammonia, phosphine, hydrogen sulphide, a variety of hydrocarbons and a long list of stable and open shell diatomics) by 2016 (Tennyson & Yurchenko, 2012). These data will therefore be available for pre-launch testing and design studies (Tennyson & Yurchenko, 2014).

### *3.3.2 Reaction / photodissociation rates*

The diversity of exoplanetary atmospheres observable with EChO spans a broad range of physical conditions. Individual reaction rates must therefore be known at temperature ranging from below room temperature to above 2500 K and – because the deep atmospheric layers are chemically mixed with the layers probed by spectroscopic observations – at pressures up to about 100 bars. Today these rates are well-known at room temperature, but only rarely determined at high temperature. The teams from University of Bordeaux and LISA Créteil, France, are measuring new photoabsorption cross-section at high temperatures, at wavelengths shorter than 200 nm (Venot et al., 2013). The first measurements for $CO_2$ have been performed at the synchrotron radiation facility BESSY, in Berlin, and at LISA, Créteil.

### *3.3.3 Optical properties of gases at high Pressure-Temperature*

Despite various measurements and theoretical models dedicated to the optical properties of gases, accurate data at different temperatures and pressures are still lacking in numerous spectral regions. Little or no data in some case are available for continuum absorption, line mixing, far wings and collision induced absorption, even for the well-studied carbon dioxide molecule. The scenario is further complicated by the need to reproduce in the laboratory very long path lengths to be able to measure weak but important absorption and/or to boost the sensitivity and accuracy of the setup. New data will become available due to experiments performed in support to operational or planned solar system missions. In particular, measurements are available from the laboratory at INAF-IAPS Rome (http://exact.iaps.inaf.it) performed for Venus Express orbiting around Venus (Stefani et al., 2013), and more measurements are planned for JUNO presently in cruise to Jupiter. Finally, the increasing availability of new tunable lasers in the EChO spectral range makes possible the use of the cavity ring down technique, which has been proven to be very effective e.g. in the continuum measurements of the Venus' atmospheric windows (Snels et al., 2014).

## 3.4 Dealing with systematic & astrophysical noise

### *3.4.1 EChO performance requirements*

EChO's top-level requirement is that the photometric stability over the frequency band of interest shall not add significantly to the photometric noise from the astrophysical scene (star, planet and zodiacal light). The frequency band over which the requirement applies is between $2.8\times10^{-5}$ Hz and 3.7 mHz, or ~5 minutes to 10 hours (Puig et al., 2014; Eccleston et al., 2104; Pascale et al., 2014; Waldmann and Pascale, 2014). This implies having the capability to remove any residual systematics and to co-add the elementary observations from many repeat visits to a given target.

The photometric stability budget is described by Puig et al. (2014), Pascale et al. (2014), and Waldmann and Pascale (2014). To achieve the required performance, particular attention is required to:

- the <u>design of the instrument</u>
- the <u>calibration strategy</u> to characterise all possible systematic variations in performance
- the <u>data processing pipeline</u>(s).

We briefly discuss these topics in the following sections.

### *3.4.2 Design of the instrument and knowledge of its characteristics*

The most important factor determining the final performance of the mission is the way the instrument is designed. Even though the whole wavelength range is divided into bands observed using different physical spectrometer modules, the instrument is designed to operate as a single entity within the same thermal, optical, electrical and mechanical environment.

Particular care has been given to the way the modules are designed in order to share similar technological solutions for each module. For example, the detector technology is similar among all the modules and the readout units and the common electronics are designed as a single unit to simplify the electro-magnetic compatibility. All the modules as well as the Fine Guiding Sensor (FGS) share a common field of view and telescope optical train with specific dichroics mounted on the same optical bench. They are thus at the same temperature and see the same mechanical environment. In this way optical path errors between modules and the common optics are reduced to a minimum and thermo-mechanical drift within the instrument is eliminated by having an isothermal design of the optical modules. Any pointing jitter is seen directly by both the FGS and the spectrometer instrument and this information can be accounted for in the data processing. Likewise, through calibration, performance monitoring and use of the FGS data, changes in optical path between the telescope and the instrument (such as "breathing" of the point spread function or changes in telescope focus) can be identified and calibrated out of the data.

During the development phase all the critical components, particularly the detectors, will be intensively tested to determine their intrinsic characteristics. This will include determining their sensitivity to environmental variations such as temperature variations, pointing jitter, high-energy particles, electro-magnetic contamination etc. The aim is to understand and predict the evolution of the instrument response when the environmental conditions vary, and therefore to optimise the correction pipeline and the housekeeping monitoring needed as input to the pipeline. The overall instrument will thus be fully calibrated and its performance verified at subsystem and system level before launch in order to check its global behaviour and evaluate its performance using laboratory calibration sources.

### 3.4.3 The calibration strategy

As described in Eccleston et al., (2014), photovoltaic detectors based on MCT (mercury cadmium telluride) will be used for EChO. They are known to have various non-linear behaviours both in regard to responsivity and dark current. Whilst we have designed an instrument that allows to monitor continuously this behaviour during observation phases, it will also be necessary to verify the behaviour of the detectors and instrument in flight over a number of timescales (in-flight calibration). These will range from determining the short-term response of the detectors through to slow changes in the instrument performance due to the effects of the space environment and component ageing. It is therefore necessary to consider regular calibration phases between the observations and, possibly, during them. Depending on the final temporal stability of the instrument, several parameters will be checked at different timescales from several hours to days. The calibration strategy includes the use of both an internal calibration unit within the instrument and a list of stable stars (known to be stable to $10^{-5}$ over the necessary timescales) spread all over the sky.

### 3.4.4 The data processing

It is crucial to correct the raw observed signal time series to account for variations in the signal which are not directly linked to the planetary transit or occultation. The methods for doing this will be encapsulated in the data processing algorithms to be employed in the data pipeline; the final data quality and performance of EChO are highly dependent on the performance of these algorithms. There may be many systematic variations to account for, most of which will be negligible, but we highlight two areas requiring particular attention:

- *The astrophysical scene contributions*: the stellar variability, the local zodiacal cloud contribution, the exozodiacal cloud contribution and any contaminating stars. These are independent of the instrument performance but may add systematic signals resembling the transiting planet.

- *The instrument drifts, pointing jitter, detector non-linearity and any dependence on environmental variations and ageing.* These effects will be highly correlated between the spectral bands and many of the effects will be monitored by, for example, off axis detectors, thermistors, the Fine Guidance Sensor and will ultimately be assessed through dedicated calibration observations.

These issues will be addressed by data reduction techniques validated on current instruments as described in sections 3.4.5 and 3.4.6. These techniques use the inherent redundancy in the data, knowledge of the target planetary orbital phase and secondary information from the instrument and satellite to remove unwanted systematic effects.

### 3.4.5 Decorrelating instrument systematics

Detecting the atmospheric signal of an exoplanet requires high precision measurements. Limitations to said precision come from the systematic noise associated with the instrument with which the data are observed. This is particularly true for general, non-dedicated observatories. In the past, parametric models have been used extensively by most teams in the field of exoplanet spectroscopy/differential band photometry to remove instrument systematics (Agol et al. 2010; Beaulieu et al. 2008, 2010, 2011; Charbonneau et al. 2005, 2008; Crouzet et al., 2012, 2014; Deming et al. 2013; Grillmair et al. 2008; Knutson et al. 2007, 2014; Kreidberg et al., 2014, 2014b; Stevenson et al. 2010; Swain et al. 2008, 2009a,b; Tinetti et al., 2007b, 2010; Todorov et al., 2014). Parametric models approximate systematic noise via the use of auxiliary information about the instrument, the so called Optical State Vectors (OSVs). Such OSVs often include the X and Y-positional drifts of the star or the spectrum on the detector, the focus and the detector temperature changes, as well as positional angles of the telescope on the sky. By fitting a linear combination of OSVs to the data, the parametric approach derives its systematic noise model. We refer to this as the “linear, parametric” method. In many cases precisions of a few parts in 10000 with respect to the stellar flux were reached.

In the case of dedicated missions, such as Kepler (Borucki et al. 1996; Jenkins et al. 2010), the instrument response functions are well characterised in advance and conceived to reach the required $10^{-4}$ to$10^{-5}$ photometric precision. EChO aims at reaching the same level of photometric precision. For general purpose instruments, not calibrated to reach this required precision, poorly sampled OSVs or a missing parameterisation of the instrument often become critical issues. Even if the parameterisation is sufficient, it is often difficult to determine which combination of these OSVs may best capture the systematic effects of the instrument. This approach has caused some debates for current instruments regarding the use of different parametric choices for the removal of systematic errors.

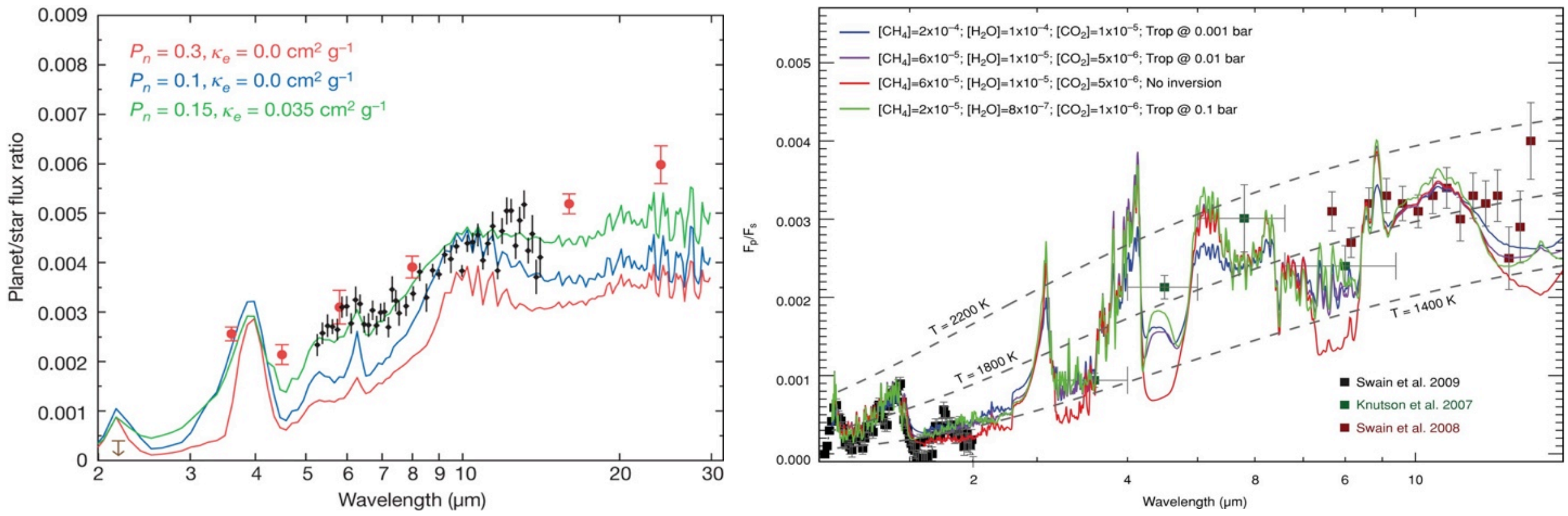

*Figure 20: Eclipse spectra and photometric data for hot-Jupiters observed with Hubble (NICMOS) and Spitzer (IRS & IRAC). Left: MIR observations of HD 189733b. Simulated spectra of water vapour are overlapped (Grillmair et al., 2008). Right: NIR and MIR observations compared to synthetic spectra for three models that illustrate the range of temperature/composition possibilities consistent with the data (Swain et al., 2009). For each model case, the molecular abundance of $CH_4$, $H_2O$, and $CO_2$ and the location of the tropopause is given. Note that the mid-infrared data are not contemporaneous with the near-infrared data, and attempting to "connect" these data sets with a model spectrum is potentially problematic if significant variability is present.*

Given the potential intricacies of a parametric approach, in the past years alternative methods have been developed to de-correlate the data from instrumental and stellar noise. The issue of poorly constrained parameter spaces is not new in astrophysics and has given rise to an increased interest in unsupervised (and supervised) machine learning algorithms (e.g. Wang et al. 2010). Unsupervised machine learning algorithms do not need to be trained prior to use and do not require auxiliary or prior information on the star, instrument or planet but only the observed data themselves. The machine learning approach will then (from observations) 'learn' the characteristics of an instrument and allows us to de-trend systematics from the astrophysical signal. This guarantees the highest degree of objectivity when analysing observed data. In Waldmann (2012, 2014), Waldmann et al. (2013) and Morello et al. (2014, 2015), Independent Component Analysis – ICA (Hyvarinen 1999) has been adopted as an effective way to decorrelate the exoplanetary signal from the instrument in the case of Hubble-NICMOS and Spitzer/IRS data or to decorrelate the stellar activity from the exoplanet transit lightcurve, in Kepler data. The error-bars for non-parametric approaches can be sometimes larger than those reported by parametric approaches. This difference is due to the higher amount of auxiliary information injected in the parametric approach. Ultimately, it is a trade-off between a higher degree of objectivity for the non-parametric methods and smaller errors for the parametric detrending.

For the EChO data, both methods will be used to correct instrumental systematics and astrophysical noise. Very thorough tests and calibration of the instrument before launch (especially detector performances), will substantially help to constrain the auxiliary information of the instrument hence the decorrelation process.

### *3.4.6 Correcting for stellar activity*

The impact of stellar activity on the EChO data has been carefully evaluated by many teams working on EChO. Results from the Kepler mission (Basri et al. 2013) indicate that most G dwarfs have photometric dispersions less than 50 ppm over a period of 6 hours, while most late-K and M dwarfs vary at a level of some 500 ppm. Note that Kepler operates in the visible where stellar photometric variability is over a factor of 2 higher than in the "sweet spot" of EChO – the NIR and MIR – because of the contrast between spots and the stellar photosphere. The effects of stellar activity on EChO's observations will vary for transit and eclipse observations. Alterations in the spot distribution across the stellar surface can modify the transit depth (because of the changing ratio of photosphere and spotted areas on the face of the star) when multiple transit observations are combined, potentially giving rise to spurious planetary radius variations. The situation is simpler for occultations, where the planetary emission follows directly from the depth measurement. In this case, only activity-induced variations on the timescale of the duration of the occultation need to be corrected for to ensure that the proper stellar flux baseline is used. The EChO mission has been designed to be self-sufficient in its ability to correct for the effects of stellar activity. This is possible thanks to the instantaneous, broad-wavelength coverage and the strong chromatic dependence of light modulations caused by stellar photospheric inhomogeneities (star-spots and faculae). We have explored several possible approaches to evaluate the effect of stellar activity and developed methodologies to prove the performance of EChO data in reaching the required precision (Herrero et al., 2014; Micela, 2014; Danielski et al., 2015; Scandariato et al., 2014).

***Method 1*** – We have investigated a direct method of correlating activity-induced variations in the visible with those in the IR. The underlying hypothesis is that variations of the transit depth in the visible are solely caused by stellar activity effects and not influenced by the atmosphere of the transiting planet. To test this approach, a realistic stellar simulator has been developed that produces time series data with the same properties as the measurements from EChO. The simulator considers surface inhomogeneities in the form of (dark) starspots and (bright) faculae, takes into account limb darkening (or brightening in the case of faculae), and includes time-variable effects such as differential rotation and active region evolution. We have generated series of transits at wavelengths 0.8, 2.5, and 5.0 μm. Then, we have measured the transit depths and calculated the variations of those depths with time. We have found that there is a well-defined correlation between activity-induced transit depth variations in the visible (0.8 μm) and the IR (2.5 and 5.0 μm). An illustration of the transit light curves generated by the simulator and the correlation between visible and IR transit depth variations (TDV) can be seen in Figure 21 (left & middle). In practice, the correction of EChO data for stellar activity using, for example, a series of measurements in the visible and an IR band can be done using the following expression: $d_{IR}^{corr} = d_{IR} + a_0 + a_1 \cdot (d_{VIS} - \langle d_{VIS} \rangle)$, where $d$ stands for the transit depth, and $a_0$ and $a_1$ are the coefficients of a linear fit that can be determined from simulations.

A number of combinations of stellar photospheres and active region parameters (size and location of spots, temperature contrast) were considered to obtain a statistical view of the method. The results can be seen in Table 7. The cases that we have analysed represent standard stars of GKM spectral types with filling factors of 1-7%, i.e., corresponding to stars that are ~4-30 times more spotted than the active Sun. The case in row 1 has parameters similar to HD 189733. As can be seen from Table 7, the direct procedure provides a correction of the transit data to a few times $10^{-5}$, and thus is fully compliant with EChO noise requirements.

| $T_{eff}$ (K) | $\Delta T_{sp}$ (K) | Filling Factor | $rms_T$ (0.8μm) | $rms_T$ (2.5μm) | $rms_T$ (5.0μm) | $rms_{T(corr)}$ (2.5μm) | $rms_{T(corr)}$ (5.0μm) | Corr. fact (2.5μm) | Corr. fact (5.0μm) |
|---|---|---|---|---|---|---|---|---|---|
| 5060 | 500 | 0.061 | 9.0e-3 | 3.9e-3 | 3.0e-3 | 1.7e-5 | 2.3e-5 | 2.3e2 | 1.3e2 |
| 5850 | 500 | 0.053 | 7.3e-3 | 2.9e-3 | 2.9e-3 | 4.0e-5 | 2.5e-5 | 7.3e1 | 1.2e2 |
| 6200 | 550 | 0.049 | 4.4e-3 | 1.7e-3 | 1.8e-3 | 5.3e-6 | 5.9e-6 | 3.2e2 | 3.1e2 |
| 3580 | 400 | 0.055 | 1.1e-2 | 6.2e-3 | 4.7e-3 | 3.8e-5 | 2.2e-5 | 1.6e2 | 2.1e2 |
| 4060 | 400 | 0.035 | 7.1e-3 | 5.3e-3 | 2.6e-3 | 4.4e-5 | 3.4e-6 | 1.2e2 | 7.6e2 |
| 5850 | 500 | 0.008 | 1.9e-4 | 1.4e-4 | 1.5e-4 | 8.9e-6 | 9.8e-6 | 1.6e1 | 1.5e1 |
| 5850 | 500 | 0.060 | 6.3e-3 | 2.6e-3 | 2.7e-3 | 3.2e-5 | 2.7e-5 | 8.1e1 | 1.0e2 |
| 3580 | 400 | 0.066 | 1.5e-2 | 8.3e-3 | 6.4e-3 | 3.0e-5 | 2.2e-5 | 2.8e2 | 2.9e2 |
| 5850 | 500 | 0.020 | 2.0e-3 | 9.2e-4 | 9.7e-4 | 1.9e-5 | 2.4e-5 | 4.8e1 | 4.0e1 |
| 5060 | 500 | 0.074 | 5.1e-3 | 2.2e-3 | 1.7e-3 | 1.4e-5 | 1.5e-5 | 1.6e2 | 1.1e2 |

*Table 7: Results for the simulations of 10 cases of star-planet systems randomly selected from a set of 6 stellar models and 4 different possible active region maps, and with a rotation period of 15 days. The planet parameters were fixed to $R_p$=0.05 $R_{star}$, $P_{planet}$=2.54 days, b=0.2 (impact parameter). The facula temperature contrast and the facula-to-spot area ratio (Q) were fixed to $\Delta T_{fac}$=+100 K and Q=7.0, respectively. The first three columns indicate the temperature for the quiet photosphere, the spot contrast and the spot filling factor. The following three columns list the rms of the in-transit sections at 0.8, 2.5, and 5.0 μm. The next two columns give the rms of the in-transit sections at 2.5 and 5.0 μm after correcting for activity effects using the procedure described in the text. The final two columns give the correction factor at 2.5 and 5.0 μm.*

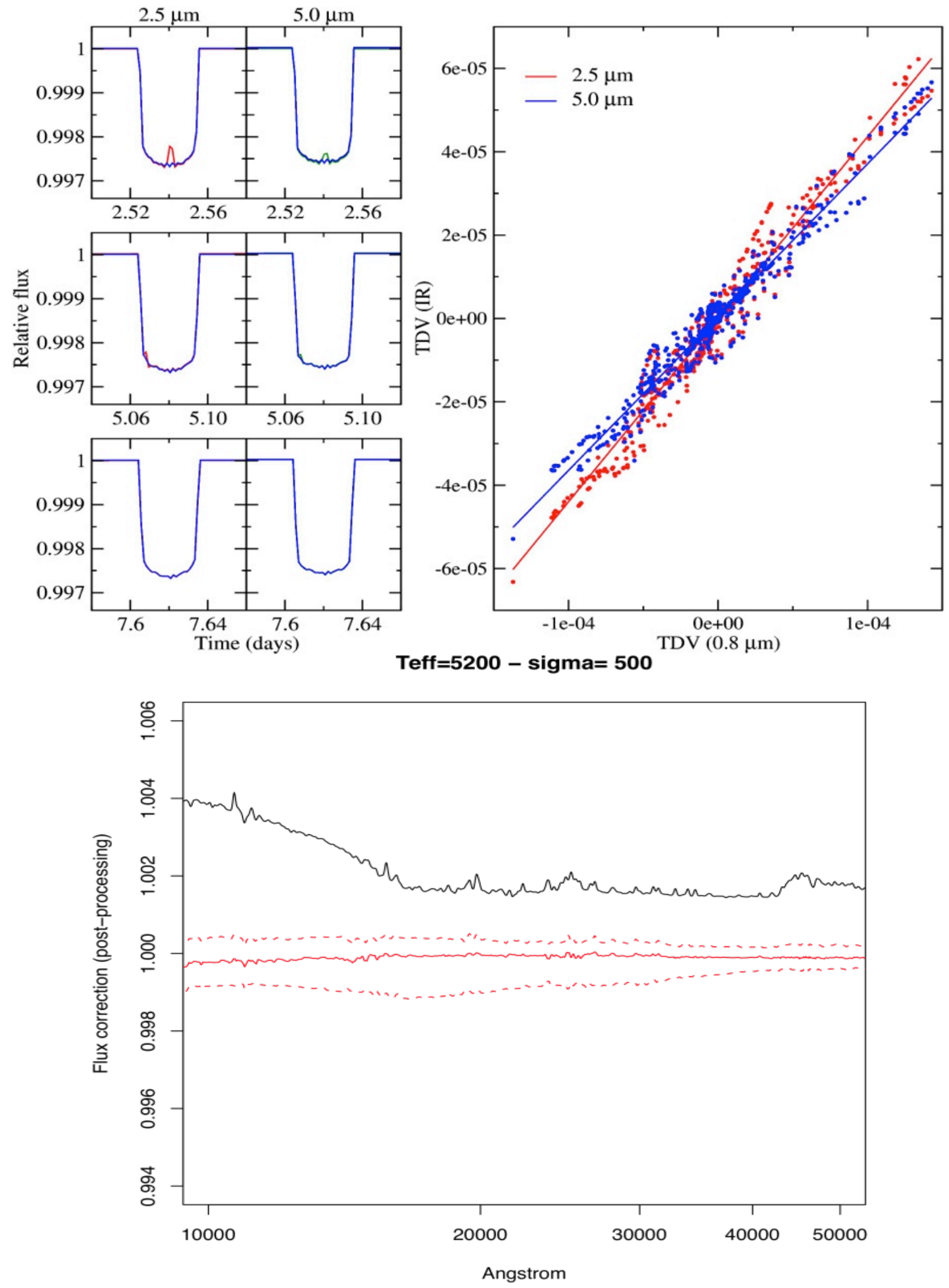

*Figure 21: Top left: Transit light curves at 2.5 µm (red) and 5.0 µm (green) for one of the cases generated in the sample, compared with the transit light curve of an immaculate star. Note the small systematic deviations and the more apparent spot crossing events. Top right: Correlation of activity-induced transit depth variations (TDV) in the visible (0.8 µm) and the IR (2.5 and 5.0 µm). Bottom: Spectrum distortion without corrections (solid black line), residual distortion after correction with method 2 (median and 25 - 75% percentiles of simulations).*

***Method 2 –*** A complementary method has been developed to reconstruct the spectral energy distribution of the target stars in the IR using the visible spectrum (0.55-1 µm) as an instantaneous calibrator. Having a sufficient number of spectra of a given stars observed at different levels of activity, it is possible to calibrate the method for each star. The approach has been developed on a grid (in spot temperature and filling factor) of models of active stars and has been tested through simulations taking into account for photon noise. The method is based on principal component analysis. Since the new variables are chosen to maximize the variance, it is possible to reduce the dimensionality of the space, eliminating the dependences among the original variables and noise. In all the explored cases the first two components are retained: the first component is related to the slope of the spectrum while higher order components are related to features of the spectrum.

The procedure involves the following steps: 1) generation of 1000 simulations of the input model assuming an average SNR per resolution element; 2) projection of the simulated spectra into the space of the first two components; 3) identification of the best fit spectrum in the principal component space and selection of the corresponding NIR spectrum as the "best estimate" of the NIR stellar spectrum; and 4) comparison between the spectral distortion with no correction (assuming an unspotted star) and the residual after adopting the best estimate. Figure 21 shows as an example the median correction of the 1000 simulations and the 25% and 75% quartiles for $T_{eff}$=5200 K and stellar SNR=500. To quantify the correction we compare the distortion before applying our method, measured as the average value in the 1-2 μm band (where the effect is larger), and the equivalent average of the median and 25-75% quartiles of the residuals after the correction. Table 8 shows that the method allows for a significant reduction of the spectral distortion.

| | | Residual distortion after correcting | | |
|---|---|---|---|---|
| **$T_{eff}$ (K)** | **No correction (1-2 μm)** | **SNR=200 (±25-75% quartiles)** | **SNR=500 (±25-75% quartiles)** | **SNR=1000 (±25-75% quartiles)** |
| 6000 | 2.2e-3 | -6.4e-5 [-9e-4 / 7e-4] | 8.6e-5 [-5e-4 / 4e-4] | 1.2e-4 [-4e-4 / 4e4] |
| 5200 | 2.5e-3 | -2.9e-4 [-1e-3 / 3e-4] | -6.8e-5 [-5e-4 / 2e-4] | 5.2e-5 [-2e-4 / 2e-4] |
| 4200 | 4.8e-3 | 3.0e-6 [-2e-3 / 1e-3] | 4.0e-6 [-1e-3 / 9e-4] | 0 [-3.8e-4 / 5e-4] |

*Table 8: Results of the comparison between spectral distortion before applying the corrections and the residuals after correcting, as a function of stellar effective temperature and SNR. The average values in the 1-2 μm band and the 25-75% percentiles derived from 1000 simulations are given.*

***Method 3*** **–** A further approach has focused on statistical methods to de-correlate astrophysical noise from the desired science signal. Whilst the statistical fundamental of these methods are very different and often complementary, they all try to disentangle the astrophysical signal from various noise sources using the coherence of the exoplanetary transit/eclipse signature over time and/or frequencies of light. Figure 22 shows two examples of such a decorrelation. Given single time series on an active star with various modes of pulsation obtained by the Kepler space telescope, Waldmann (2012) showed that a randomly chosen pulsation mode of the star could be isolated and the remaining autocorrelative noise of the star suppressed, resulting in a strong reduction of the stellar noise component (Figure 22 left). Similar concepts apply to periodic exoplanetary lightcurves observed over multiple transits and/or wavelengths.

The results were repeated successfully for a sample of Kepler stellar light curves, spanning from M to G types. In all cases a correction of the order of $10^{-5}$ to 5 $10^{-4}$ depending on the frequency of the sampling (i.e. 10 hours continuous observations every day or 10 hours once a week), was obtained (Danielski et al., 2015).

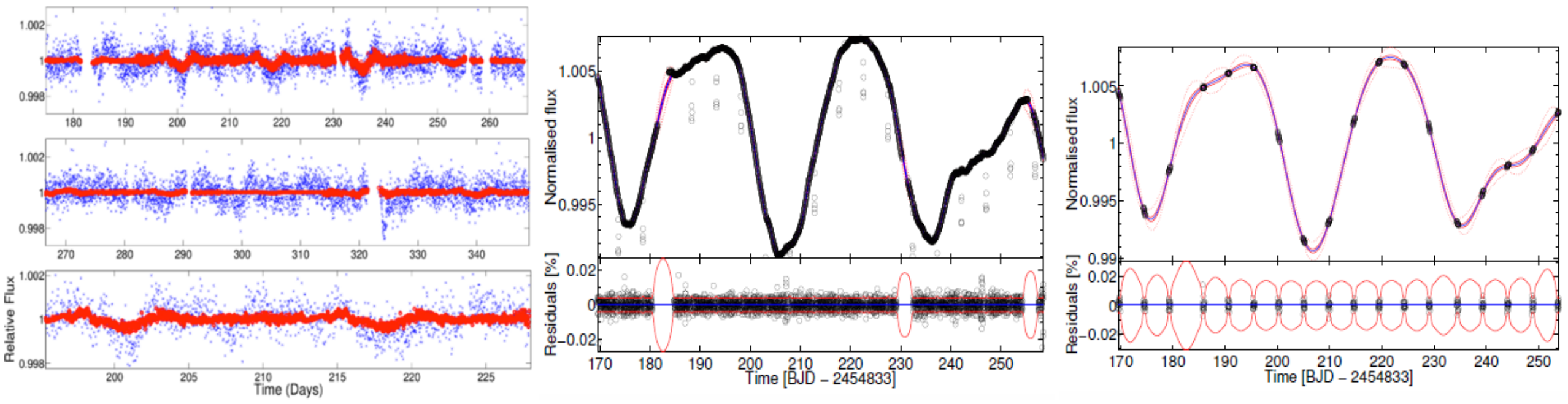

*Figure 22: Left: Kepler time series of an active M0 star (blue dots). Using Independent Component Analysis, the periodic pulsation filter at t=202, 218 and 235 was filtered from other correlated noise in the time series. The filtered signal is shown in red (Waldmann, 2012). Centre and Right: Kepler time series of another active K4 star. Using a Gaussian Process based method the stellar activity was successfully filtered out, with residuals as small as $10^{-5}$ when considering daily observations of 10 hours (Centre) and $10^{-4}$ when data are acquired for 10 hours every 5 days (Right) (Danielski et al., 2015).*

## 3.5 Evaluation of EChO performance

In this section, we briefly describe the strategy developed for the evaluation of EChO's performance, and the evaluation done during the Phase A study.

### *3.5.1 Performance evaluation tools*

The performance of EChO has been assessed using computational models based on two approaches. The first approach taken is based on a static radiometric model that takes the required performance figures for the payload to 'size' the mission. This model has been used to calculate the number of transit/occultation revisits necessary to achieve a specified SNR and the possible revisits during a given mission lifetime (Puig et al., 2014). The second approach is to construct a model that simulates the actual performance of the mission as realistically as possible (EChOSim). This end-to-end simulation is fully dynamic and accounts for the major systematic influences on the performance such as pointing jitter, internal thermal radiation sources, detector dark current and noise etc (Pascale et al., 2014; Waldmann and Pascale, 2014). Both models have been used to calculate the observation duration needed for the targets in the EChO sample. We find that a nominal mission lifetime of four years is sufficient to fulfill the science requirements and a mission of six years would fulfill the most ambitious EChO goals. The use of separate performance models with similar results gives confidence that the mission can be undertaken as planned and can deliver the science described in this paper.

### *3.5.2 Overall noise allocation*

Using the EChOSim tool, we can evaluate the performance by calculating the overall noise allocation and comparing this to the scientific requirements. The procedure is extensively described in (Pascale et al., 2014) and here we only summarise the main results.

- Noise associated to the astrophysical scene:

  The number of detected photons from the planet and star, $N_0$, and the zodiacal background photons in a sampling interval $\Delta t$, *Zodi*, are used to estimate the level of photon noise from the astrophysical scene:

$$\sigma_N^S = \sqrt{N_0 + Zodi} \qquad \frac{e^-}{pixel} - rms \tag{1}$$

  It is convenient to refer the noise in one sampling interval to the noise per unit time:

$$\sigma_N = \sigma_N^S / \sqrt{\Delta t} \qquad [e^- pixel^{-1} s^{-1/2} - rms] \tag{2}$$

- Noise associated with the instrument:

  All sources of instrumental noise contribute to the total system noise level, $\sigma_{SN}$. The system noise level is then given by the sum in quadrature of all individual noise components:

$$\sigma_{SN} = \sqrt{\sigma_{RO}^2 + \sigma_{DC}^2 + \sigma_{Tel}^2 + \sigma_{Opt}^2 + \sigma_{RPE+PDE}^2} \qquad [e^- pixel^{-1} s^{-1/2} - rms] \tag{3}$$

  $\sigma_{RO}$ is the detector readout noise, $\sigma_{DC}$ is the dark current noise, $\sigma_{Tel}$ is the combined photon noise associated to the thermal emission of all optical surfaces in the line of sight, $\sigma_{Opt}$ is the photon noise associated to the thermal emission of the module enclosure, and $\sigma_{RPE+PDE}$ expresses the photometric noise associated to the pointing jitter.

### *3.5.3 Simulations of EChO planetary spectra*

Using the EChOSim tool, we can simulate the observation of key targets and see how the overall requirements translate into reconstructed spectra. We show here two cases: the transit of a warm Neptune around a faint object (GJ 3470b) in the Rosetta Stone tier, and the eclipse of this same object in the Origin tier.

GJ 3470b is a 0.0437 $M_J$ planet with a radius of 0.374 $R_J$ (where $M_J$ and $R_J$ are respectively the mass and the radius of Jupiter), orbiting at 0.036 AU with a period of 3.3367 days around its parent star (M1.5V star, mV=12.27, $T_{eff}$ = 3600 K at 30.7 pc). The transit and eclipse duration is about 1 hour and 45 minutes. The effective temperature of the planet, assuming the thermal equilibrium is 615 K. The atmospheric composition used for our simulation is taken from Venot et al. (2013).

The transit observation of GJ3470b in the Rosetta Stone tier requires the co-addition of 21 transits, assuming the current design of the mission and the known parameters of the planetary system. We estimated the observable $(R_p/R_s)^2$ (the transit depth, where $R_p$ and $R_s$ are the planetary and the stellar radius respectively) as a function of the wavelength (Figure 23). The associated error bars are computed using a dynamical fitting method implemented in the observation pipeline. This figure clearly shows that the transit depth chromatic variations associated with atmospheric absorptions can be detected in the IR spectral range even with a limited number of transit observations. The SNR decreases over 12 μm due to the increase noise in the detection chain and the contribution from thermal noise. The transit spectrum exhibits various spectral features associated not only with water vapour but numerous other molecules.

Similar simulations have been done for the eclipse spectra, which provide the planet to star contrast as a function of wavelength (Figure 23)

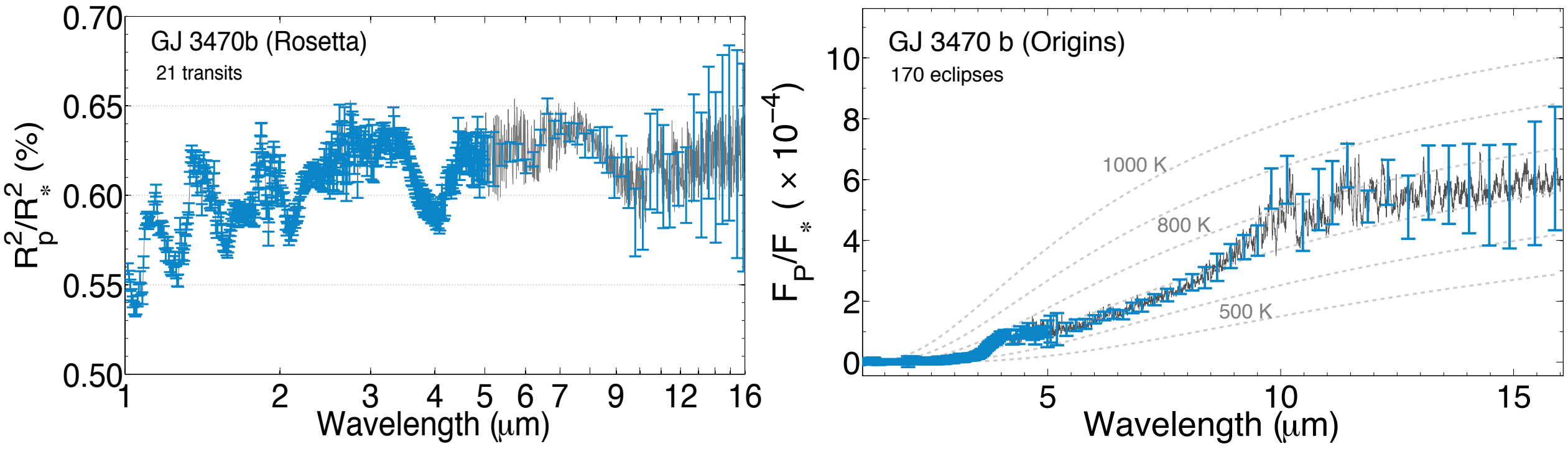

*Figure 23: Left: Transit depth as a function of wavelength for GJ3470b as observed with EChO. We assumed the co-addition of 21 transits and the current design of the instrument and data processing pipeline. Right: Planet to star contrast as a function of wavelength assuming the co-addition of 170 eclipses.*

Using both transit and eclipse spectra together, one can determine the mean molecular weight of the atmosphere, the atmospheric components, the temperature as a function of pressure, the presence and type of clouds. These calculations have been repeated for all the targets observable by EChO (see next section).

# 4. Mission strategy

In this section we describe the list of currently available targets for EChO (> 150), and we discuss the foreseen developments for the future, given the large number of ground and space dedicated facilities to discover new exoplanets in the next decade. The final list is expected to include hundreds of exoplanets, with a variety of sizes, temperatures, stellar hosts and orbital parameters.

## 4.1 EChO's current Core Sample

To produce a sample of potential targets for EChO using known systems we first drew up a "long list" of known targets with well characterised stellar and planetary parameters. This list has been generated using the EChO Target List Observation Simulator (ETLOS) (Varley et al., 2015) and will be continuously updated. ETLOS extracts the star/planet information from the Open Exoplanet Catalogue (Rein, 2015); further verification is done using SIMBAD, the 2MASS catalogue and exoplanet.eu (Schneider, 2015) where appropriate. The Core Survey targets were then selected to ensure as diverse range stellar types, metallicities and temperatures as possible to fulfil the requirements of the Chemical Census. Suitable targets for the Origin and Rosetta Stone tiers were further selected to fulfil the requirements expressed in Table 3.

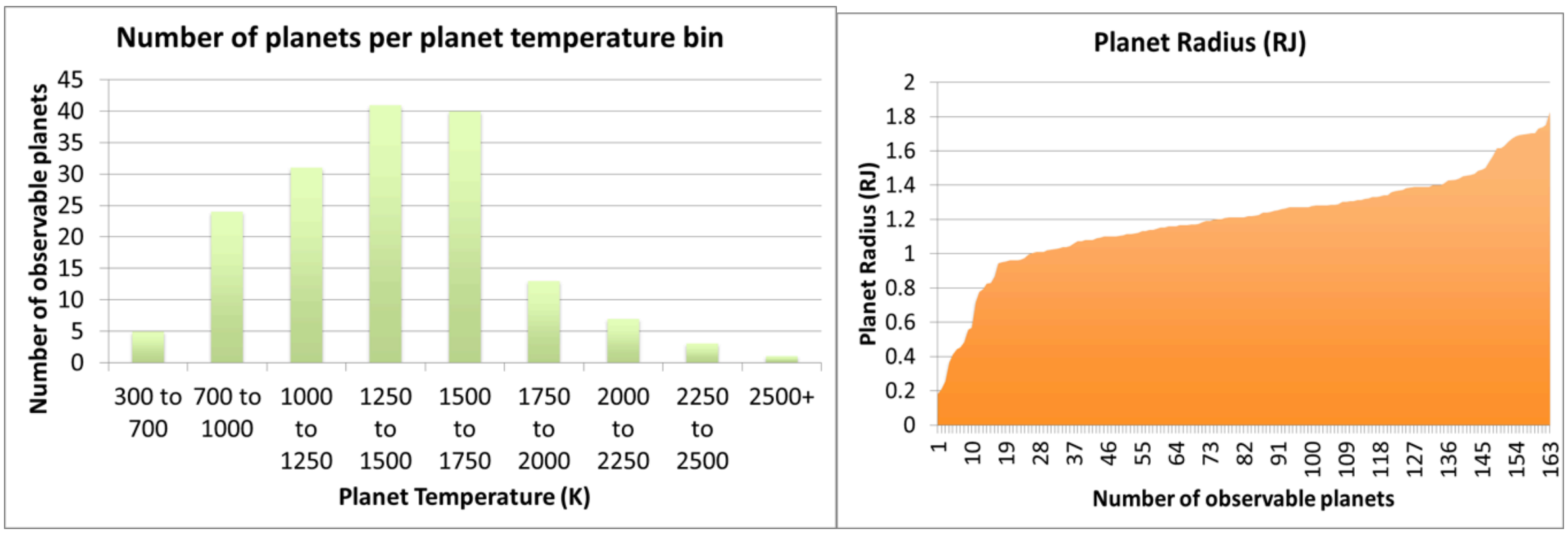

*Figure 24: Left: known planets observable by EChO classified as function of the planetary temperature in K. Right: known planets observable by EChO classified as function of the planetary radius.*

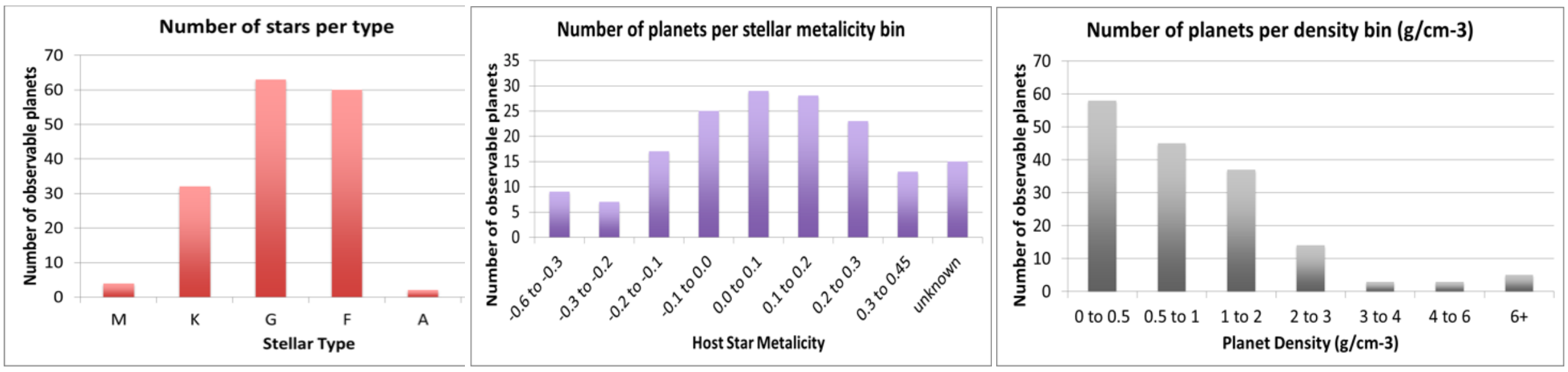

*Figure 25: Known planets observable by EChO classified as function of the: Left: Stellar Type, Middle: Stellar Metallicity, Right: Planetary density in g/cm$^3$.*

To assess the total time needed to observe the required number of targets in the three survey tiers we have undertaken simulations of the mission and instrument performance. As explained in section 3.5, two rather different approaches were taken for this. The first is a static model built using more generic assumptions about the instrument and mission performance (ESA Radiometric Model, Puig et al., 2014). The second approach models the instrument as designed and uses a dynamic approach to the performance simulation using realistic stellar and planetary parameters to model to actual time domain signal from the observation (*EChOSim,* Pascale et al., 2014). The list of known targets was run through the ESA–RM and *EChOSim* performance models. Although some differences are expected due to the different parameterisation of the instrument and other model assumptions, the results spread over the Core Survey are consistent, and the discrepancies for specific targets are understood and traceable. We are therefore confident of the robustness of the estimates obtained.

The integration times needed for each observing mode and the detectability for key molecular species are reported in (Varley et al., 2015): http://www.ucl.ac.uk/exoplanets/echotargetlist. The diversity of the selection is shown in Figures 24 and 25 where we show how current select targets are distributed between stellar type, metallicity, orbit type, density and temperature.

## 4.2 The future EChO Core Sample

A comprehensive exercise has been run to establish a target statistical sample of transiting targets for EChO that would cover the widest possible range of exoplanet/host star parameter space (Ribas and Lovis, 2014). As a first step, star counts were estimated using (a) new catalogues (Lepine et al., 2013, Frith et al., 2013) making cuts based on spectral type and magnitude directly, and (b) using the combination of the stellar mass function derived from the 10-pc RECONS sample and the mass-luminosity-K-band relationship from (Baraffe et al., 1998). Estimates were then made of the maximum number of exoplanets of a given exoplanet class (mean radius/mass: Jupiter-like 10 $R_{Earth}$/300 $M_{Earth}$; Neptune-like 4 $R_{Earth}$/15 $M_{Earth}$; Small Neptune 2.6 $R_{Earth}$/6 $M_{Earth}$; Super-Earth-like 1.8 $R_{Earth}$/7 $M_{Earth}$) and fiducial equilibrium temperature ($T_{hot}$ = 1500 K; $T_{warm}$ = 600 K; $T_{temperate}$ = 320 K) that transit a selection of stellar spectral types from K to M. This was done using statistics from the Kepler mission and adopting a methodology similar to that described in a recent paper by Fressin et al. (2013).

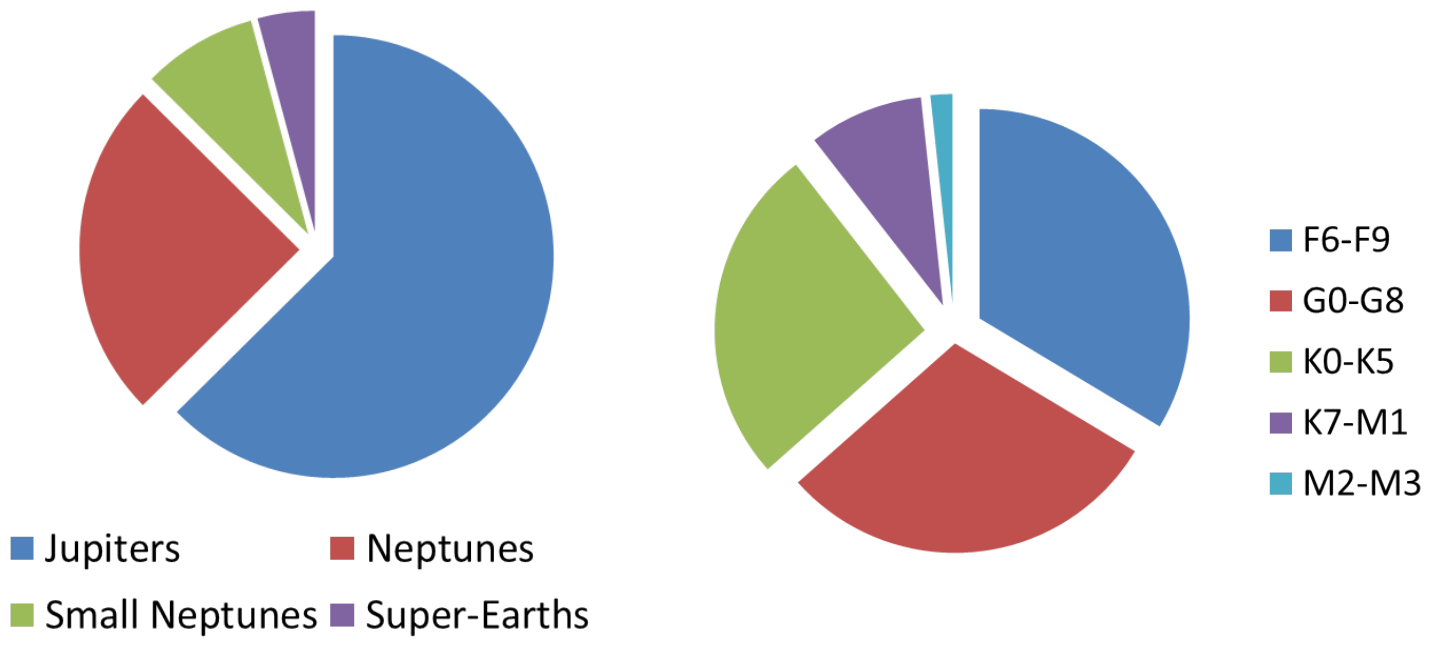

*Figure 26: Pie charts illustrating the different planetary classes considered for the future core sample*

Planet occurrence rates based on Kepler results were calculated for all spectral types. These rates are weighted towards solar-like stars because of the predominance of FGK hosts in the Kepler survey itself. An analysis of the planet occurrence rates for M hosts observed by Kepler indicates that the rates are consistent with those found for earlier spectral types, albeit at low statistical significance (e.g. (Dressing et al., 2013)).

Star counts, planet temperatures and types, and the transiting planet occurrence rate were then used to determine the numbers and types of transiting exoplanets around host stars down to a K-band magnitude of 9, with the overall total number in good agreement with estimates from HARPS (Mayor et al., 2011) as well as other estimates based on Kepler data (Howard et al., 2012). Figure 29 illustrates a possible parameter space that EChO may observe in the Chemical Census and Origin surveys according to current SNR requirements and conservative assumptions on instrument performance. These predictions are in line with the expected science yield from the future surveys (see Table 9 and Figure 29).

## 4.3 Sky visibility/source accessibility

EChO will visit a large and well-defined set of targets (see Section 4.1 & 4.2). Repeated visits may be required to build up the SNR of individual target spectra. The maximum duration of a visit to a target system will be ~10 hours – the time of the transit itself, plus half that time before and then after the transit. The time between successive transit observations will depend on orbital period and scheduling, and could be as little as a day, to as long as a few tens of days. In principle, the targets may be in any part of the sky, and as such the satellite needs a large field of regard, with minimal constraints (due to Earth/Sun) on the direction in which it can be pointed. The most challenging targets for EChO will be temperate super-Earths around M-type stars. Given the orbital radius and so period of a typical temperate planet ($T_p$~300K), a maximum number of a couple of hundred transits (depending on the effective temperature/spectral type of the host star) would occur during a mission lifetime of 4 years. The complete sky shall be accessible within a year, with a source at the ecliptic observable for 40% of the mission lifetime. Shown in Figure 27 is a plot of the sky visibility for EChO, superposed on which are targets from the different tires of the EChO core survey.

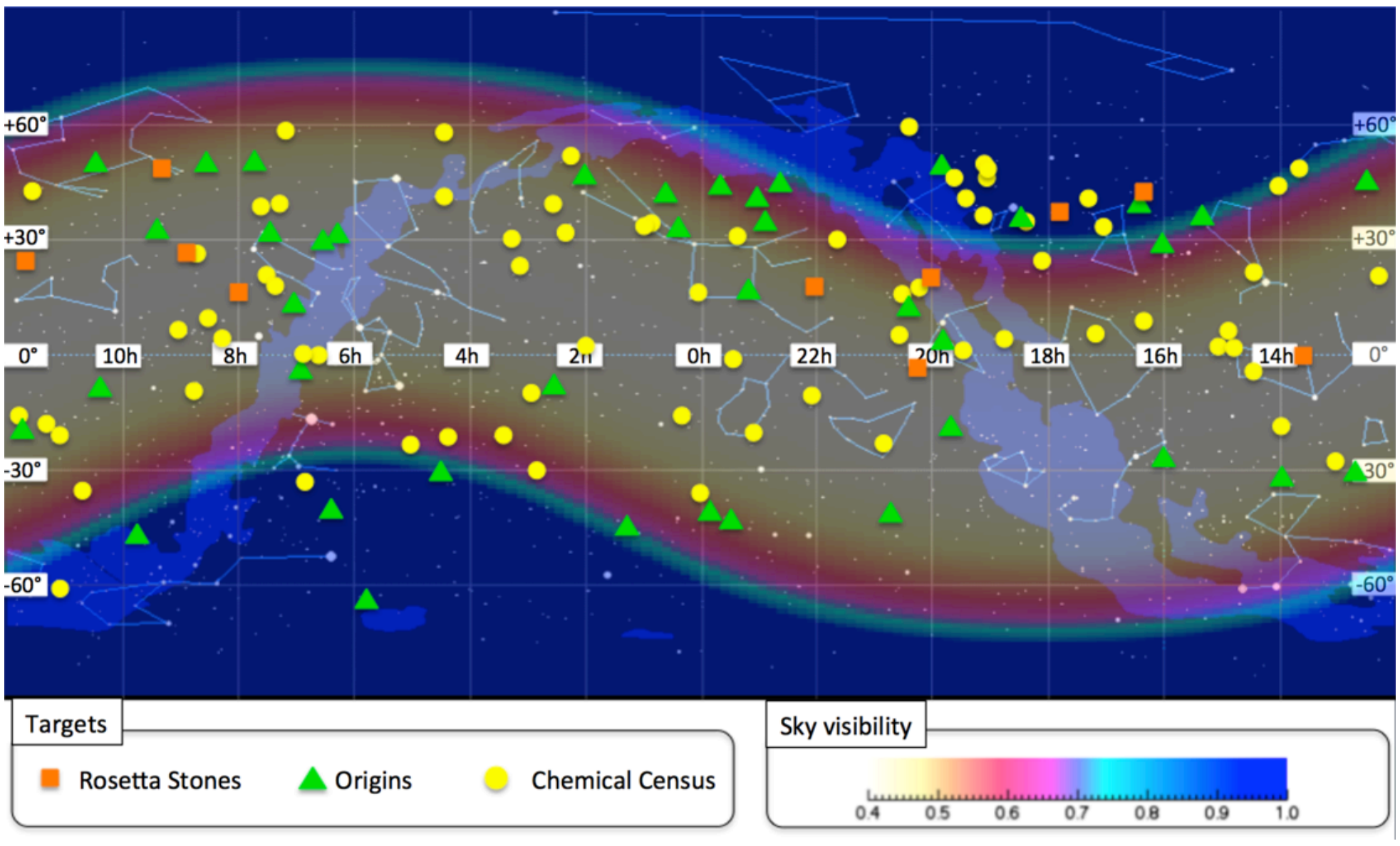

*Figure 27: A plot illustrating the fraction of the year for which a given location in the sky (in equatorial coordinates) is visible to EChO (courtesy of M. Ollivier), as seen from a representative operational orbit of EChO at L2. Superposed are known exoplanets that would be targets in the EChO Core Survey, as described in Section 4.1). Each target is accessible for at least 5 months (40% of time).*

## 4.4 New targets for EChO

Target selection is a key aspect of EChO. The choice of the targets will determine the planetary parameter space we will explore. The scientific outcome of the mission clearly depends on the observed sample.

There is no need to select the sample more than ten years before launch but we need a good plan to select the best sample immediately prior to launch. In the present phase we are defining the primary physical planetary parameters that define the "diversity" of planet population. These include:

- *Stellar metallicity, age, temperature,*
- *Planetary temperature, mass and density.*

A sub-space of this parameter space will be explored by EChO. The mission is designed to fill such space. Several surveys both from ground and from space will provide targets with the necessary characteristics to meet the objectives of the mission. Table 9 summarises the most important surveys from which we expect a significant contribution to the final core sample. The list is not exhaustive.

| Name of Survey/Mission | Key characteristics | Target stars relevant for EChO | Expected planets relevant for EChO | Notes |
|---|---|---|---|---|
| **WASP/SuperWASP**<br>(Pollacco et al., 2006, PASP, 118, 1407) | • Ground photometric survey - broad band<br>• All sky<br>• Ongoing | G-early K | 100 *J*<br>Few *N* | $P_{orb}$ < 10 days;<br>> 70 J already discovered |
| **K2**<br>(Beichman et al., 2014) | • Space survey<br>• Survey in the ecliptic plane<br>• Ongoing | All | ~ 500 *J*<br>~ 500 *SE, N* | $P_{orb}$ < 5 days |
| **HATNet/HATSouth**<br>(Bakos et al., 2002, PASP, 114, 974; 2013, PASP, 125, 154) | • Ground photometric survey - broad band<br>• All sky<br>• Ongoing | G/K | 100 *J*<br>Few *N* | $P_{orb}$ < 10 days;<br>> 50 J already discovered |
| **HARPS, HARPS-N, Keck, ESPRESSO, CARMENES, SPiROU** | • Ground Doppler surveys - VIS/IR<br>• Transit search through photometric follow-up<br>• All sky, bright stars<br>• Ongoing/being built | G/K/M | See below | Discovered the brightest targets in each category |
| **CHEOPS**<br>(Broeg et al., 2013, EPJWC, 47, 3005) | • Space photom. follow-up<br>• 2017-2021 (3.5yr)<br>• Monitoring of bright stars with Doppler-detected planets | G/K/M | 10 *N*<br>5 *SE* | Also used to refine parameters of planets detected by ground-based transit surveys |
| **NGTS**<br>(Chazelas et al., 2012, SPIE, 8444) | • Ground photometric survey – broad band<br>• Coverage 1,920°<br>• -50 < dec < -30<br>• 2014 – 2019 | G/K/M | 100 *J*<br>20 *N*<br>20 *SE* | $P_{orb}$ < 16 days |
| **APACHE**<br>(Sozzetti et al, 2013, EPJWC,47, 3006) | • Ground photom. survey<br>• Monitoring of 3,000 M<br>• 2012-2017 | M | 5 *SN/SE* | $P_{orb}$ < 10 days |
| **GAIA**<br>(Lindegren, 2010, IAUS, 261, 296) | • Space astrometric survey<br>• All sky<br>• 2014-2019 | All | 10-15 *J* | Around M stars<br>0.5-3 AU |
| **MEarth**<br>(Nutzman et al., 2008, PASP, 120, 317) | • Ground photom. survey<br>• Ongoing | Late-M | 5 *SN/SE* | $P_{orb}$ < 10 days;<br>GJ 1214b |
| **TESS**<br>(Ricker et al., 2010, AAS, 42, 459) | • Space photometric survey<br>• 45,000 sq degree<br>• 2017- | G/K/M | 650 *J*<br>1000 *N*<br>700 *SN*<br>300 *E & SE* | $P_{orb}$ < 50 days |
| **PLATO**<br>(Rauer et al., 2014) | • Space photometric survey<br>• sq degree<br>• 2024 | All | ~1000 J<br>1200 N<br>700 SN<br>600 SE | |

*Table 9: Summary of the main surveys/projects that will provide targets for EChO in the next ten years. The columns on target stars and expected planets refer specifically to the observations relevant for EChO. Legend: (J=Jupiters, N=Neptunes, SN=sub-Neptunes, SE= Super-Earths, E=Earths).*

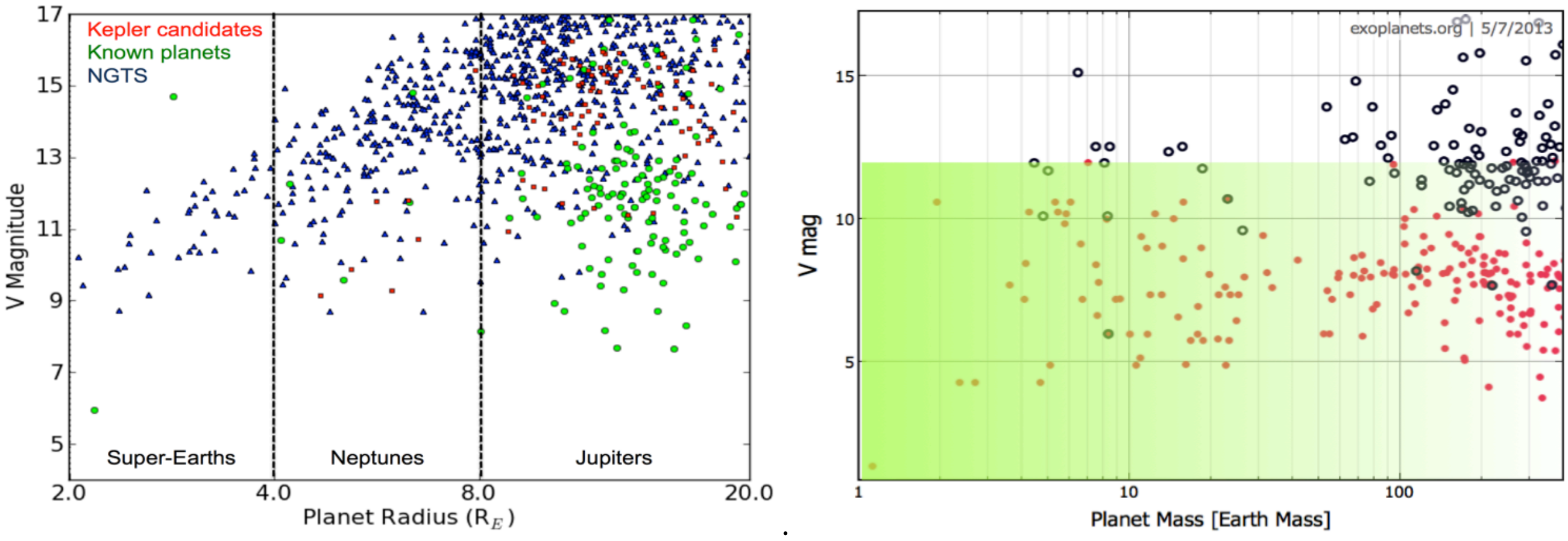

*Figure 28: Left: Simulated planet population from NGTS. This assumes a survey of 1920 square degrees over five years. Each of the plotted simulated planets can be confirmed with HARPS or ESPRESSO in less than 10h exposure time. This instance of the simulation shows 39 confirmable super-Earths and 231 Neptunes. Of these, 23 super-Earths and 25 Neptunes orbit stars brighter than I=11. These planets will be the optimal targets for EChO. Right: Planets with measured mass from RV survey (red dots). Planets with measured radius from transit survey (black circles). The green shaded area is where CHEOPS will provide accurate radius measurements*

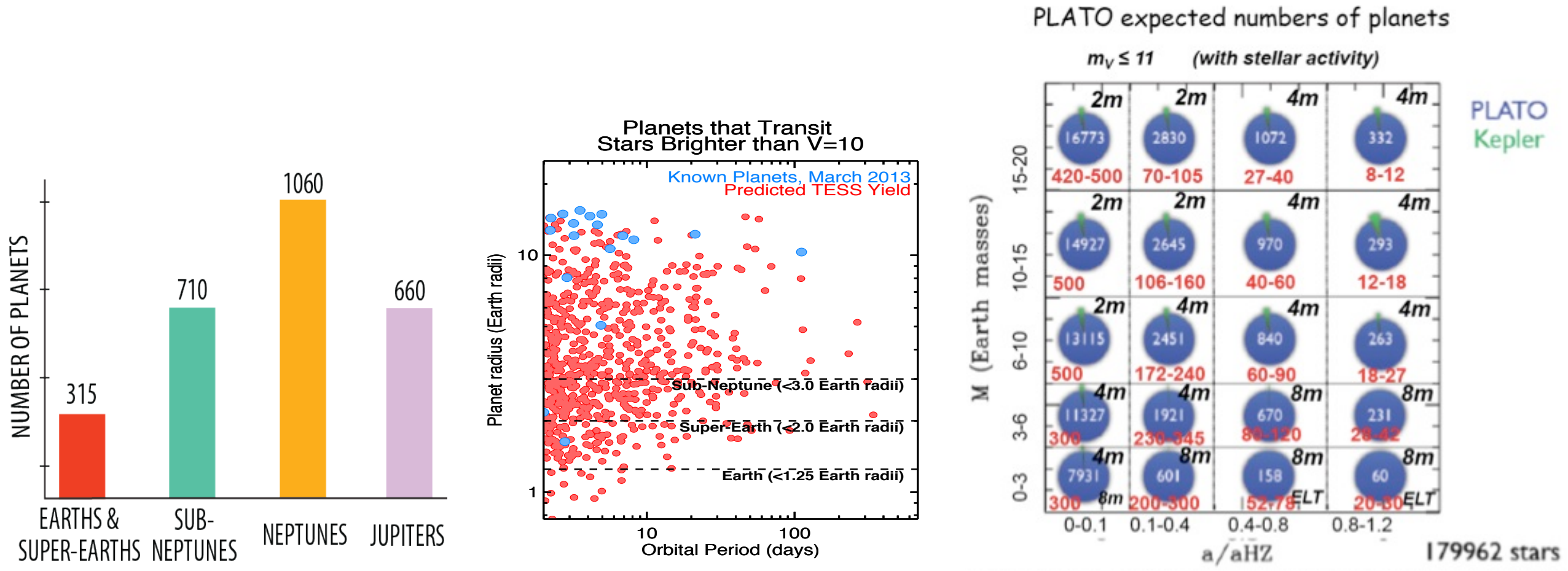

*Figure 29: Left: Expected science yield from the TESS mission. Centre: Radius-Orbital period distribution of transiting exoplanets found around nearby stars brighter than V=10 as of March 2013 (blue dots), versus the number of such planets expected to be discovered by TESS (red dots). These planets will be the optimal targets for EChO. Right: Expected science yield from the PLATO mission.*

## 4.5 The optimization of the observation program

The ability to fulfil the scientific program strongly depends on the optimization of the observation program. Because the planetary transits and occultations happen at specific epochs (given by ephemerides), the observation program, the data transfer sequences and the on-board calibration phases have to be well-defined and are time critical. The final performance evaluation of EChO also needs to take into account the way the observation and calibration/data transfer phases are optimized.

We have simulated an observing programme with an assumed target reference sample using scheduling simulation tools (Garcia-Piquer et al., 2014; Morales et al., 2014). These tools aim to check the feasibility and efficiency of the observation program. They include optimisation routines that allow the scheduling assuming knowledge of the visibility of the objects, the transit/occultation ephemerides, the expected spacecraft performance and some assumed calibration and data transfer phases. The net result of the overall process is that, using the target lists described in Section 4.1 & 4.2, the EChO mission would meet its scientific objectives.

# 5. Synergy with other facilities

EChO, JWST and E-ELT observations are highly complementary and mutually beneficial. While JWST will provide state-of-the-art measurements for a few tens of planets, the E-ELT will provide targeted observations for a few tens of planets at ultra-high spectral resolving power at specific wavelengths. *The role of EChO is to provide the broad picture by performing a systematic and uniform survey of hundreds of exoplanets*. EChO instantaneous broad wavelength coverage is also essential to correct for the stellar activity (see Section 3.4.6). The three observatories together would deliver transformational science.

## 5.1 EChO & the JWST

JWST is the largest space telescope ever conceived, with an equivalent telescope diameter of 5.8 m and 22 $m^2$ collecting area. It is designed to operate over the visible (~0.6 μm) to mid-IR waveband (28 μm) providing very high sensitivity imaging and spectroscopy of faint astronomical targets. It is a true observatory with multiple capabilities, instruments and operating modes, optimised for background limited observations. JWST is scheduled for launch in late 2018. Although primarily designed for observations of very faint targets (in the μJy range), JWST will do a great deal of ground breaking exoplanetary science. Table 10 summarises the JWST instruments and operating modes that will be useful for exoplanet transit spectroscopy. Studies of the performance of the instruments for transit spectroscopy have been carried out notably for NIRISS and NIRSpec (Dorner Phd Thesis Universite de Lyon 2012, Clampin 2010, http://www.cosmos.esa.int/web/jwst/exoplanets). Both transit & eclipse measurements over the full waveband from 0.6 to 28 μm are possible with the combination of the instruments and modes on JWST. However, both its extremely high sensitivity and observatory nature mean there are significant restrictions on the type and number of targets that will be observable (see Table 10 and Figure 30).

| Instrument | Mode | Resolving power | Wavelength range (μm) | Comments |
|---|---|---|---|---|
| NIRISS | Grism, cross-dispersed, slit-less | 700 | 0.6 - 2.5 | Saturates at K<9 at some part of band |
| NIRCam | Grism, slit-less | 2000 | 2.4 - 5.0 | Not proposed for transit spectroscopy in SODRM |
| NIRSpec | Prism, wide slit (1.6”) | 100 | 0.6 - 5.0 | Saturates at J<11 (see Fig. 30) Wavelength range covered using 3 separate orders |
| NIRSpec | Grating, wide slit (1.6”) | 1000 or 2700 | (0.7)1.0 - 1.8<br>1.7 - 3.0<br>2.9 - 5.0 | Uses three grating settings to cover wavelength range. Effective SW cut on is 0.9 μm |
| MIRI | Prism, 0.6” slit or slit-less | 100 | 5.0 - 11.0 | |
| MIRI | IFU<br>(0.2” - 0.27”/pixel) | 2400–3600 | 5.0 - 7.7<br>7.7 - 11.9<br>11.9- 18.3<br>18.3 - 28.3 | Each band uses 3 sub-bands with separate gratings (Glasse et al., 2014). |

*Table 10: JWST instruments and observing modes useful for transit spectroscopy*

In addition to transits, there are a number of direct imaging possibilities using JWST – for a full summary see the exo-planet “white papers” (see http://www.stsci.edu/jwst/doc-archive/white-papers). A first cut, notional observing program for the JWST is encompassed in the Science Observations Design Reference Mission (SODRM - http://www.stsci.edu/jwst/science/sodrm/jwst/science/sodrm/): this consists of a number of observing programs built around seven science themes designed to allow the mission team test the observation planning tools.

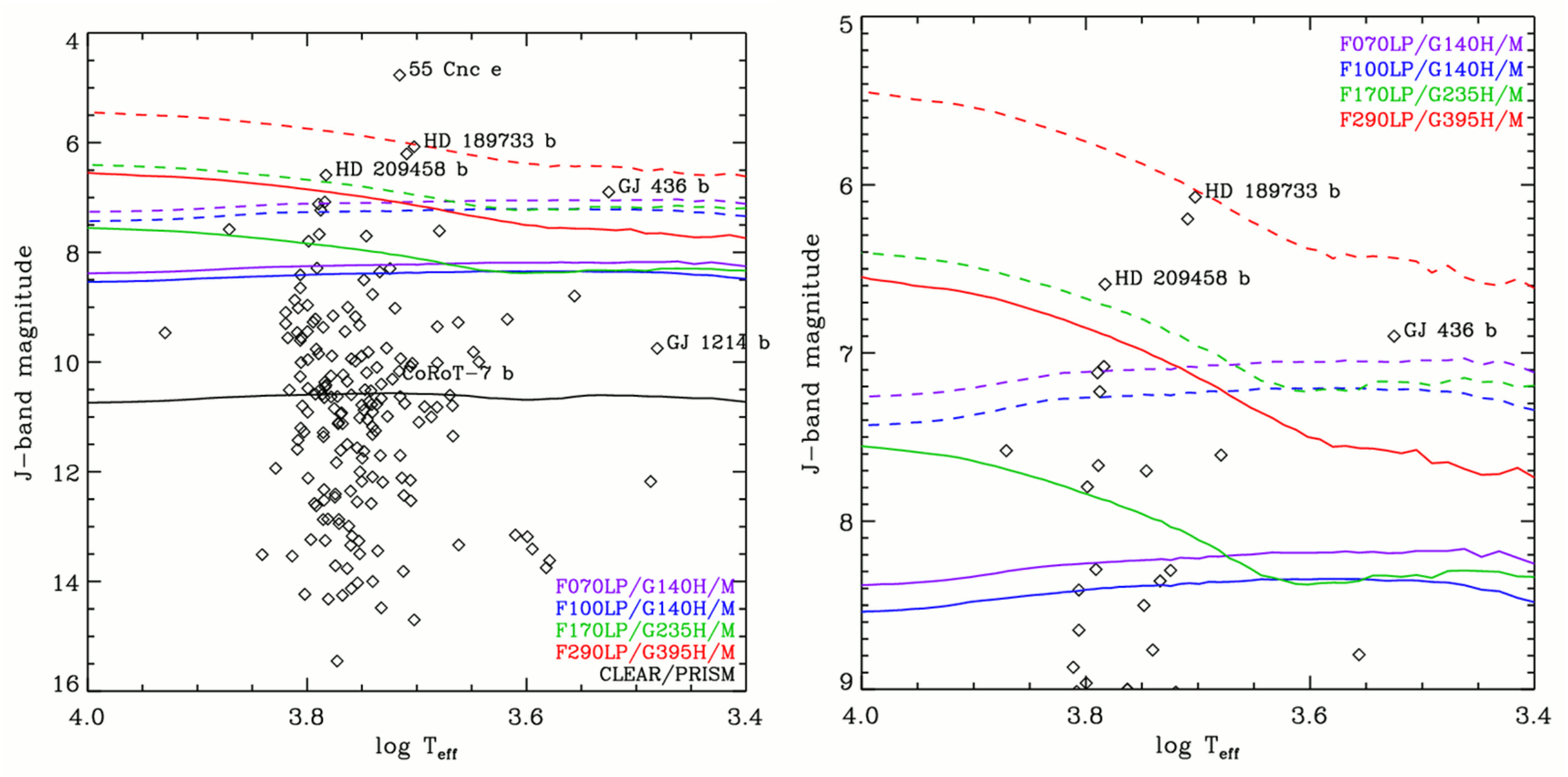

*Figure 30: J-band limiting magnitudes for the different NIRSpec modes as a function of host star temperature (http://www.cosmos.esa.int/web/jwst/exoplanets). The colored dashed lines are for the high resolution gratings, the coloured solid lines for the medium resolution gratings, and the solid black line for the prism. Sources below the lines can be observed in the full wavelength range of the given mode as specified in the table above. The black symbols denote the host stars of known transiting exoplanet.*

## 5.2 EChO & the E-ELT

E-ELT and EChO observations will be highly complementary and mutually beneficial. Ground-based observations of exoplanet atmospheres have many challenges and limitations. Large parts of the electromagnetic spectrum are blocked from view due to absorption and scattering in the Earth's atmosphere. In addition, the thermal background from the sky and telescope are strongly variable, making high-precision ground-based transit or eclipse spectroscopy practically impossible from the ground at >5 micron. However, the E-ELT will be very valuable in specific ways. One particularly successful observing strategy makes use of spectroscopy at a very high dispersion of R=100,000. At this resolution, molecular bands in exoplanet spectra are resolved into hundred(s) to thousands of individual lines, whose signals can be combined to secure a more robust molecular detection. Only astrophysical information over small wavelength scales is preserved, hence the line-contrast is being measured with respect to a local pseudo-continuum. This technique has been used very successfully using the VLT, for both exoplanet transmission spectroscopy (Snellen et al. 2010) and emission spectroscopy (Brogi et al. 2012), and will be more powerful on the next-generation of extra-large telescopes.

E-ELT observations will be highly complementary to EChO. The EChO spectra, which will be obtained over a large instantaneous wavelength range, are crucial for measuring the most important planetary atmosphere parameters − the temperature-pressure profile and the main molecular abundances. With these parameters determined by EChO, high-resolution E-ELT observations, providing planet differential transmission and day-side spectra at specific wavelengths, can be calibrated and used to target other, specific aspects of the planetary atmospheres. For the best observable targets, e.g. those targeted by EChO in the Origin and Rosetta tiers, the E-ELT can provide information on the rotation of the planet and high-altitude wind speeds using the absorption line profiles – important ingredients for global circulation models (e.g. see Showman et al. 2013 for theoretical simulations). Using the high-dispersion technique, the line-contrasts can be measured for a large part of the planet orbit, meaning that variations in molecular abundance ratios (when linked to EChO observations) and/or the atmospheric temperature-pressure profile could be traced from the night, morning, to evening-side of the planet, revealing the influences of possible photo-chemical processes.

| Telescope | Diameter | Instrument | Spectral Range | Instant coverage | spectral dispersion |
|---|---|---|---|---|---|
| E-ELT | 39 m | METIS | 2.9-5.3 μm | 0.1 μm | R=100,000 |
| | | HIRES | 0.4-2.3 μm | 0.4-2.3 μm | R=100,000 |
| | | MOS | 0.4-1.7 μm | 0.4-1.7 μm | R<30,000 |
| GMT | 24.5 m | MOS | 0.4-1.0 μm | 0.4-1.0 μm | R<5000 |
| | | NIR-HRS | 1.0-5.0 μm | TBD | R~50-100,000 |
| | | G-CLEF | 0.4-1.0 μm | 0.4-1.0 μm | TBD |
| TMT | 30 m | WFOS | 0.3-1.0 μm | 0.3-1.0 μm | R<7,500 |
| | | HROS | 0.3-1.0 μm | 0.3 -1.0 μm | R~50-90,000 |
| | | IRMOS | 0.8 - 2.5 μm | 0.3 μm | R=2,000-10,000 |
| | | MIRES | 9-18 μm | 8-14 μm | R=100,000 |
| | | NIRES | 1-5 μm | ~2 μm | R=100,000 |

*Table 11: Planned next-generation telescopes and their instrumentation relevant to transiting exoplanet characterization science. Currently, three next generation telescopes are on the drawing board, the European Extra Large Telescope (E-ELT - http://www.eso.org/public/teles-instr/e-elt.html), the Giant Magellan Telescope (GMT - http://www.gmto.org), and the Thirty-Meter Telescope (TMT - http://www.tmt.org/). Note that at the time of writing, funding has not been completely secured for any of the three telescope projects. The earliest deployment for any of these will be the early 2020s. Also, the instrumentation for the telescopes has by no means been finalised, and a significant fraction of these instruments may never be developed, or change.*

# 6. EChO science beyond exoplanets

In addition to the science of exoplanets, EChO has the capability to make important observations in the field of planetology, stellar physics, disks and brown dwarf studies, exploring a continuum of objects between planets and stars, in particular:

(i) **Stellar physics –** A relevant part of stellar science will come from the activity analysis that is needed to extract the planetary signal. So most of the material is described in the main activity plan.

(ii) **Physics of circumstellar disks around young stars –** a list of accessible objects shows that tens of T Tauri stars are potentially accessible for EChO. Physics of circumstellar disks with spectral variability in the 0.4/11 micron range is of interest for disk astrophysics and planetary systems' formation.

(iii) **Solar System objects –** Planetary objects can be observed with EChO (even with the slit aperture of 2x10 arcsec in the visible channel limiting the FOV) mainly for calibration purpose. Planetary satellites are also good reference objects to observe. This can be done with limited pointing accuracy (~1 arcsec). Comets (if a bright comet is available) can also be observed with EChO.

(iv) **Stellar occultations on Solar System Kuiper Belt Objects** – Planetary occultations can search for atmospheric perturbations during occultation. An occurrence of ~1 event/year for large KBO objects (Pluto, Quaoar, Eris...) is expected. Nevertheless, these occultations are rare.

(v) **Planetary seismology –** Due to the EChO aperture, only Uranus and Neptune are observable. Search for planetary oscillations through long duration continuous spectral observations in the infrared.

(vi) **Brown dwarf observations** – Homogenous sample of brown dwarfs (K=10-15), spanning the range of known spectral types, each observed during one rotational period (typically 10 hours) (Morales et al., 2015).

# 7. Conclusions

Our knowledge of planets other than the eight "classical" Solar System bodies is in its infancy. We have discovered over a thousand planets orbiting stars other than our own, and yet we know little or nothing about their chemistry, formation and evolution. Planetary science therefore stands at the threshold of a revolution in our knowledge and understanding of our place in the Universe: just how special are the Earth and our Solar System? It is only by undertaking a comprehensive chemical survey of the exo-planet zoo that we will answer this critical question.

**Acknowledgments**
We would like to thank all the National Space Agencies who supported the EChO phase-A study.